\documentclass[trackchanges, twocolumn]{aastex701}

\usepackage{amsmath}
\usepackage{graphicx}
\usepackage[caption=false]{subfig}
\usepackage{float}
\usepackage{multirow,array,booktabs}
\usepackage{bm}

\newcommand{\nrc}{NRC Herzberg Astronomy and Astrophysics,
5071 West Saanich Road,
Victoria, BC, V9E 2E7, Canada}
\newcommand{\ucsd}{Department of Astronomy \& Astrophysics,  University of California, San Diego, La Jolla, CA 92093, USA}
\newcommand{\nmsu}{Department of Astronomy, New Mexico State University, 1320 Frenger Mall, Las Cruces, NM 88003, USA}
\newcommand{\uvic}{Department of Physics and Astronomy, University of Victoria,
3800 Finnerty Road, Elliott Building,
Victoria, BC, V8P 5C2, Canada}
\newcommand{\mac}{Department of Physics \& Astronomy, McMaster University, 1280 Main Street West, Hamilton, ON, L8S 4L8, Canada}
\newcommand{\caltech}{Department of Astronomy, California Institute of Technology, Pasadena, CA 91125, USA}

\newcommand{\ucsc}{Department of Astronomy \& Astrophysics, University of California, Santa Cruz, CA 95064, USA}
\newcommand{\ucsb}{Department of Physics, University of California, Santa Barbara, CA 93106, USA}
\newcommand{\ucla}{Department of Earth, Planetary, and Space Sciences, University of California, Los Angeles, CA 90095, USA}

\usepackage{xcolor}

\begin{document}

\title{Astrometric Evidence for Outer Giant Planets in Known Exoplanet Systems}

\author[0000-0001-5173-2947]{Clarissa R. Do \'O}
\affiliation{\caltech}
\email[show]{cdoo@caltech.edu}

\author[0000-0001-5684-4593]{William Thompson}
\affiliation{\nrc}
\email{william.thompson@nrc-cnrc.gc.ca}

\author[0000-0001-9582-4261]{Dori Blakely}
\affiliation{\uvic}
\affiliation{\nrc}
\email{dblakely@uvic.ca}  

\author[0000-0002-1838-4757]{Aniket Sanghi}
\altaffiliation{NSF Graduate Research Fellow}
\affiliation{\caltech}
\email{asanghi@caltech.edu}

\author[0000-0002-6618-1137]{Jerry W. Xuan}
\altaffiliation{Heising-Simons Foundation 51 Pegasi b Fellow}
\affiliation{\ucsb}
\affiliation{\ucla}
\email{jerryxuan@g.ucla.edu}

\author[0000-0002-8895-4735]{Dimitri Mawet}
\affiliation{\caltech}
\affiliation{Jet Propulsion Laboratory, California Institute of Technology, Pasadena, CA 91109, USA}
\email{dmawet@astro.caltech.edu}

\author[0000-0003-3430-3889]{Jessica Speedie}
\altaffiliation{Heising-Simons Foundation 51 Pegasi b Fellow}
\affiliation{Department of Earth, Atmospheric, and Planetary Sciences, Massachusetts Institute of Technology, Cambridge, MA 02139, USA}
\email{jspeedie@mit.edu}

\author[0000-0002-3199-2888]{Sarah Blunt}
\affiliation{\ucsc}
\email{sarah.blunt.3@gmail.com}

\author[0000-0002-9820-1884]{Simon Blouin}
\affiliation{\uvic}
\email{sblouin@uvic.ca}

\author[0000-0002-2696-2406]{Jingwen Zhang}
\affiliation{\ucsb}
\email{jwzhang@ucsb.edu}

\author[0000-0002-6773-459X]{Doug Johnstone}
\affiliation{\nrc}
\affiliation{\uvic}
\email{Douglas.Johnstone@nrc-cnrc.gc.ca}

\author[0000-0003-2233-4821]{Jean-Baptiste Ruffio}
\affiliation{\ucsd}
\email{jruffio@ucsd.edu}

\author[0000-0001-6975-9056]{Eric Nielsen}
\affiliation{\nmsu}
\email{nielsen@nmsu.edu}

\author[0000-0003-2649-2288]{Brendan P. Bowler}
\affiliation{\ucsb}
\email{bpbowler@ucsb.edu}

\author[0000-0001-7402-8506]{Alexandre Bouchard-C\^ot\'e}
\affiliation{Department of Statistics, University of British Columbia, Vancouver, BC V6T 1Z4, Canada}
\email{bouchard@stat.ubc.ca}

\author[0000-0003-4557-414X]{Kyle Franson}
\altaffiliation{NHFP Sagan Fellow}
\affiliation{\ucsc}
\email{kfranson@ucsc.edu}

\author[0009-0008-9687-1877]{William Roberson}
\affiliation{\nmsu}
\email{wcr@nmsu.edu}

\author[0000-0001-5383-9393]{Ryan Cloutier}
\affiliation{\mac}
\email{ryan.cloutier@mcmaster.ca}

\author[0009-0007-6766-2040]{Andre Fogal}
\affiliation{\uvic}
\email{afogal@uvic.ca}

\author[0009-0009-2223-2404]{Kaitlyn Hessel}
\affiliation{\uvic}
\email{khessel@uvic.ca}

\author[0000-0002-4164-4182]{Christian Marois}
\affiliation{\nrc}
\affiliation{\uvic}
\email{Christian.Marois@nrc-cnrc.gc.ca}

\author[0009-0008-1229-3230]{Alexandra Rochon}
\affiliation{\mac}
\email{rochoa3@mcmaster.ca}

\begin{abstract}
Outer giant planets can influence the formation and dynamical evolution of their planetary systems. However, their long orbital periods make them difficult to detect using transit and radial velocity (RV) techniques. Measurements from \textit{Hipparcos}, \textit{Gaia} DR2 and DR3 (G23H) can detect the stellar reflex motion induced by distant companions before the release of \textit{Gaia} DR4 epoch astrometry. Using the G23H catalogue, we identify 29 outer companion candidates ($a_{median}$ $>$ 8.4 AU) in known transiting and RV systems. Starting from 170 exoplanet hosts with
$P(\mathrm{companion})>0.75$ and posterior support for $0.3$--$13\,M_{\rm J}$ companions, we retain 29 systems where each selected component has a median semi-major axis beyond the outermost known planet with every known planet outside its 95\% highest posterior density region in mass and semi-major axis. We incorporate published RV and imaging data from the literature to further constrain the outer candidate's mass and semi-major axis. In eight systems present in the sample, an RV variation or trend has already been reported in the literature. The reported candidates could potentially explain these RV variations, but joint astrometric and RV fits are needed to establish whether the same companions produce both signals. These results provide a set of high priority targets for further RV monitoring and joint orbit analysis with \textit{Gaia} DR4. If these companions are confirmed, their orbital
constraints will provide a starting point for studies of their influence on the dynamical evolution of the known
inner planets.
\end{abstract}

\section{Introduction} \label{sec:intro}
The fourth data release from the \textit{Gaia} spacecraft (hereafter GDR4) is expected to reveal thousands of giant exoplanets \citep{2014ApJ...797...14P,2026AJ....171...18L}. GDR4 is planned for release in December 2026 and will include epoch astrometry for the first 5.5 years of the mission (July 2014 -- January 2020). It will be particularly sensitive to super-Jupiters with masses in the range $3-13 M_{\rm J}$ on orbits between 2--5 AU \citep{2026AJ....171...18L}. 
\par
Giant planets can have a strong impact on the architectures of their planetary systems due to their large masses. Inner planets can have their eccentricities and inclinations dynamically excited and sometimes be destabilized by outer giant companions \citep{2021MNRAS.508..597P,VanZandt2025, 2026ApJ..1006..159L}. Similar processes involving the giant planets have played an important role in shaping the architecture of our own Solar System \citep{2016MNRAS.455.3561K}. The detection of outer giant planets in systems with known inner companions can therefore provide important constraints on the formation and dynamical evolution of planetary systems \citep[e.g.][]{2018AJ....156...92Z,2019AJ....157...52B,Xuan2020,Zhang2025,VanZandt2025}.
\par
Ahead of GDR4, existing measurements from \textit{Gaia} DR2 \citep{2018AA...616A...1G}, \textit{Gaia} DR3 \citep{2023AA...674A...1G}, and \textit{Hipparcos} \citep{1997ESASP1200.....E} can provide evidence for giant companions around nearby stars. These calibrated astrometric and radial velocity measurements can reveal deviations from the motion expected for a single star, providing hints of the presence of companions. These  measurements include the \textit{Hipparcos}--\textit{Gaia} proper motion anomaly \citep{2018ApJS..239...31B,2019AA...623A..72K,2021ApJS..254...42B}, the \textit{Hipparcos} intermediate astrometric data \citep{2007AA...474..653V,2020AJ....159...71N}, differences between the GDR2 and GDR3 astrometric solutions, \textit{Gaia} radial velocity variability \citep{2025ApJ...992..131C}, and \textit{Gaia} astrometric excess noise modeling \citep{2025AA...702A..76K}.
\par
Several recent surveys have specifically searched for sub-stellar companions using these calibrated data. For instance, \citet{2024AJ....167...89Z,2026AJ....171...77Z} selected \textit{Kepler}, K2, and TESS planet hosts with significant \textit{Hipparcos}--\textit{Gaia} proper motion anomalies and used follow-up high contrast imaging and radial velocity observations to determine whether the astrometric signals were caused by additional companions. \citet{2025AJ....169..235V} conducted a homogeneous radial velocity survey of 47 Sun-like stars hosting inner transiting planets, incorporating HGCA astrometry and imaging constraints to classify companions inferred from long-term radial velocity trends. Other surveys have searched for cold Jupiters using radial velocities alone. For example, \citet{2023AA...677A..33B} analyzed nearly ten years of HARPS-N observations for 38 \textit{Kepler} and K2 small planet systems. These surveys have begun to characterize the occurrence and properties of outer companions in systems with inner planets, but their samples are selected through significant HGCA accelerations or the availability of long baseline radial velocity observations.
\par
Most recently, \citet{Thompson2026} cross-calibrated the measurements of Gaia and Hipparcos and combined them into the \textit{Gaia} DR2, DR3 and \textit{Hipparcos} catalog (G23H). The G23H framework jointly models measurements spanning different observing baselines to produce posterior constraints on companion masses and orbits using existing published data. This provides a complementary approach to previous surveys. Rather than selecting targets from HGCA proper motion anomalies alone or obtaining a uniform set of new radial velocity observations, we use the joint G23H posteriors to search for possible outer companions across a broad population of known exoplanet hosts. We then use published radial velocity and imaging data to further constrain the parameter space provided by the G23H.
\par
In this work, we identify 29 known exoplanet systems detected through transit or radial velocity for which the G23H catalogue identifies an astrometric signal consistent with a possible additional outer giant planet. We consider each system individually and compare the G23H posterior distribution of companion mass and semi-major axis with the published planetary architecture and constraints from the literature, particularly the time baselines, precision, and any detected trends of existing radial velocity observations. We aim to separate the signals induced by known planets or stellar companions from those still compatible with previously unknown outer giant planets in order to determine the most promising targets for joint astrometric and radial velocity orbit fitting and follow-up with radial velocity monitoring and direct imaging.

\section{Methods} \label{sec:methods}

Here we briefly describe the construction of the G23H
catalog and its orbital posteriors
(Section~\ref{sec:g23h}). A detailed description is
provided by \citealt{Thompson2026} and Thompson et al 2026b (in prep). We then describe how
we selected our sample from the G23H
(Section~\ref{sec:sample_selection}).
Sections~\ref{sec:rv_limits} and
\ref{sec:imaging_limits} describe the RV and imaging
sensitivity maps overlaid on the astrometric posteriors used to shade the figures presented in
Section~\ref{sec:results}.

\subsection{G23H}
\label{sec:g23h}

\citet{Thompson2026} constructed the G23H catalog by placing the astrometric measurements from \textit{Hipparcos}, GDR2, and GDR3 into a common GDR3 reference frame. This allows measurements from the three catalogs to be modeled together, yielding a combination of the long baseline between Hipparcos and Gaia with the shorter timescales probed by GDR2 and GDR3. At the \textit{Hipparcos} epoch, G23H uses the \textit{Hipparcos} proper motions and the long-baseline \textit{Hipparcos}--\textit{Gaia} measurements from the EDR3 version of the Hipparcos--Gaia Catalog of Accelerations. The GDR2 proper motions and the proper motions measured from the change in position between GDR2 and GDR3 were calibrated independently. These calibrations account for underestimated uncertainties and correlations introduced by observations shared between GDR2 and GDR3. The calibrated proper motions are then combined with the GDR3 astrometric excess noise and GDR3 radial velocity variability. The astrometric excess noise is incorporated through a calibrated likelihood based on the unbiased estimator of variance (UEVA) \citep{2025AA...702A..76K}, accounting for the noise expected for single stars.
\par
The orbital models in G23H were calculated using \texttt{Octofitter} \citep{2023AJ....166..164T}. For each proposed orbit, the motion induced by the companion was modeled over the \textit{Hipparcos}, GDR2, and GDR3 observing windows. The predicted motion was then compared with the full set of G23H measurements. \citet{Thompson2026} generated predicted observing times and scan angles and marginalized over which observations entered each solution. The resulting uncertainty is therefore already incorporated into the orbital posteriors. G23H provides posterior distributions for the companion mass and orbital parameters. \par
Recently, Thompson et al 2026b (in prep) incorporated additional capabilities to the posterior computation, including the light contributed by luminous companions in the astrometric motion modeling, modeling multiple companions, and matching companions with common proper motions. Here, we present these novel posteriors in companion mass and semi-major axis. Throughout this work, $M$ denotes the mass of the candidate companion inferred from G23H, while $M_\star$ denotes the host star mass. Subscripts identify specific known planets. For example, $M_b$ and $M_c$ denote the masses of planets b and c.
\par

\subsection{Sample Selection}
\label{sec:sample_selection}

Our initial sample comprised 170 systems in the G23H. We selected systems that are known exoplanet hosts with $P(\mathrm{companion})>0.75$ (that is, BF $>$ 3) and with posterior support for a companion mass between $0.3$ and 13$\,M_{\rm J}$. This mass requirement selects signals consistent with a giant planet candidate. The candidate did not need to have its median mass or the entire credible interval contained within this range in order to be selected. 
We cross-matched the host identifiers with the NASA Exoplanet Archive \citep{2025PSJ.....6..186C}. Published solutions for known planetary masses and semi-major axes were obtained from the Archive and the individual literature sources considered for this work. We used published minimum masses or masses (when available). In five systems, a measured mass or minimum mass was not reported, so we did not include them in our sample, leaving 165 systems with a usable mass measurement. Because the model we adopted from Thompson et al 2026b (in prep) can fit more than one companion in a system, we assessed each companion from the fits separately. For each qualifying companion, we used all of the valid posterior draws, including those with a mass larger than the giant planet range. The comparison therefore preserves posteriors in the brown dwarf or stellar mass regimes.
\par
An astrometric signal in a known planetary system may come from a previously known planet instead of an additional companion. We compared the joint mass and semi-major axis posterior of each eligible G23H component with the published masses and semi-major axes of the known planets in order to constrain whether the signal is consistent with a known planet. For planets with measured minimum masses, we calculated $M\sin i$ using the mass and inclination of each G23H posterior draw. When a published true mass was available, we compared it directly with the G23H mass posterior. We did not include the known planet parameters' uncertainties.
\par
We estimated the posterior density in
$(\log a,\log M\sin i)$ space for minimum mass
measurements and in $(\log a,\log M)$ space for
true mass measurements. For each known planet $k$,
$C_k$ is the fraction of the posterior enclosed by
the contour that passes through its published location.
For example, a planet outside the 95\% contour
has $C_k>0.95$. Intuitively, this region begins at the densest part of the posterior and ends at the location of the known planet. A small $C_k$ places the known planet in a high posterior density region, while a value near 1 places it in the low density tail of the posterior.
\par
For each fitted component, the known planet with
the greatest agreement with the posterior determines
the association metric:
\begin{equation}
    C_{\rm known}=\min_k C_k.
\end{equation}
We required $C_{\rm known}>0.95$, such that every
assessed known planet remained outside the component's
95\% highest posterior density region. At least one companion component satisfied this condition in 38 of the 165 remaining systems. An example of this density calculation (along with a visualization plot) is presented in Appendix~\ref{app:known_planet_association}.
\par
To select signals exterior to a known inner planet,
we also required the median semi-major axis
of the candidate to exceed the outermost known planet's semi-major axis. Four systems failed this condition, leaving 34 systems that satisfied the selection.

\begin{figure*}[t]
    \centering
    \includegraphics[width=\textwidth]{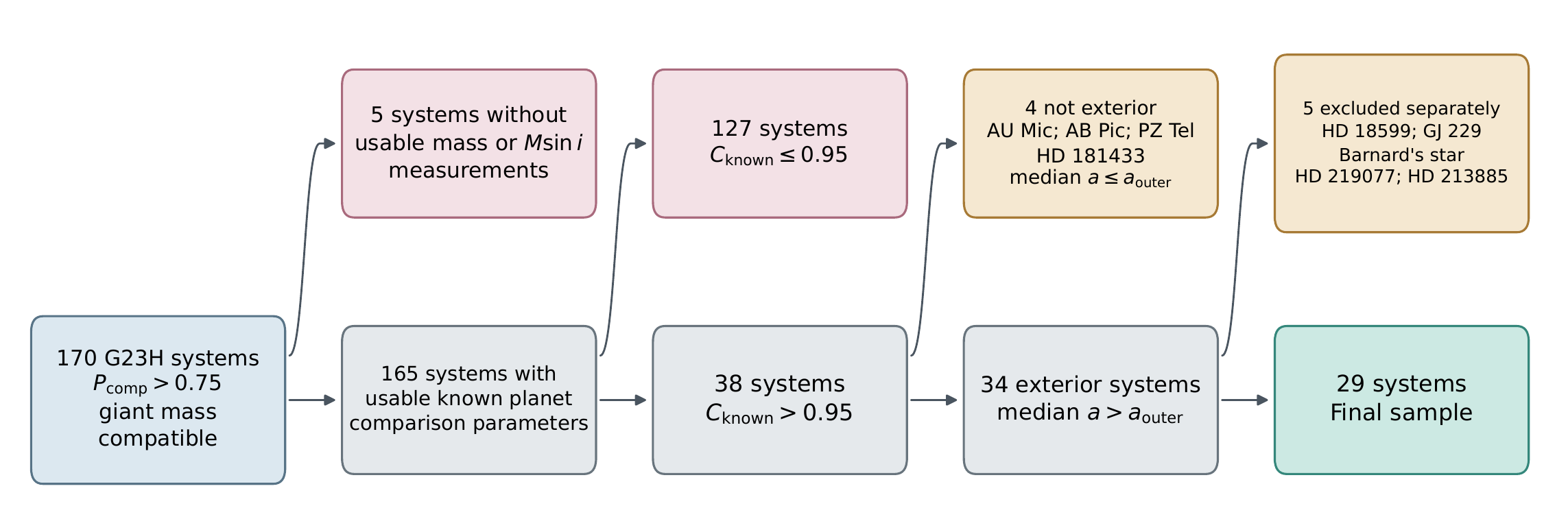}
    \caption{Selection of the final sample of 29 systems from
170 G23H systems with evidence for a giant planet companion. Lower branches show the systems retained at each step and upper branches
show those removed. Systems proceeding to the
posterior comparison have at least one known planet
with a measured mass or minimum mass and semi-major axis. At least one eligible fitted component must satisfy both $C_{\rm known}>0.95$ and
$\mathrm{median}(a)>a_{\rm outer}$, where $a_{\rm outer}$ is the semi-major axis of the outermost assessed known planet. Additional exclusions were applied separately from these numerical criteria.}
    \label{fig:sample_selection}
\end{figure*}
\par
Finally, we omitted five systems after assessing the  posteriors in context with their known architectures. HD~18599 hosts an imaged companion near the boundary between a brown dwarf and a low-mass star \citep{2023AA...675A.158D}. For that reason, its contribution must be accounted for before interpreting the G23H signal as evidence for an additional planet. HD~213885 has two reported stellar companions at projected separations of approximately 23.6 and 55.1~AU \citep{2021AJ....162...75L,2024AA...689A.302G}. These companions were not included in our comparison with the known planets and could contribute to the detected astrometric signal. GJ~229 has a known brown dwarf binary companion \citep{2024Natur.634.1070X} with a visual orbit listed in the Sixth Catalog of Orbits of Visual Binary Stars (ORB6). HD~181433 contains two known outer giant planets \citep{2009AA...496..527B}. Our posteriors do not test whether their combined astrometric perturbations could explain the recovered signal, so we defer this system to a later analysis that accounts for both giants separately. For HD~219077, joint radial velocity and astrometric fits already associate the acceleration with a known massive companion \citep{2023AA...670A..65P}, so we do not include the system in this analysis. We also excluded Barnard's star from the sample as its astrometric signal requires additional analysis, which will be deferred to a later study.
\par
This sequence of exclusions leaves 29 systems in the final sample. Table~\ref{tab:sample_summary} lists their names, alternative identifiers, companion probabilities, and inferred companion masses and semi-major axes. The sample selection is summarized in
Figure~\ref{fig:sample_selection}'s schematic.

\startlongtable
\begin{deluxetable*}{lllccccccccc}
\tablecaption{The 29 systems in the final sample\label{tab:sample_summary}}
\tablewidth{0pt}
\tabletypesize{\scriptsize}
\setlength{\tabcolsep}{2.5pt}
\tablehead{
\colhead{System} &
\colhead{Alt. ID} &
\colhead{Sp. type} &
\colhead{$\varpi$} &
\colhead{$V$} &
\colhead{$M_\star$} &
\colhead{$N_{\rm pl}$} &
\colhead{$P_{\rm comp}$} &
\colhead{$C_{\rm known}$} &
\colhead{$M$} &
\colhead{$a$} &
\colhead{Fig.} \\
\colhead{} &
\colhead{} &
\colhead{} &
\colhead{(mas)} &
\colhead{(mag)} &
\colhead{($M_\odot$)} &
\colhead{} &
\colhead{} &
\colhead{} &
\colhead{($M_{\rm J}$)} &
\colhead{(AU)} &
\colhead{} \\
\colhead{(1)} &
\colhead{(2)} &
\colhead{(3)} &
\colhead{(4)} &
\colhead{(5)} &
\colhead{(6)} &
\colhead{(7)} &
\colhead{(8)} &
\colhead{(9)} &
\colhead{(10)} &
\colhead{(11)} &
\colhead{(12)}
}
\startdata
BD$+55\,362$ & HIP~7441 & K3 & $19.030\pm0.014$ & \nodata & $0.91^{+0.05}_{-0.08}$ & 1 & 0.9009 & 0.9545 & $8.8^{+84.9}_{-6.0}$ & $15.9^{+56.2}_{-13.0}$ & \ref{fig:HIP7441} \\
HD~45652 & HIP~30905 & K5 & $28.648\pm0.023$ & \nodata & $0.97^{+0.05}_{-0.09}$ & 1 & 1.0000 & 0.9924 & $4.9^{+34.1}_{-2.4}$ & $17.6^{+44.1}_{-12.7}$ & \ref{fig:HIP30905} \\
HD~63454 & HIP~37284 & K3Vk: & $26.554\pm0.011$ & $9.36\pm0.02$ & $0.82^{+0.04}_{-0.09}$ & 1 & 1.0000 & 0.9976 & $9.2^{+52.6}_{-5.8}$ & $20.7^{+41.4}_{-14.6}$ & \ref{fig:HIP37284} \\
$\nu^2$~Lupi & HIP~75181 & G2-V & $67.847\pm0.060$ & $5.65$ & $1.05^{+0.07}_{-0.10}$ & 3 & 1.0000 & 0.9998 & $2.6^{+8.7}_{-1.1}$ & $8.4^{+30.3}_{-6.2}$ & \ref{fig:HIP75181} \\
GJ~682 & HIP~86214 & M3.5V & $199.694\pm0.031$ & $10.946$ & $0.31^{+0.02}_{-0.02}$ & 2 & 0.8318 & 0.9789 & $0.58^{+9.17}_{-0.42}$ & $8.8^{+61.0}_{-8.4}$ & \ref{fig:HIP86214} \\
HIP~91258 & BD$+61\,1762$ & G5 & $21.763\pm0.017$ & \nodata & $0.97^{+0.05}_{-0.08}$ & 1 & 1.0000 & 0.9980 & $10.7^{+60.0}_{-6.2}$ & $20.0^{+36.5}_{-14.7}$ & \ref{fig:HIP91258} \\
BD$+14\,4559$ & HIP~104780 & K5 & $20.284\pm0.017$ & \nodata & $0.87^{+0.04}_{-0.07}$ & 1 & 0.9990 & 0.9580 & $7.9^{+67.9}_{-5.0}$ & $21.6^{+56.4}_{-16.0}$ & \ref{fig:HIP104780} \\
HD~208897 & HIP~108513 & K0 & $14.959\pm0.023$ & $6.510\pm0.009$ & $1.25^{+0.31}_{-0.18}$ & 1 & 1.0000 & 0.9658 & $11.8^{+111.5}_{-6.6}$ & $21.3^{+65.4}_{-15.4}$ & \ref{fig:HIP108513} \\
HIP~5763 & BD$+15\,176$ & K6V & $31.237\pm0.032$ & $9.91\pm0.05$ & $0.75^{+0.04}_{-0.11}$ & 1 & 1.0000 & 0.9990 & $7.0^{+1.6}_{-1.3}$ & $2.5^{+3.2}_{-0.7}$ & \ref{fig:HIP5763} \\
BD$+45\,564$ & HIP~10245 & K1 & $18.742\pm0.016$ & \nodata & $0.87^{+0.05}_{-0.09}$ & 1 & 0.9990 & 0.9531 & $9.5^{+68.7}_{-5.7}$ & $19.7^{+64.1}_{-14.9}$ & \ref{fig:HIP10245} \\
GJ~96 & HIP~11048 & M0V & $83.661\pm0.026$ & \nodata & $0.62^{+0.02}_{-0.10}$ & 1 & 0.9761 & 0.9842 & $1.3^{+10.2}_{-0.8}$ & $6.3^{+55.1}_{-5.4}$ & \ref{fig:HIP11048} \\
HD~16417 & HIP~12186 & G1V & $39.295\pm0.030$ & $5.79$ & $1.29^{+0.06}_{-0.09}$ & 1 & 0.8579 & 0.9804 & $1.6^{+12.4}_{-1.0}$ & $17.0^{+75.5}_{-14.1}$ & \ref{fig:HIP12186} \\
HD~20794 & HIP~15510 & G6V & $165.524\pm0.078$ & $4.27$ & $0.95^{+0.07}_{-0.09}$ & 3 & 1.0000 & 0.9973 & $1.4^{+5.3}_{-0.6}$ & $8.3^{+47.2}_{-6.2}$ & \ref{fig:HIP15510} \\
HD~28471 & HIP~20625 & G5V & $22.870\pm0.019$ & $7.91$ & $1.08^{+0.06}_{-0.09}$ & 3 & 0.9546 & 0.9987 & $4.2^{+19.1}_{-2.2}$ & $6.3^{+41.7}_{-4.3}$ & \ref{fig:HIP20625} \\
Kapteyn's star & HIP~24186 & M1VIp & $254.199\pm0.017$ & $8.853$ & $0.31^{+0.02}_{-0.02}$ & 1 & 0.9995 & 0.9934 & $1.5^{+17.8}_{-1.4}$ & $31.4^{+79.8}_{-26.1}$ & \ref{fig:HIP24186} \\
LHS~1903 & HIP~34730 & M0.5V & $28.052\pm0.024$ & $12.21$ & $0.57^{+0.02}_{-0.10}$ & 4 & 0.7961 & 0.9607 & $7.6^{+31.3}_{-5.7}$ & $6.9^{+39.2}_{-6.7}$ & \ref{fig:HIP34730} \\
HD~63433 & HIP~38228 & G5V & $44.685\pm0.023$ & $6.91\pm0.01$ & $0.97^{+0.07}_{-0.09}$ & 3 & 0.8579 & 0.9849 & $3.5^{+39.0}_{-2.7}$ & $21.0^{+93.3}_{-18.7}$ & \ref{fig:HIP38228} \\
HD~67200 & HIP~39242 & F6V & $18.370\pm0.014$ & $7.70\pm0.01$ & $1.23^{+0.06}_{-0.13}$ & 2 & 0.9700 & 0.9924 & $5.2^{+55.8}_{-2.9}$ & $19.4^{+76.3}_{-15.3}$ & \ref{fig:HIP39242} \\
HD~90156 & HIP~50921 & G5V & $45.563\pm0.021$ & $6.92$ & $0.95^{+0.08}_{-0.10}$ & 1 & 0.9546 & 0.9918 & $2.6^{+15.4}_{-1.9}$ & $11.6^{+61.6}_{-10.6}$ & \ref{fig:HIP50921} \\
HD~96735 & HIP~54491 & F8 & $16.069\pm0.019$ & $8.97\pm0.02$ & $0.99^{+0.08}_{-0.10}$ & 1 & 0.8823 & 0.9950 & $9.0^{+90.8}_{-6.4}$ & $13.1^{+79.2}_{-12.1}$ & \ref{fig:HIP54491} \\
GJ~514 & HIP~65859 & M1.0V & $131.101\pm0.027$ & $9.029$ & $0.54^{+0.02}_{-0.08}$ & 1 & 0.7981 & 0.9645 & $0.55^{+3.60}_{-0.34}$ & $8.9^{+58.2}_{-7.3}$ & \ref{fig:HIP65859} \\
HD~149143 & HIP~81022 & G3V & $13.628\pm0.025$ & $7.89\pm0.01$ & $1.34^{+0.09}_{-0.13}$ & 1 & 1.0000 & 0.9919 & $12.5^{+109.1}_{-6.5}$ & $19.2^{+60.6}_{-13.7}$ & \ref{fig:HIP81022} \\
GJ~674 & HIP~85523 & M3V & $219.646\pm0.026$ & $9.407$ & $0.40^{+0.02}_{-0.02}$ & 1 & 0.9954 & 0.9990 & $0.39^{+2.10}_{-0.23}$ & $3.3^{+31.5}_{-2.4}$ & \ref{fig:HIP85523} \\
HD~177565 & HIP~93858 & G6V & $58.986\pm0.038$ & $6.16$ & $1.03^{+0.05}_{-0.10}$ & 1 & 0.9587 & 0.9893 & $1.8^{+7.7}_{-1.3}$ & $4.1^{+43.4}_{-3.4}$ & \ref{fig:HIP93858} \\
HD~179079 & HIP~94256 & G5IV & $14.319\pm0.025$ & $7.96\pm0.01$ & $1.30^{+0.13}_{-0.12}$ & 1 & 0.9294 & 0.9829 & $7.2^{+71.7}_{-4.4}$ & $16.1^{+73.7}_{-13.2}$ & \ref{fig:HIP94256} \\
HIP~109384 & BD$+70\,1218$ & G5 & $16.912\pm0.012$ & \nodata & $0.88^{+0.07}_{-0.08}$ & 1 & 1.0000 & 0.9573 & $9.2^{+54.6}_{-3.9}$ & $13.5^{+44.6}_{-9.9}$ & \ref{fig:HIP109384} \\
HD~216770 & HIP~113238 & G9VCN+1 & $27.292\pm0.025$ & $8.10$ & $0.98^{+0.06}_{-0.08}$ & 1 & 0.9553 & 0.9670 & $4.6^{+33.3}_{-2.6}$ & $13.4^{+52.0}_{-10.6}$ & \ref{fig:HIP113238} \\
GJ~887 & HIP~114046 & M2V & $304.135\pm0.020$ & $7.39\pm0.01$ & $0.53^{+0.02}_{-0.03}$ & 4 & 0.8386 & 0.9840 & $0.34^{+2.52}_{-0.27}$ & $14.1^{+81.6}_{-13.2}$ & \ref{fig:HIP114046} \\
HD~224693 & HIP~118319 & G2V & $10.580\pm0.028$ & $8.22\pm0.01$ & $1.41^{+0.10}_{-0.15}$ & 1 & 0.9534 & 0.9698 & $9.4^{+120.1}_{-5.5}$ & $14.4^{+105.1}_{-11.4}$ & \ref{fig:HIP118319} \\
\enddata
\tablecomments{
Columns (3)--(5) contain the adopted values in the SIMBAD
database \citep{2000AAS..143....9W}. Seven systems lack an adopted
$V$ magnitude in SIMBAD and are marked \nodata.
(6) Stellar mass from the marginal
\texttt{M\_pri} samples in the G23H. These values
may differ from the literature estimates quoted in the
target descriptions.
(7) Number of known planets taken from the NASA Exoplanet
Archive \texttt{pscomppars} table.
(8) $P_{\rm comp}\equiv P(\mathrm{companion})$, the probability
of the most probable non-CPM signal components.
(9) $C_{\rm known}$, the minimum highest posterior density
inclusion level among the assessed known planets.
All listed systems satisfy $C_{\rm known}>0.95$.
(10) Candidate companion mass inferred from G23H.
(11) Candidate companion semi-major axis inferred from G23H.
(12) Reference to the individual posterior figure.
The values in columns (6), (10), and (11) are posterior medians
with 16th--84th percentile intervals.
}
\end{deluxetable*}

\par

\subsection{Radial Velocity Limits}
\label{sec:rv_limits}

We make RV sensitivity contours to show which portions
of the astrometric posteriors are constrained by published
radial velocity observations. These contours are overlaid
on the mass and semi-major axis distributions in
Section~\ref{sec:results} and indicate where an additional
companion would be expected to produce detectable
velocity variations. We compiled the published observations for each
system from the literature in order to estimate which astrometric companions could have produced detectable radial velocity variations. We recorded the instrument used, observing epochs, and time baseline. We adopted the post-fit residual RMS ($\sigma_{\rm eff}$) as the effective uncertainty when available. If that value was not available, we used another explicitly identified noise proxy from the references in the literature. 
\par
Our RV sensitivty proxy calculation is inspired by \citet{2019AJ....157..252K}, who translated an observed radial velocity range into limits on companion mass and semi-major axis by treating $\Delta{\rm RV}/2$ as a lower limit on the RV semi-amplitude and marginalizing over unknown orbital geometry (see their Section~5). We forward model the eccentric Keplerian signal of a hypothetical companion at the published observing epochs and compare the sampled velocity variation to the reported RV scatter.
We evaluated a grid in companion mass, $M$, and
semi-major axis, $a$. The orbital period at each grid point was calculated
using the full two-body form of Kepler's third law,
\begin{equation}
P =
1~{\rm yr}
\left(\frac{a}{\rm AU}\right)^{3/2}
\left(\frac{M_\star+M}{M_\odot}\right)^{-1/2}.
\label{eq:rv_proxy_period}
\end{equation}
We retained the companion mass in the total system mass rather than assuming
$M\ll M_\star$so the calculation can be applied to
all posterior samples, including those in the brown dwarf and stellar regimes.
\par
At every mass--semi-major axis grid point, eccentricities were drawn from a uniform distribution (0, 0.99). We also drew an
initial mean anomaly, $\mathcal{M}_0$, and argument of periastron, $\omega$,
uniformly between zero and $2\pi$. Inclinations were drawn isotropically ($\cos i$ uniform). The full eccentric RV semi-amplitude was
calculated as
\begin{equation}
\begin{split}
K ={}& 28.4329~\mathrm{m\,s^{-1}}
\left(\frac{M}{M_{\mathrm{J}}}\right)
\left(\frac{P}{\mathrm{yr}}\right)^{-1/3} \\
&\times
\left(\frac{M_\star+M}{M_\odot}\right)^{-2/3}
\frac{\sin i}{\sqrt{1-e^2}} .
\end{split}
\label{eq:rv_proxy_amplitude}
\end{equation}
At each published observing time, $t_j$, the mean anomaly was propagated
according to
$\mathcal{M}_j=\mathcal{M}_0+2\pi(t_j-t_0)/P$. We solved Kepler's equation,
$\mathcal{M}_j=E_j-e\sin E_j$, converted the eccentric anomaly $E_j$ to the
true anomaly $\nu_j$, and evaluated
\begin{equation}
v_j =
K\left[\cos(\nu_j+\omega)+e\cos\omega\right].
\label{eq:rv_proxy_signal}
\end{equation}
For every grid point, we evaluated 10,000 realizations of $e$,
$\mathcal{M}_0$, $\omega$, and $i$ to account for different orbital plane
configurations and for observations sampling different portions of the
orbit.
\par
For each realization, we measured the largest velocity difference sampled
within a single instrument. Measurements from different instruments were not
compared directly because their independent velocity zero points could mimic
a long period variation. The sensitivity assigned to a grid point was the
fraction of realizations where
\begin{equation}
\begin{split}
f_{\mathrm{sens}}(M,a)
    &= \frac{N_{\mathrm{det}}}{N_{\mathrm{geom}}},\\
N_{\mathrm{det}}
    &= N\!\left(
    \Delta\mathrm{RV}
    > \sqrt{2}\,\sigma_{\mathrm{eff}}
    \right).
\end{split}
\label{eq:rv_proxy_sensitivity}
\end{equation}
The factor of $\sqrt{2}$ represents the uncertainty of a difference between
two independent measurements with the same uncertainty. We display regions
with $f_{\rm sens}=0.1$--$0.5$, $0.5$--$0.9$, and
$f_{\rm sens}\geq0.9$ using progressively darker shading. These percentages
represent the fraction of eccentricities and orbital geometries that produce
a sampled velocity difference exceeding the adopted threshold.
\par
When individual observing epochs and a published RV noise estimate were
available, we evaluated every simulated orbit at the actual observing dates
and calculated a cadence-specific sensitivity map. When the recoverable observing dates
represented only a subset of the complete published data set, we used that
subset and identified the resulting map as a conservative published
subset.
\par
The resulting sensitivity contours will be overlaid on
the astrometric posteriors in Section~\ref{sec:results}.
They indicate the fraction of trial configurations whose
sampled RV variation exceeds the adopted threshold based
on the effective scatter. They should not be interpreted
as regions rigorously excluded by the published data.
Formal completeness limits would require injection and
recovery using the original RV measurements and likelihood
model \citep[e.g.,][]{2004MNRAS.354.1165C}.
The adopted radial velocity inputs and their literature
sources are summarized in Appendix~\ref{app:rv_inputs}
and Table~\ref{tab:rv_inputs_two_systems}. A worked example
of the sensitivity calculation, along with a visualization plot, is presented in Appendix~\ref{app:rv_sensitivity_example}.
\par
\subsection{Imaging Limits}
\label{sec:imaging_limits}
We incorporated quantitative imaging constraints for six
of the 29 systems in our sample. We only calculated an imaging sensitivity map when the literature provided either a numerical contrast curve or a
companion mass completeness map that could be recovered quantitatively.
The adopted sensitivity products and references are summarized in
Table~\ref{tab:imaging_inputs}. The presence of archival imaging without
a quantitative sensitivity product was not treated as a constraint.
\par
When a numerical contrast curve was available, we converted companion
mass into predicted contrast using the adopted system age, distance, and
host star magnitude. We used the Sonora Red Diamondback v2 evolutionary
models for substellar companions and the BHAC15 models above
$0.08\,M_\odot$
\citep{2025ApJ...994..198D,2015AA...577A..42B}. We only evaluated the imaging sensitivity above the lower mass boundary of the available model grid. At a system age $\tau_\star$, the host star absolute
magnitude and predicted companion contrast are
\begin{equation}
\begin{split}
M_{\lambda,\star}
    &= m_{\lambda,\star}
       -5\log_{10}\left(\frac{d}{10~{\rm pc}}\right),\\
\Delta m_\lambda(M,\tau_\star)
    &= M_{\lambda,\mathrm{c}}(M,\tau_\star)
       -M_{\lambda,\star},
\end{split}
\label{eq:imaging_contrast}
\end{equation}
where $M_{\lambda,\mathrm{c}}$ is the model absolute magnitude of the
companion. We used the closest available photometric band when the exact observing filter was not available in the evolutionary grids. The filter approximations are reported in Table \ref{tab:imaging_inputs}. We tested the sensitivity of the contrast curve on the same logarithmic grid of companion mass and semi-major axis as was used for the radial velocity calculation.
At each grid point, 10,000 orbital geometries were sampled. Mean anomaly, argument of periastron, and $\cos i$ were drawn uniformly to represent isotropic orbital orientations. Eccentricity was uniformly drawn between $0\leq e\leq0.99$. For each realization we also drew an age within the reported uncertainties. After numerically solving Kepler's equation, the projected angular
separation was calculated as
\begin{equation}
\begin{aligned}
\rho_j &= \frac{a(1-e_j\cos E_j)}{d} \\
&\quad \times
\left[
\cos^2(\omega_j+\nu_j)
+\cos^2 i_j\,\sin^2(\omega_j+\nu_j)
\right]^{1/2},
\end{aligned}
\label{eq:imaging_projected_separation}
\end{equation}
where $a$ is in AU, $d$ is in pc, and $\rho_j$ is in
arcseconds. 
\par
We define the imaging sensitivity as the fraction of trial
configurations that would have been detectable:
\begin{equation}
f_{\rm img}(M,a)
    = \frac{N_{\rm detectable}}{N_{\rm geom}},
\label{eq:imaging_sensitivity}
\end{equation}
where $N_{\rm geom}=10{,}000$. A trial is detectable when
the astrometric companion falls above the contrast curve limit at its predicted magnitude and separation,
$\Delta m_{\lambda,j}\leq
\Delta m_{\lambda,\mathrm{lim}}(\rho_j)$.
We used published completeness maps directly when available. In particular, the MESS3 maps of \citet{2023AA...680A..64D} already incorporate the parameter degeneracies and evolutionary models. We digitized the published completeness
boundaries using the \texttt{plotdigitizer} Python package and retained their original companion mass and semi-major axis coordinates. 
In Section~\ref{sec:results}, we overlay the imaging
sensitivity regions on the astrometric posteriors using
aqua shading. Progressively darker shades indicate sensitivities of 10--50\%, 50--90\%, and $\geq90\%$. Where only the
50\% and 90\% boundaries are available, we show the
50--90\% and $\geq90\%$ regions. A worked example of the imaging sensitivity calculation
is presented in Appendix~\ref{app:imaging_sensitivity_example}.

\section{Results} \label{sec:results}

\begin{figure*}[p]
    \centering
    \includegraphics[width=\textwidth,height=0.8\textheight,keepaspectratio]{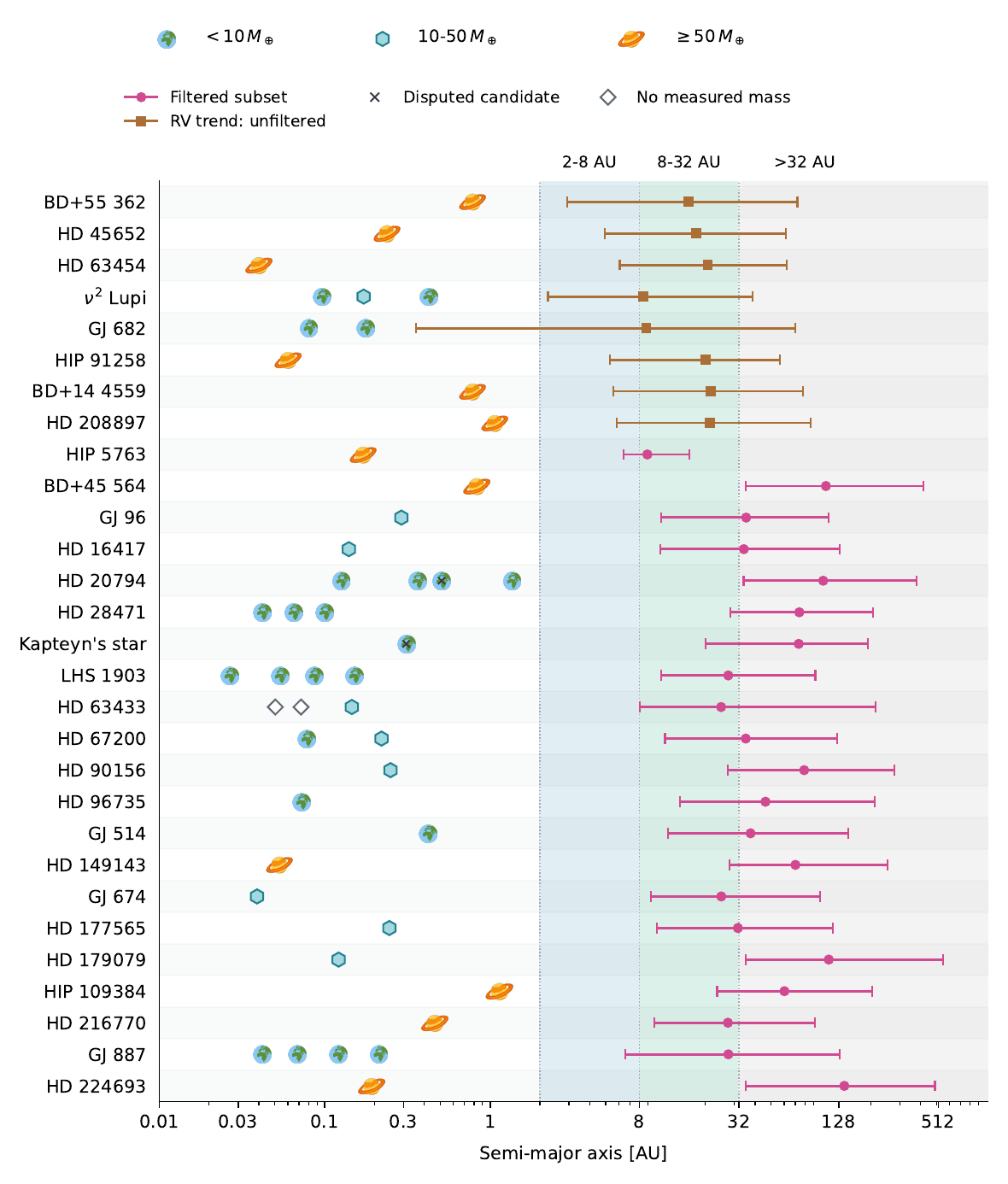}
    \caption{Orbital scales of the 29 astrometric candidates and previously reported planets. Pink circles and bars indicate medians and 16th--84th percentiles after keeping draws with RV sensitivity below 50\% and imaging sensitivity below 50\%. Brown squares and bars show the medians and 16th–84th percentiles of the full astrometric posteriors for the eight systems with RV trends. Planet icons indicate published semi-major axes for the known planets from the Exoplanet Archive. Disputed candidates are marked with crosses. The blue and green bands represent the 2--8 and 8--32~AU CLS comparison intervals \citep{2021ApJS..255...14F}. The gray band denotes $a>32$~AU.}
    \label{fig:sample_sma_comparison}
\end{figure*}

We present the individual systems in two groups. We first
discuss the eight systems with known or tentative
RV variations from the literature. We then discuss the
remaining 21 system where no such variation has
been identified in the literature.
Table~\ref{tab:sample_summary} provides an overview of
the sample and identifies the corresponding figure
for each system.
Figure~\ref{fig:sample_sma_comparison} compares the
orbital scales of the candidate companions with those
of the previously reported planets.
For the 21 systems without reported RV trends, we
show the astrometric samples that fall outside regions
with at least 50\% RV or imaging sensitivity (when available).
For the eight systems with reported RV trends,
we show the full astrometric posterior.
Systems in both the table and the overview figure
follow their order of appearance in the following sub-sections.
\subsection{Systems with Known or Tentative RV Trends}
\label{sec:rv_trends}

Eight of the 29 systems in our sample have known or
tentative long-term RV variations reported in the
literature. These variations may indicate
additional companions. We discuss the astrometric
posteriors of these systems in the context of their
reported RV trends. Joint astrometric and RV fits
are required to establish whether the same companion
can explain both signals.

\subsubsection{BD+55 362} 
\label{obj:HIP7441}
The BD+55 362 system comprises a K3 star of mass $0.91 \pm 0.10\ M_\odot$ at a distance of $\sim$ 53 pc hosting one exoplanet, BD+55 362 b. BD+55 362 b has a measured $M\sin i = 0.72 \pm 0.08\ M_{\rm J}$ and is on an eccentric orbit of $e = 0.27 \pm 0.06$ with semi-major axis $a_b = 0.78 \pm 0.05$~AU \citep{2021AA...651A..11D}, found via radial velocity with the SOPHIE spectrograph. \citealt{2021AA...651A..11D} additionally detect a quadratic RV drift, indicating an outer companion with $P \gtrsim 3600$~days and $M \gtrsim 2.1\ M_{\rm J}$ ($\gtrsim$4.5 AU).
\par
\begin{figure}[htbp]
    \centering
    \includegraphics[width=1.15\linewidth]{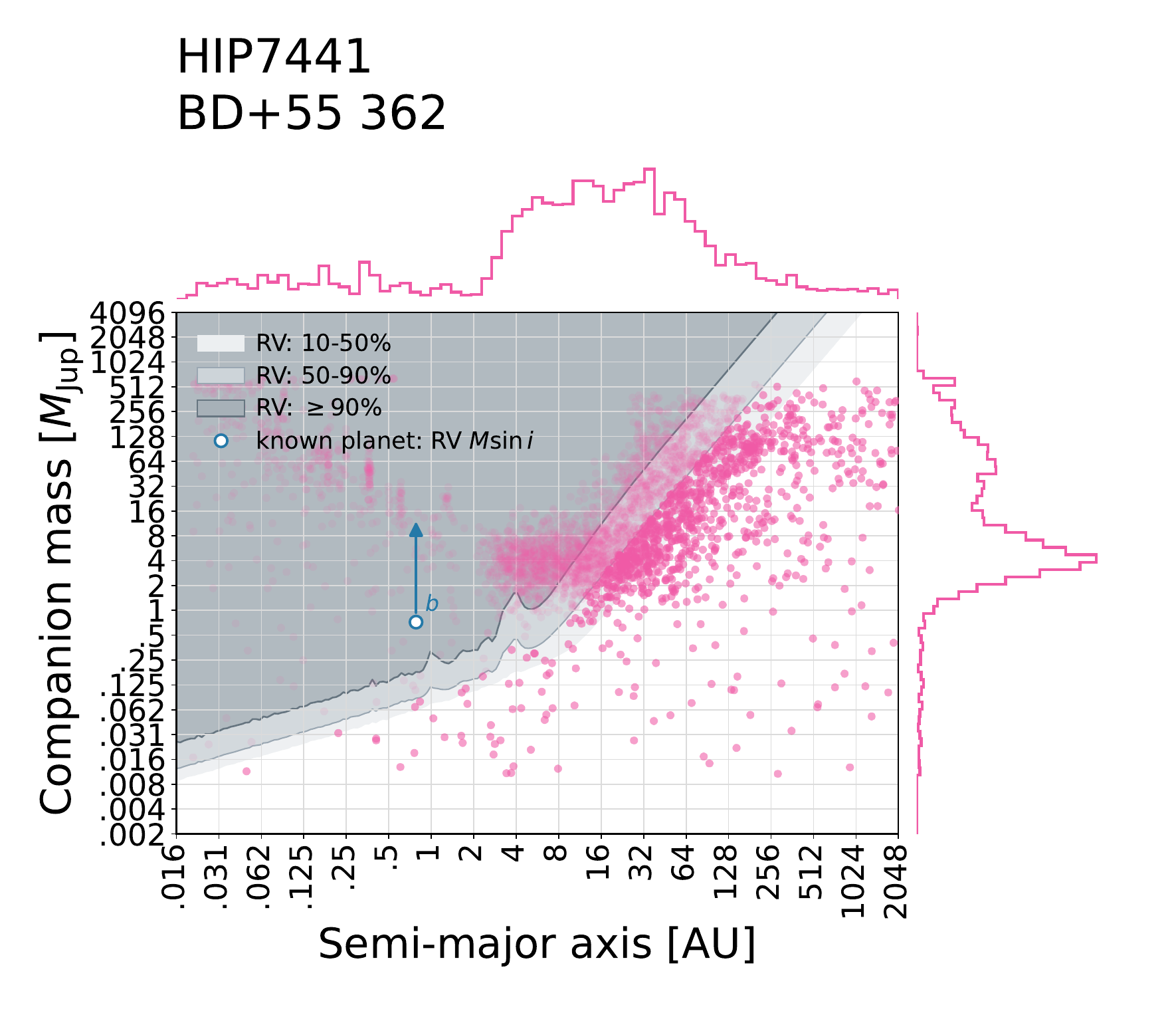}
    \caption{Posterior distribution of companion mass and
semi-major axis for BD$+55\,362$ (HIP~7441). Pink points show the G23H posterior draws, and the upper and right histograms show their marginal distributions. The blue open circle marks the published semi-major axis and minimum mass, $M_b\sin i$, of the known planet. Progressively darker gray shading indicates RV
sensitivities of 10--50\%, 50--90\%, and $\geq90\%$. These percentages give the fraction of trial orbital configurations for which sampled RV variation exceeds the adopted threshold at a given mass and semi-major axis. Points in regions with sensitivity $\geq50\%$ are drawn more transparently. Because a quadratic RV trend has been reported for this system, the shading should not be considered an exclusion region. A joint fit would be required to test whether the astrometric and RV signals share a common origin.}
    \label{fig:HIP7441}
\end{figure}

Figure~\ref{fig:HIP7441} shows a broad posterior for a candidate outer
companion, with a true mass of
$M = 8.8^{+84.9}_{-6.0}\,M_{\rm J}$ and a semi-major axis of
$a = 15.9^{+56.2}_{-13.0}\,\mathrm{AU}$, where the uncertainties denote
the 16th--84th percentile interval. \citet{2021AA...651A..11D} analyzed 22 SOPHIE+ radial velocities spanning 3050.81 days, or 8.35 yr, and measured a post-fit residual RMS of $3.75\ {\rm m\,s^{-1}}$. Their preferred quadratic drift model was favored over a single planet model at 96\% confidence. The electronic table contains 26 SOPHIE+ timestamps over the same baseline, which we use as a cadence proxy because the four measurements excluded from the published fit are not identified. The median astrometric solution is compatible with the RV lower limits, although the extended tails in mass and semi-major axis show that the companion remains poorly constrained and could extend into the brown dwarf or stellar regime.

\subsubsection{HD 45652}
\label{obj:HIP30905}
The HD 45652 system comprises a G8--K0 star of mass
$0.92\pm0.13\,M_\odot$ \citep{2018AJ....156..213M} at a distance of approximately 35 pc.
The system hosts HD 45652 b, which was discovered using ELODIE
radial velocities and confirmed with CORALIE and SOPHIE
\citep{2008AA...487..369S}. A subsequent analysis incorporating
Keck/HIRES observations revised the planet parameters to
$M\sin i=0.433\pm0.076\,M_{\rm J}$,
$e=0.607\pm0.026$, and
$a_b=0.237\pm0.011$ AU \citep{2018AJ....156..213M}.
Although \citet{2008AA...487..369S} searched the residuals for
another periodic signal, they found no significant evidence for one.
The longer-baseline analysis of \citet{2018AJ....156..213M},
however, measured a linear RV trend of
$2.83\pm0.21$ m\,s$^{-1}$\,yr$^{-1}$.

\par
\begin{figure}[htbp]
    \centering
    \includegraphics[width=1.15\linewidth]{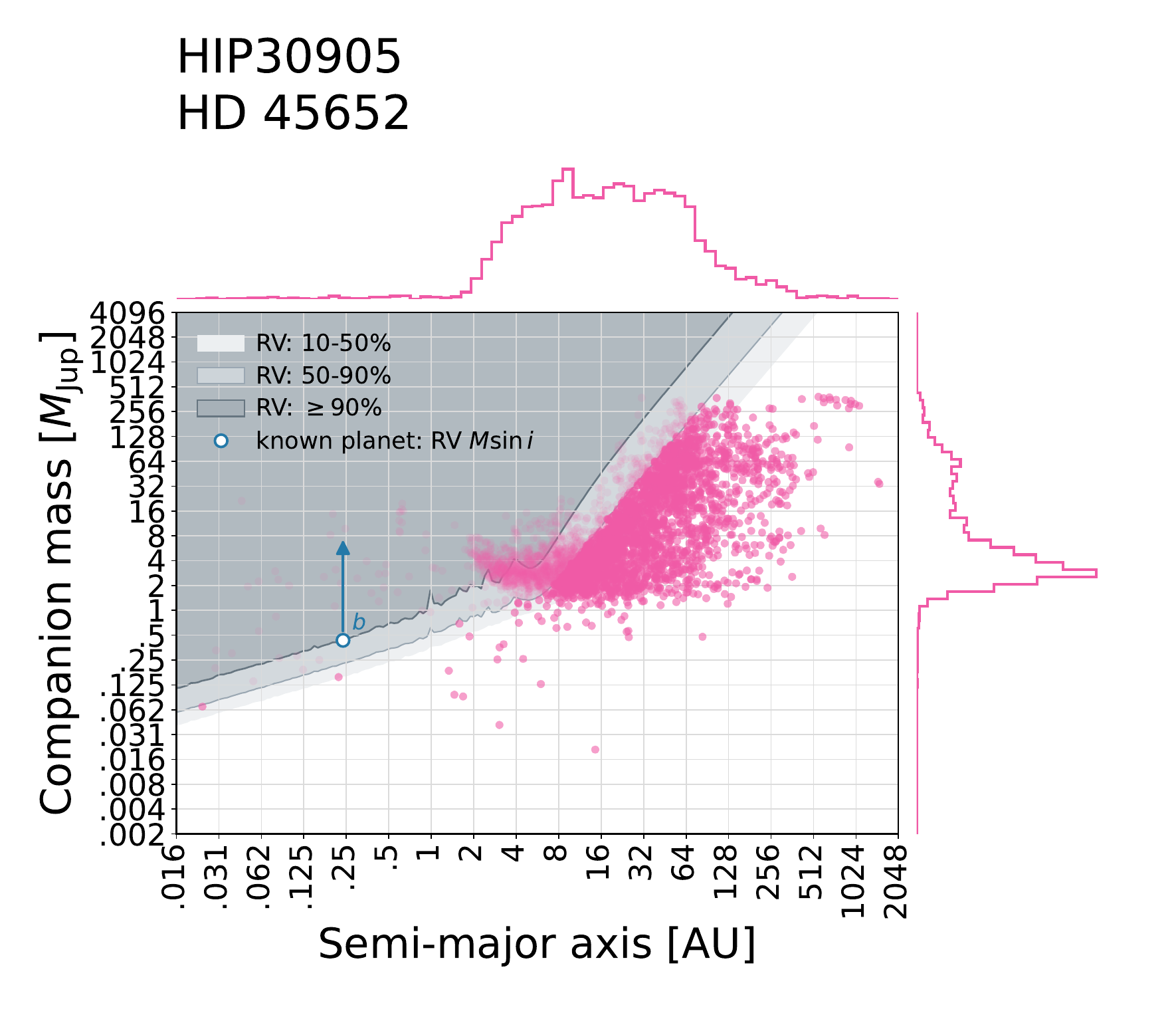}
    \caption{Posterior distribution of companion mass and semi-major axis for HIP 30905 (HD 45652). The gray shaded regions show our RV sensitivity proxy based on the published observations.}
    \label{fig:HIP30905}
\end{figure}

Figure~\ref{fig:HIP30905} shows a broad astrometric posterior
for an outer companion with true mass
$M=4.9^{+34.1}_{-2.4}\,M_{\rm J}$ and semi-major axis
$a=17.6^{+44.1}_{-12.7}$ AU. The RV sensitivity calculation uses the published post-fit residual RMS of
$\sigma_{\rm eff}=18.16$ m\,s$^{-1}$ and 50 recoverable nightly epochs spanning 9.87 yr across ELODIE, CORALIE, SOPHIE, and Keck/HIRES. \citet{2018AJ....156..213M} report 49 fitted RVs, but the published tables reconstruct 50 observing nights. Because the omitted measurement cannot be identified, all 50 epochs were retained as a cadence proxy. Approximately 36.6\% of the astrometric posterior lies where at least half of the simulated orbital geometries would produce detectable RV variation, while approximately 11.4\% lies above 90\% sensitivity. The astrometric companion is a plausible source of the published acceleration, but a joint fit of the RV and astrometric measurements is required to establish whether the two signals arise from the same body.

\subsubsection{HD 63454}
\label{obj:HIP37284}
The HD 63454 system comprises a young ($<$1 Gyr) K4V star of mass $0.8\,M_\odot$ at a distance of $\sim$37 pc hosting one known exoplanet, HD 63454 b. For the known planet comparison, we adopt
$M\sin i=0.25\pm0.01\,M_{\rm J}$ and $a_b\simeq0.04$~AU from \citet{2017AJ....153..136S}.
The HARPS orbital solution is consistent with a circular orbit, with $e=0.000\pm0.022$
\citep{2011ApJ...737...58K}. The planet was discovered from 26 HARPS radial velocity measurements \citep{2005AA...439..367M}. Photometric monitoring subsequently ruled out transits \citep{2011ApJ...737...58K}.
\par
\begin{figure}[htbp]
    \centering
    \includegraphics[width=1.15\linewidth]{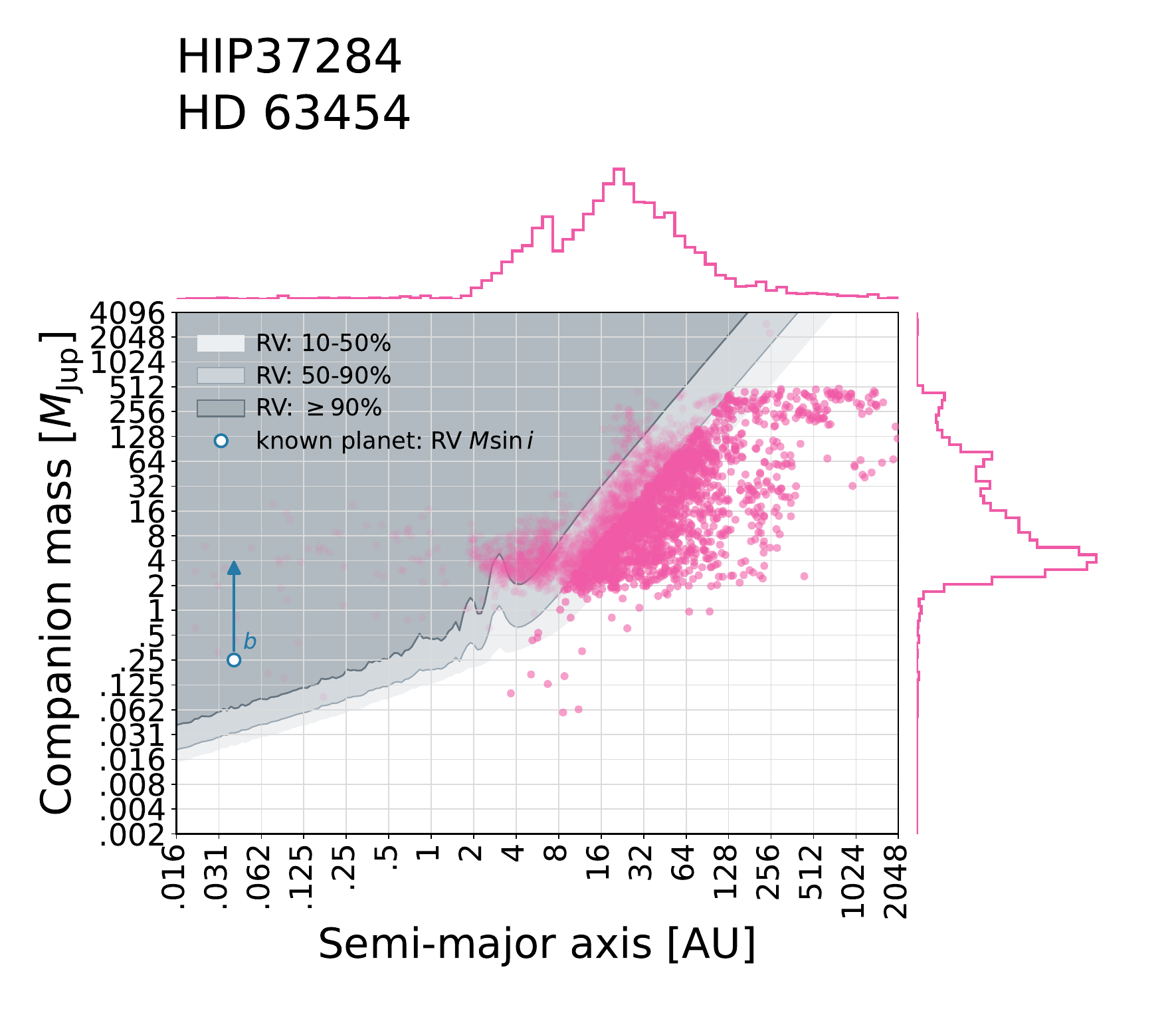}
    \caption{Posterior distribution of companion mass and semi-major axis for HIP 37284 (HD 63454). The gray shaded regions show our RV sensitivity proxy based on the published observations.}
    \label{fig:HIP37284}
\end{figure}

Figure~\ref{fig:HIP37284} shows an astrometric hint of an outer companion with mass $M=9.2^{+52.6}_{-5.8}\,M_{\rm J}$ and semi-major axis $a=20.7^{+41.4}_{-14.6}$~AU. \citet{2011ApJ...737...58K} combined the 26 discovery measurements with eight additional HARPS observations, yielding 34 velocities spanning 6.064~yr. Their fit measured a linear trend of $-3.95\pm0.95$~m\,s$^{-1}$\,yr$^{-1}$. Including this trend reduced $\chi_{\rm red}^2$ from 30.68 to 10.14 and the residual RMS from 10.87 to 6.84~m\,s$^{-1}$. They favored a genuine trend over an instrumental offset but could not distinguish between an outer companion and the magnetic cycle of the active host star. Our sensitivity calculation uses the 34 HARPS epochs and adopts the post-fit residual RMS, $\sigma_{\rm eff}=6.84$~m\,s$^{-1}$. The astrometric candidate is therefore a plausible source of the reported acceleration.


\subsubsection{$\nu^2$ Lupi}
\label{obj:HIP75181}
The HIP~75181 ($\nu^2$ Lupi, HD~136352) system comprises a G4V
star of mass $0.87\pm0.04\,M_\odot$ at a distance of $\sim$14.7~pc
with three transiting planets: HD~136352~b
($M_b=4.72\pm0.42\,M_\oplus$,
$R_b=1.664\pm0.043\,R_\oplus$, and
$a_b=0.0964\pm0.0028$~AU), c
($M_c=11.24^{+0.65}_{-0.63}\,M_\oplus$,
$R_c=2.916^{+0.075}_{-0.073}\,R_\oplus$, and
$a_c=0.1721\pm0.0050$~AU), and d
($M_d=8.82\pm0.93\,M_\oplus$,
$R_d=2.562^{+0.088}_{-0.079}\,R_\oplus$, and
$a_d=0.425\pm0.012$~AU) \citep{2021NatAs...5..775D}.
The planets were discovered using HARPS radial velocities
\citep{2019AA...622A..37U}. The two inner planets were
later found to transit by \emph{TESS}
\citep{2020AJ....160..129K}, and planet d was shown to transit
with \emph{CHEOPS} \citep{2021NatAs...5..775D}. A subsequent full transit of planet d observed with \emph{CHEOPS} refined its ephemeris \citep{2023AA...671A.154E}.
\par
\begin{figure}[htbp]
    \centering
    \includegraphics[width=1.15\linewidth]{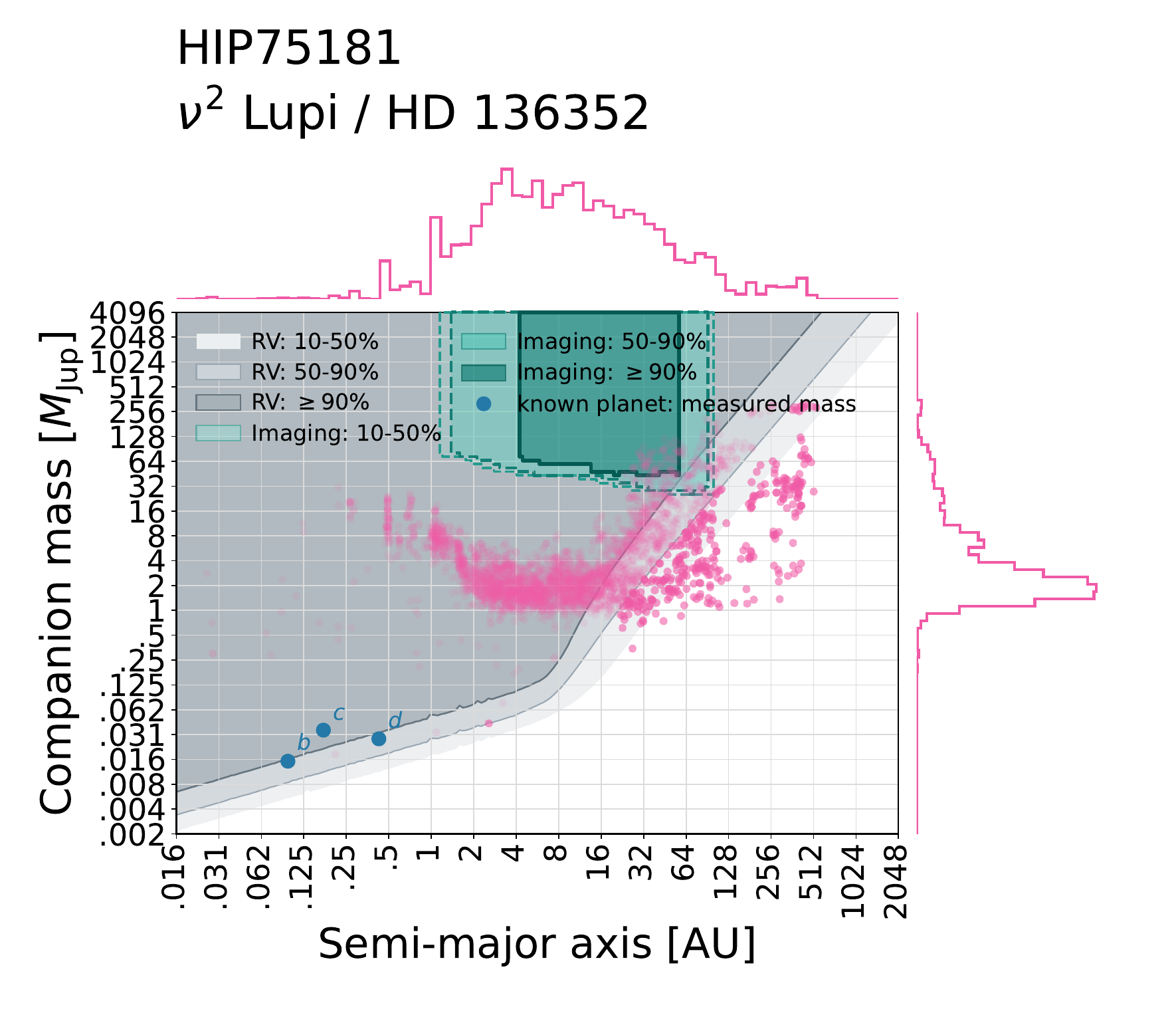}
\caption{Posterior distribution of the semi-major axis and companion mass for HIP 75181 ($\nu^2$ Lupi/HD 136352). The G23H astrometric posterior is plotted as pink points and marginal histograms. The blue circles are the published masses and semi major axs of the 3 known planets. Gray shading indicates our RV sensitivity proxy and aqua shading indicates regions of imaging completeness digitized from the VLT/SPHERE MESS3 map in Figure~H.2 of \citealt{2023AA...680A..64D}. The darker shaded regions are the 10--50\%, 50--90\%, and $\geq 90\%$ sensitivity regions for the two techniques. These percentages are the fraction of orbital configurations that would be detectable at a given semi-major axis and mass. The imaging map was already computed using the projection onto the sky and the conversion of brightness to mass, so we used it directly as published. We vertically shaded the same ranges of semi-major axes above the published upper mass limit of $100\,M_{\rm J}$, assuming the companions of higher mass at that distance are still detectable.}
    \label{fig:HIP75181}
\end{figure}

Figure~\ref{fig:HIP75181} shows a broad astrometric posterior
for an additional companion with true mass
$M=2.6^{+8.7}_{-1.1}\,M_{\rm J}$ and semi-major axis
$a=8.4^{+30.3}_{-6.2}$~AU. For the RV sensitivity calculation,
we use the data available in \citet{2025AJ....170..343H}, comprising 7 post-upgrade HARPS,
234 pre-upgrade HARPS, 29 HIRES, 3 post-upgrade PFS,
21 pre-upgrade PFS, and 169 UCLES measurements spanning
21.9~yr. We adopted the published jitter values of 1.00, 1.163, 2.31,
1.30, 1.00, and 3.72~m\,s$^{-1}$, respectively.
\citet{2025AJ....170..343H} recover all three known planets and fit a small linear trend of $-0.00023\pm0.00011$~m\,s$^{-1}$\,day$^{-1}$. 
\par
The resulting RV proxy has an RV sensitivity of 50\% or greater for 92\% of the G23H posteriors and sensitivity of 90\% or greater for $\sim$75\% of the posteriors. The system was also observed with
VLT/SPHERE by \citet{2023AA...680A..64D}. We used their MESS3 direct imaging detection map which already provides the sensitivity in mass and semi-major axis. About
4.4\% and 1.1\% of the posterior are in regions with imaging
sensitivities of at least 50\% and 90\%. The combined
fractions remain approximately 92\% and 75.3\%, since the samples
constrained by imaging are already located in regions of high RV
sensitivity. Although the three known planets are transiting, and therefore in an edge-on configuration, we did not assume that the candidate outer
companion is coplanar in these sensitivity proxies.

\subsubsection{GJ 682}
\label{obj:HIP86214}
The HIP~86214 (GJ~682) system comprises an M3.5V star of mass
$\sim0.27\,M_\odot$ at a distance of approximately 5.0~pc hosting two
reported radial velocity planets. GJ~682~b has
$M\sin i=4.4^{+3.7}_{-2.4}\,M_\oplus$ and semi-major axis
$a_b=0.080^{+0.014}_{-0.004}$~AU, placing it
within the habitable zone, while GJ~682~c has
$M\sin i=8.7^{+5.8}_{-4.6}\,M_\oplus$ and
$a_c=0.176^{+0.030}_{-0.009}$~AU
\citep{2014MNRAS.441.1545T}. The two signals were identified using 49 UVES and 12 HARPS radial velocities.
\citet{2014MNRAS.441.1545T} also report evidence for a significant
linear radial velocity trend in this system (see their Table 2). The HGCA adopted a systemic RV of $-60$~km\,s$^{-1}$, whereas modern ground-based measurements give approximately $-34.9$~km\,s$^{-1}$
\citep{2018AA...616A...7S,2026arXiv260114459B}.
Using the latter value in the calculation of
\citet{2026arXiv260114459B} increases the HGCA
$\chi^2$ from approximately 1.5 to 12,
which is approximately $3\sigma$ for
two degrees of freedom. Therefore, the updated perspective acceleration correction strengthens the detected anomaly.
\par
\begin{figure}[htbp]
    \centering
    \includegraphics[width=1.15\linewidth]{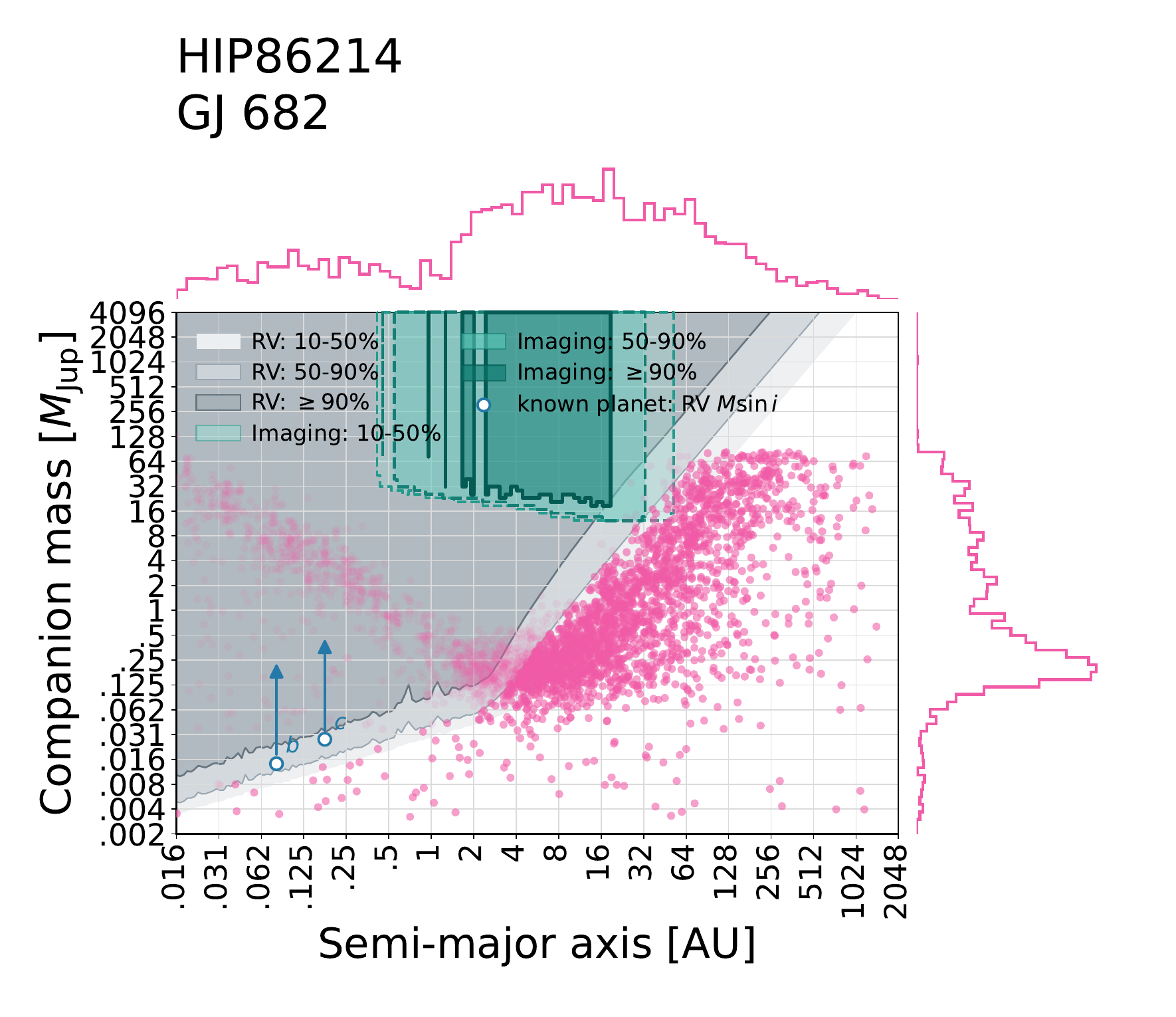}
    \caption{Posterior distribution of companion mass and semi-major axis for HIP 86214 (GJ 682). The gray shaded regions show our RV sensitivity proxy based on the published observations.}
    \label{fig:HIP86214}
\end{figure}

Figure~\ref{fig:HIP86214} shows a broad astrometric posterior for an
additional companion with true mass
$M=0.58^{+9.17}_{-0.42}\,M_{\rm J}$ and semi-major axis
$a=8.79^{+61.02}_{-8.44}$~AU. For the RV sensitivity calculation, we
used the 49 published UVES exposure times from
\citet{2009AA...505..859Z} and the 12 HARPS epochs from
\citet{2014MNRAS.441.1545T}, giving 61 measurements spanning
2251.01~days (6.16~yr). We adopted the published UVES RV scatter of
4.0~m\,s$^{-1}$ and an RMS of 2.28~m\,s$^{-1}$, which we calculated directly
from the published HARPS velocities in Table B17 of \citet{2014MNRAS.441.1545T}. Approximately 37\% of the
astrometric posterior lies in a region with
$f_{\rm sens}\geq0.5$ and 28\% has $f_{\rm sens}\geq0.9$. Since there is a linear trend identified for this system, it is possible that the detected signal here could be associated with the known trend.
\par
GJ~682 was also observed with VLT/SPHERE by
\citet{2023AA...680A..64D}. We used their MESS3
imaging detection map directly since it is reported in companion mass and
semi-major axis. Only 0.1\% of the astrometric
posterior falls in a region with at least 50\% imaging sensitivity and none of the posterior samples reaches 90\% imaging sensitivity.
\par

\subsubsection{HIP 91258}
\label{obj:HIP91258}

The HIP~91258 (BD$+61\,1762$) system comprises a G5V star of mass
$0.95\pm0.03\,M_\odot$ at a distance of $45.9\pm0.1$~pc hosting one
known exoplanet, HIP~91258~b. The planet is a hot Jupiter with
$M\sin i=1.068\pm0.038\,M_{\rm J}$ on a nearly circular
($e=0.024\pm0.014$) orbit with semi-major axis
$a_b=0.057\pm0.001$~AU. It was discovered using the SOPHIE+
spectrograph after its upgrade in 2011
\citep{2014AA...563A..22M}. In addition to the planetary
signal, \citet{2014AA...563A..22M} detected significant quadratic
curvature in the radial velocities, which could indicate the presence of an
undetected outer planetary or stellar companion.
\par
\begin{figure}[htbp]
    \centering
    \includegraphics[width=1.15\linewidth]{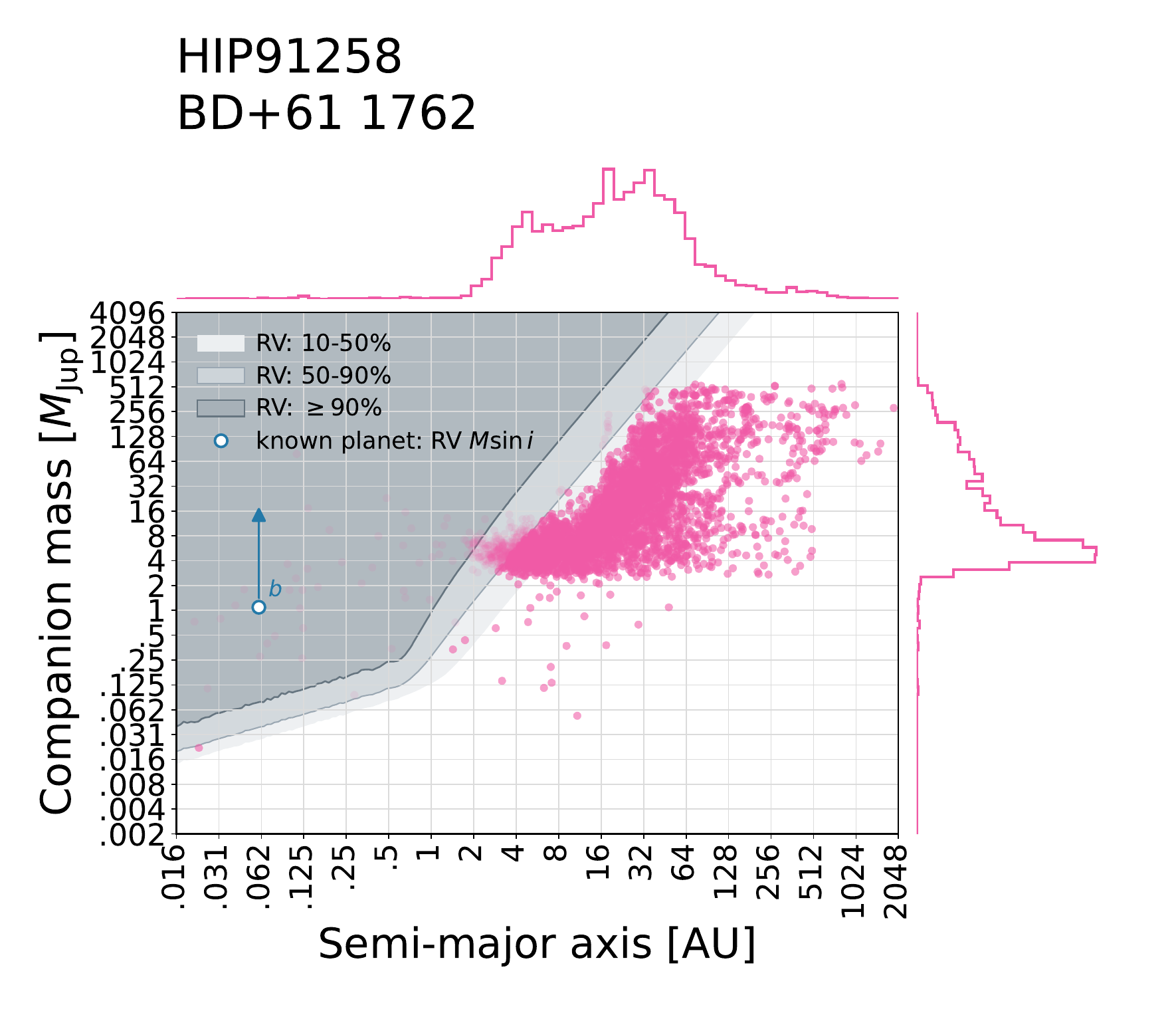}
    \caption{Posterior distribution of companion mass and semi-major axis for HIP 91258 (BD+61 1762). The gray shaded regions show our RV sensitivity proxy based on the published observations.}
    \label{fig:HIP91258}
\end{figure}

Figure~\ref{fig:HIP91258} shows a broad astrometric posterior for an
outer companion with true mass
$M=10.7^{+60.0}_{-6.2}\,M_{\rm J}$ and semi-major axis
$a=20.0^{+36.5}_{-14.7}$~AU. For the RV sensitivity calculation, we
used the 27 SOPHIE+ measurements from
\citet{2014AA...563A..22M}, which span 145.8~days (0.399~yr), and
adopted their published post-fit residual RMS of
$\sigma_{\rm eff}=5.97$~m\,s$^{-1}$. Due to this short baseline, only
8.9\% of the astrometric posterior lies in a region with
$f_{\rm sens}\geq0.5$ and only 1.5\% reaches
$f_{\rm sens}\geq0.9$. 
\par
Under a circular orbit interpretation, \citet{2014AA...563A..22M}
found that the shortest viable outer orbit, with $P\simeq150$~days and
$K\simeq100$~m\,s$^{-1}$, would require
$M\sin i\gtrsim2.5\,M_{\rm J}$. Their preferred two-Keplerian fit favored an approximately five year period companion near the
hydrogen burning limit. The G23H astrometric signal could potentially be associated with the published RV curvature, but determining whether both
signals arise from the same companion will require a joint fit of the
astrometric and radial velocity measurements.

\subsubsection{BD+14 4559}
\label{obj:HIP104780}

The HIP 104780 (BD$+14\,4559$) system comprises a K2V
star of mass $0.86\pm0.15\,M_\odot$ at a distance of
approximately 50~pc. The discovery analysis reported one
planet, BD$+14\,4559$~b, with
$M\sin i=1.47\,M_{\rm J}$, orbital period
$P=268.94\pm0.99$~days, eccentricity
$e=0.29\pm0.03$, and semi-major axis
$a_b=0.777$~AU \citep{2009ApJ...707..768N}. The planet
was detected using 43 Hobby--Eberly Telescope/HRS radial
velocities obtained over 1265~days. The same analysis
identified a statistically significant nonlinear RV trend
with an amplitude of approximately $50$~m\,s$^{-1}$ and
a false alarm probability below $0.1\%$, indicating the
possible presence of an additional companion.
\par
\begin{figure}[htbp]
    \centering
    \includegraphics[width=1.15\linewidth]{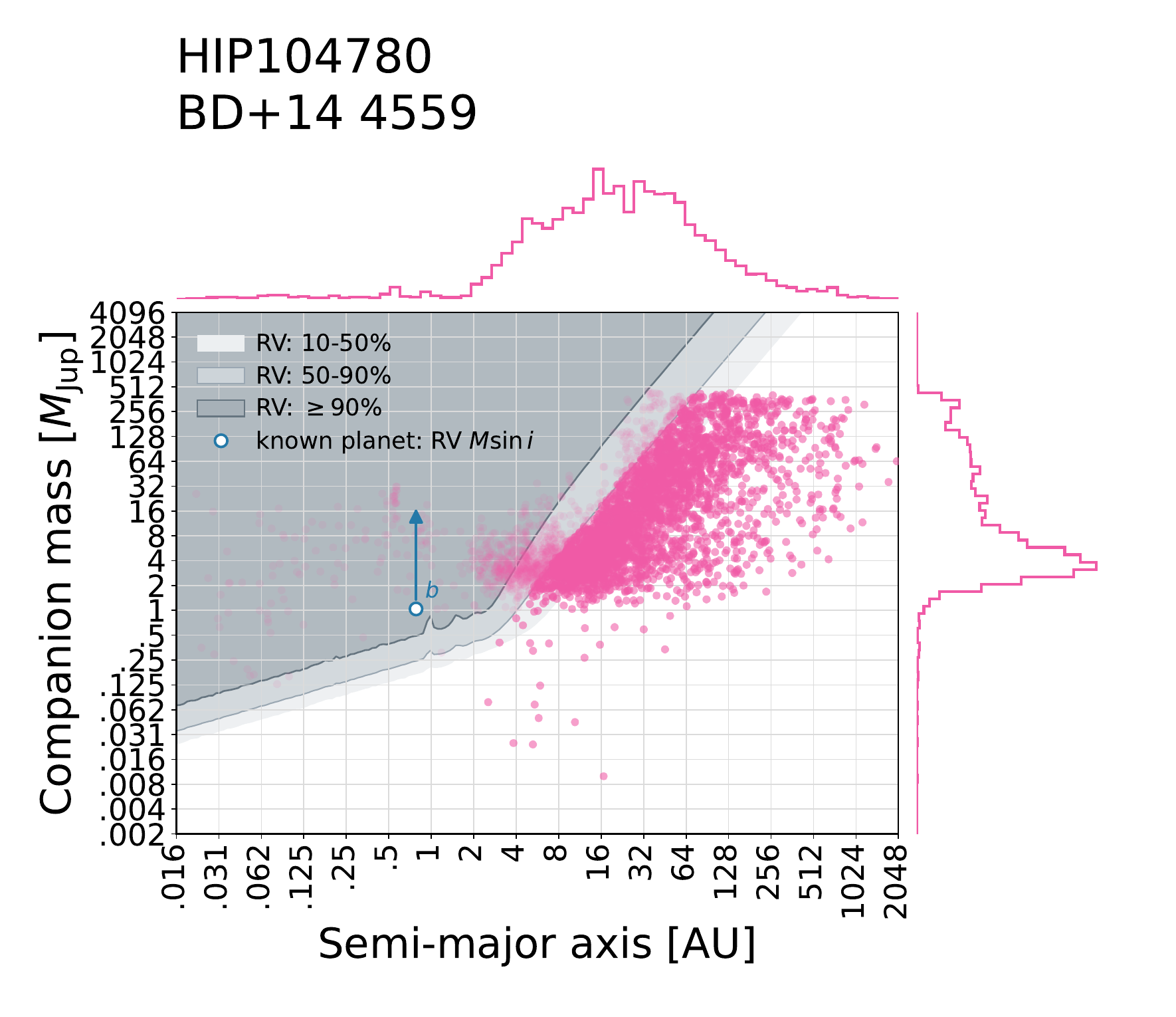}
    \caption{Posterior distribution of companion mass and semi-major axis for HIP 104780 (BD+14 4559). The gray shaded regions show our RV sensitivity proxy based on the published observations.}
    \label{fig:HIP104780}
\end{figure}

Figure~\ref{fig:HIP104780} shows a broad astrometric
posterior for an additional companion with true mass
$M=7.9^{+67.9}_{-5.0}\,M_{\rm J}$ and semi-major axis
$a=21.6^{+56.4}_{-16.0}$~AU. For the RV sensitivity
calculation, we used the exact epochs of all 43 HET/HRS
measurements from \citet{2009ApJ...707..768N},
which span 3.46~yr, and adopted their published post-fit
residual RMS of $11.43$~m\,s$^{-1}$ as
$\sigma_{\rm eff}$. Approximately 25.5\% of the
astrometric posterior lies in a region with
$f_{\rm sens}\geq0.5$, while 10.1\% has
$f_{\rm sens}\geq0.9$. The discovery analysis placed
approximate lower limits of $M\sin i>2.4\,M_{\rm J}$
and $a>2.3$~AU on the additional body. The G23H
companion is compatible with these limits, providing a plausible identification of the source of the published RV trend.




\subsubsection{HD 208897}
\label{obj:HIP108513}
The HIP~108513 (HD~208897) system comprises a K0
giant with mass $M_\star=1.25\pm0.11\,M_\odot$ at a distance of approximately 65~pc. The
star hosts one exoplanet, HD~208897~b, which was discovered using TUG/CES
and OAO/HIDES-F radial velocities
\citep{2017AA...608A..14Y}. An updated analysis revised the
planet's minimum mass and semi-major axis to
$M_b\sin i=1.194^{+0.049}_{-0.044}\,M_{\rm J}$ and
$a_b=1.063^{+0.002}_{-0.003}$~AU, respectively, on a nearly
circular orbit with
$e=0.020^{+0.049}_{-0.010}$
\citep{2023PASJ...75.1030T}. The same analysis detected a
significant linear RV trend of
$\dot{\gamma}=4.073^{+0.441}_{-0.825}$~m\,s$^{-1}$\,yr$^{-1}$,
indicating the presence of an additional wide companion.

\par

\begin{figure}[htbp]
    \centering
    \includegraphics[width=1.15\linewidth]{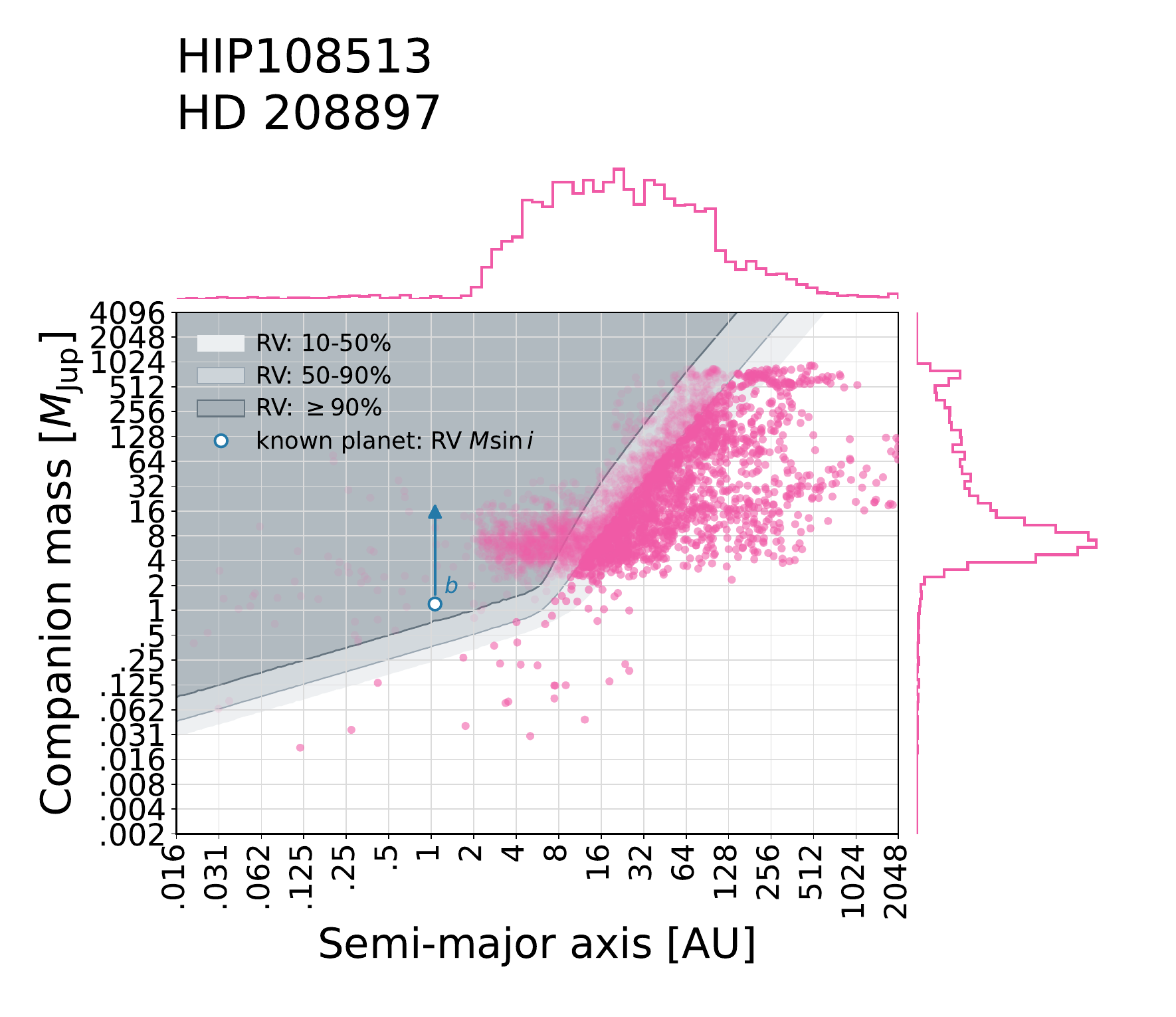}
    \caption{Posterior distribution of companion mass and semi-major axis for HIP 108513 (HD 208897). The gray shaded regions show our RV sensitivity proxy based on the published observations.}
    \label{fig:HIP108513}
\end{figure}

Figure~\ref{fig:HIP108513} shows a broad astrometric posterior
for an additional companion with true mass
$M=11.8^{+111.5}_{-6.6}\,M_{\rm J}$ and semi-major axis
$a=21.3^{+65.4}_{-15.4}$~AU. For the RV sensitivity
calculation, we used the 86 binned measurements
adopted by \citet{2023PASJ...75.1030T}: 39 TUG/CES,
30 HIDES-F1, and 17 HIDES-F2 epochs spanning 11.85~yr. We used the published global post-fit residual RMS of
$\sigma_{\rm eff}=12.776$~m\,s$^{-1}$ for all three
regimes. Approximately 58.5\% of the astrometric posterior
lies in a region with $f_{\rm sens}\geq0.5$, while 27.9\%
has $f_{\rm sens}\geq0.9$. 
\par
\citet{2023PASJ...75.1030T} infer the
approximate constraint
$(M_{\rm c}\sin i_{\rm c})/a_{\rm c}^{2}\simeq
0.023\,M_{\rm J}\,{\rm AU}^{-2}$ from the measured trend.
Calculating $M\sin i/a^2$ for each G23H posterior draw
gives a median of $0.034\,M_{\rm J}\,{\rm AU}^{-2}$
and a 16th--84th percentile interval of
$0.005$--$0.170\,M_{\rm J}\,{\rm AU}^{-2}$,
which includes the published estimate.
The astrometric posterior is therefore compatible
with the approximate mass--distance constraint
from the RV trend. A joint fit is required to test
whether the same companion orbit reproduces both data sets.

\subsection{Systems without Reported RV Trends}

The remaining 21 systems do not have an established long-term RV trend in the literature, but show strong support for an outer companion. We compare their astrometric posteriors with published RV and, where available, imaging constraints to identify regions where an additional companion would be expected to produce a detectable signal.
\par
\subsubsection{HIP 5763}
\label{obj:HIP5763}
The HIP 5763 system comprises a K6V star of mass $0.72\ M_\odot$ at a distance of 29.3 pc hosting one exoplanet, HIP 5763 b. HIP 5763 b has a measured $M\sin i = 0.51\ M_{\rm J}$ and is on a nearly circular orbit of $e = 0.054$ with semi-major axis $a_b = 0.170$~AU \citep{2021AJ....162..176P}, found via radial velocity with the CHIRON spectrograph. 
\par
\begin{figure}[htbp]
    \centering
    \includegraphics[width=1.15\linewidth]{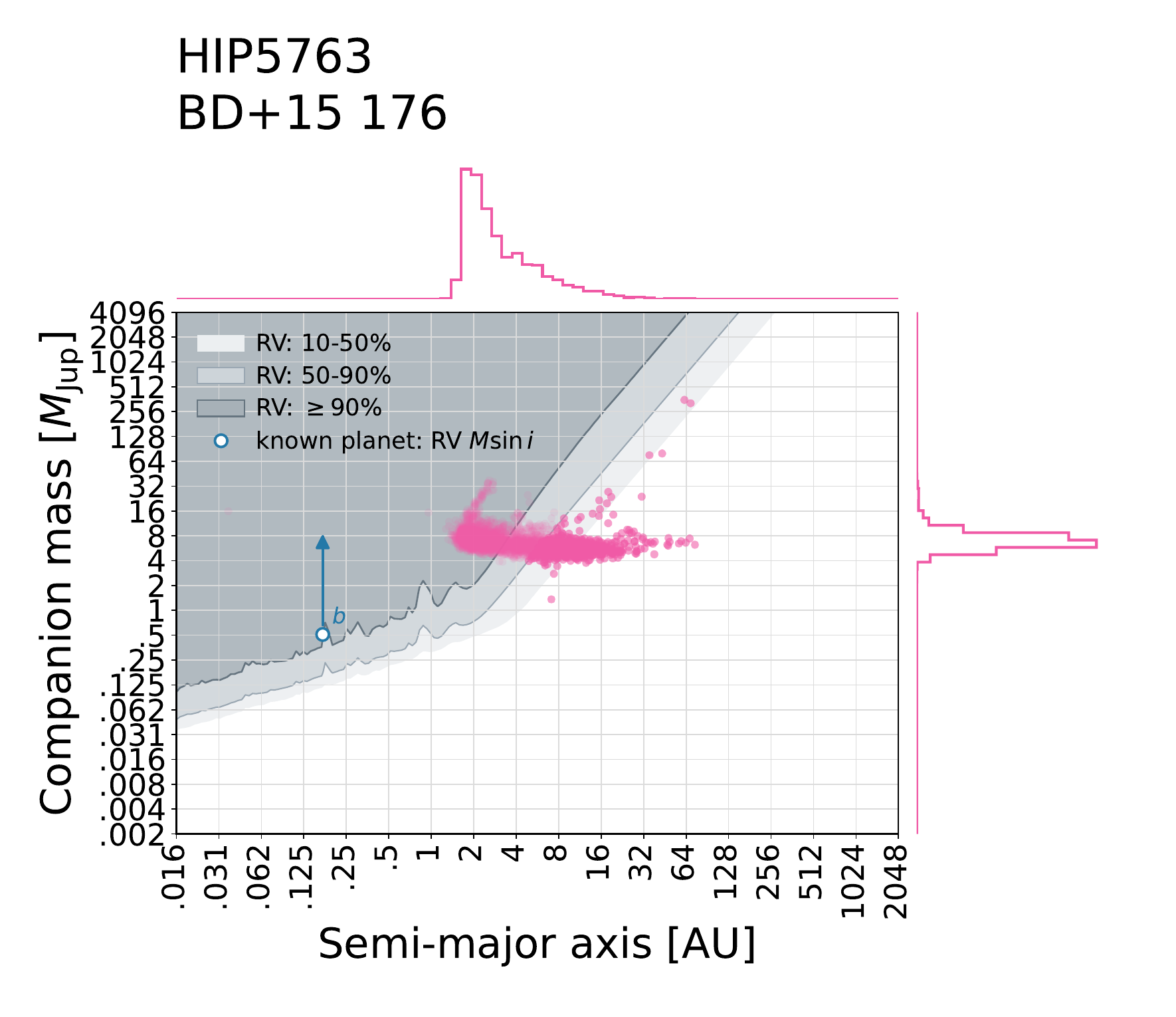}
    \caption{Posterior distribution of companion mass and
    semi-major axis for HIP~5763 (BD$+15\,176$). The gray shaded regions show our RV sensitivity proxy based on the published observations.}
    \label{fig:HIP5763}
\end{figure}

Figure~\ref{fig:HIP5763} shows a potential outer companion with
mass
$M=6.97^{+1.63}_{-1.29}\,M_{\rm J}$ and semi-major axis
$a=2.55^{+3.21}_{-0.71}$~AU. \citet{2021AJ....162..176P} obtained
19 CHIRON measurements between 2017 November 20 and 2019 December
17, corresponding to a baseline of 2.07~yr, and reported a
post-fit residual RMS of
$\sigma_{\rm eff}=16.2\,\mathrm{m\,s^{-1}}$.  Under our sensitivity proxy, the median sensitivity across the posterior is $f_{\rm sens}=0.97$;
approximately 84.7\% of the posterior samples have
$f_{\rm sens}\geq0.5$, and 65.4\% have $f_{\rm sens}\geq0.9$.
\par
The known planet lies well outside the 95\% highest
posterior density region in the projected
mass--semi-major axis comparison, with
$C_{\rm known}=0.9990$. Its published orbit at
$a_b=0.170$~AU is substantially smaller than the
astrometric concentration near 2--3~AU. Under our
association criterion, the G23H solution favors an additional companion. The relatively tight companion mass constraints are informed by significant proper motion differences
across several observing baselines.
The calibrated Gaia DR2 proper motion and the
proper motion inferred from the DR2--DR3 position
difference deviate from the long-baseline
Hipparcos--Gaia proper motion at $9.3\sigma$ and
$28\sigma$, respectively.
The Gaia DR3 proper motion also differs from the
same long-baseline measurement at $5.8\sigma$.
The mass--semi-major axis sensitivity map alone does
not establish how the RV observations would change
this distribution.

\subsubsection{BD+45 564}
\label{obj:HIP10245}
The BD$+45\,564$ system comprises a K1 star of mass
$0.81 \pm 0.07\,M_\odot$ at a distance of approximately 53~pc hosting
one known exoplanet, BD$+45\,564$~b. The planet has a measured minimum
mass of $M\sin i = 1.36 \pm 0.12\,M_{\rm J}$ and follows an orbit with
$e = 0.12 \pm 0.06$ and semi-major axis
$a_b = 0.83 \pm 0.04$~AU \citep{2021AA...651A..11D}. It was detected
using 14 SOPHIE+ radial velocity measurements. The literature does not identify a long-term RV trend. Using GASTON to analyze the \textit{Gaia} DR1 astrometric excess noise, \citet{2021AA...651A..11D} derived a
$3\sigma$ upper limit of $31.4\,M_{\rm J}$ on the true
mass of BD$+45\,564$~b.
\par
\begin{figure}[htbp]
    \centering
    \includegraphics[width=1.15\linewidth]{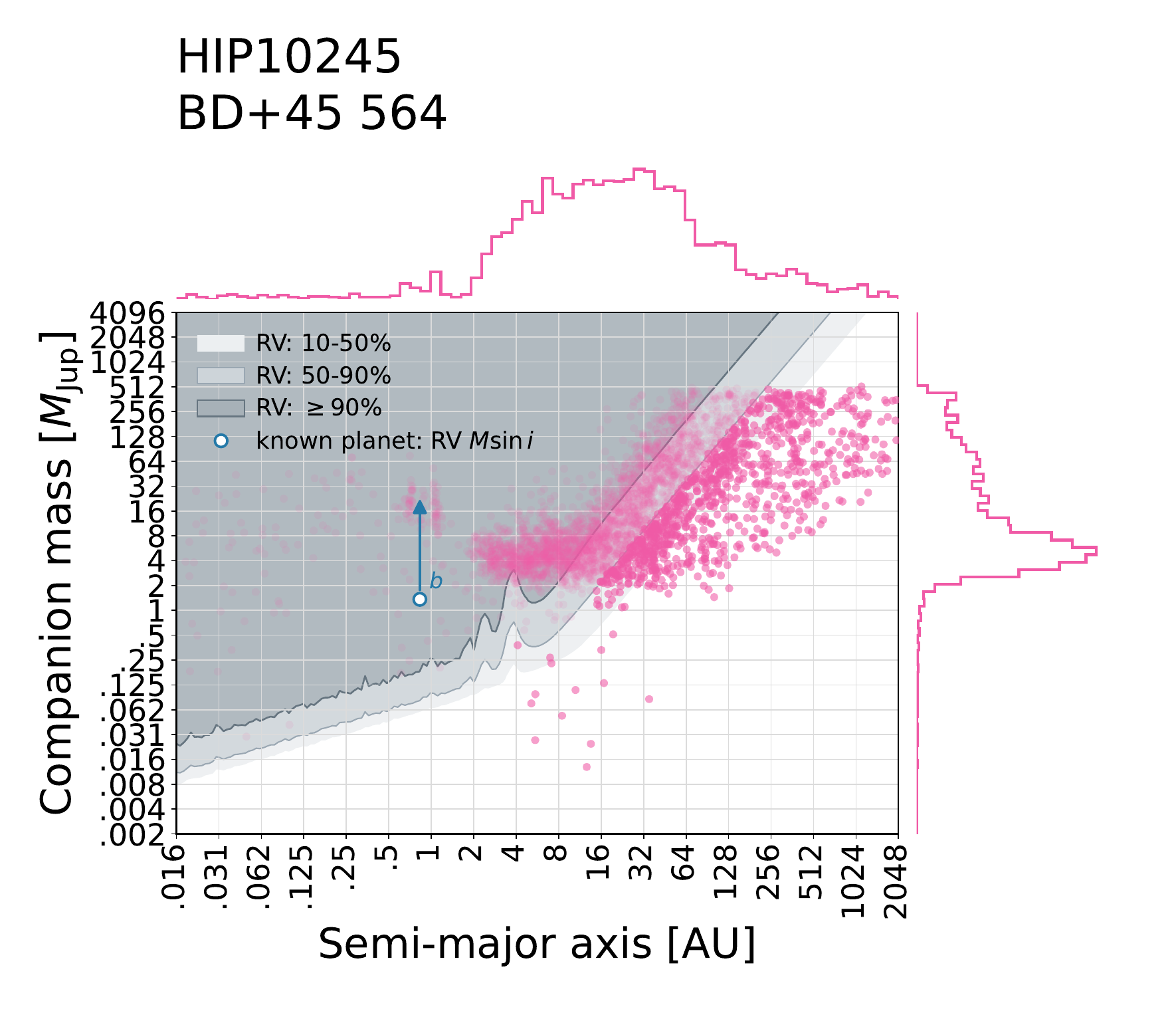}
    \caption{Posterior distribution of companion mass and semi-major axis for HIP 10245 (BD+45 564). The gray shaded regions show our RV sensitivity proxy based on the published observations.}
    \label{fig:HIP10245}
\end{figure}

Figure~\ref{fig:HIP10245} shows a broad astrometric posterior for an
additional companion, with mass
$M = 9.5^{+68.7}_{-5.7}\,M_{\rm J}$ and semi-major axis
$a = 19.7^{+64.1}_{-14.9}$~AU, where the uncertainties denote the
16th--84th percentile interval. The published SOPHIE+ observations
span 3052.81~days, or 8.36~yr, and have a post-fit residual dispersion
of $\sigma_{\rm eff}=3.30$~m\,s$^{-1}$. The electronic table contains
16 timestamps but only 14 measurements entered the published
fit. Because the two excluded observations are not identified, we use
all 16 timestamps as a cadence proxy. Under our sensitivity
calculation, approximately 77.6\% of the astrometric posterior falls in
regions with at least 50\% RV sensitivity, while 44.1\% falls in regions
with at least 90\% sensitivity. 

\subsubsection{GJ 96}
\label{obj:HIP11048}

The GJ~96 system comprises an M2 star of mass
$0.60 \pm 0.07\,M_\odot$ at a distance of 11.9~pc hosting one known
exoplanet, GJ~96~b. The planet has a measured minimum mass of
$M\sin i = 19.66^{+2.42}_{-2.30}\,M_\oplus$ and follows an eccentric
orbit with $e = 0.44^{+0.09}_{-0.11}$ and semi-major axis
$a_b = 0.291 \pm 0.005$~AU \citep{2018AA...618A.103H}. It was
detected using SOPHIE+ radial velocities. The single Keplerian model was strongly favored over the no-Keplerian and quadratic trend-only models.
\citep{2018AA...618A.103H}.
\par
\begin{figure}[htbp]
    \centering
    \includegraphics[width=1.15\linewidth]{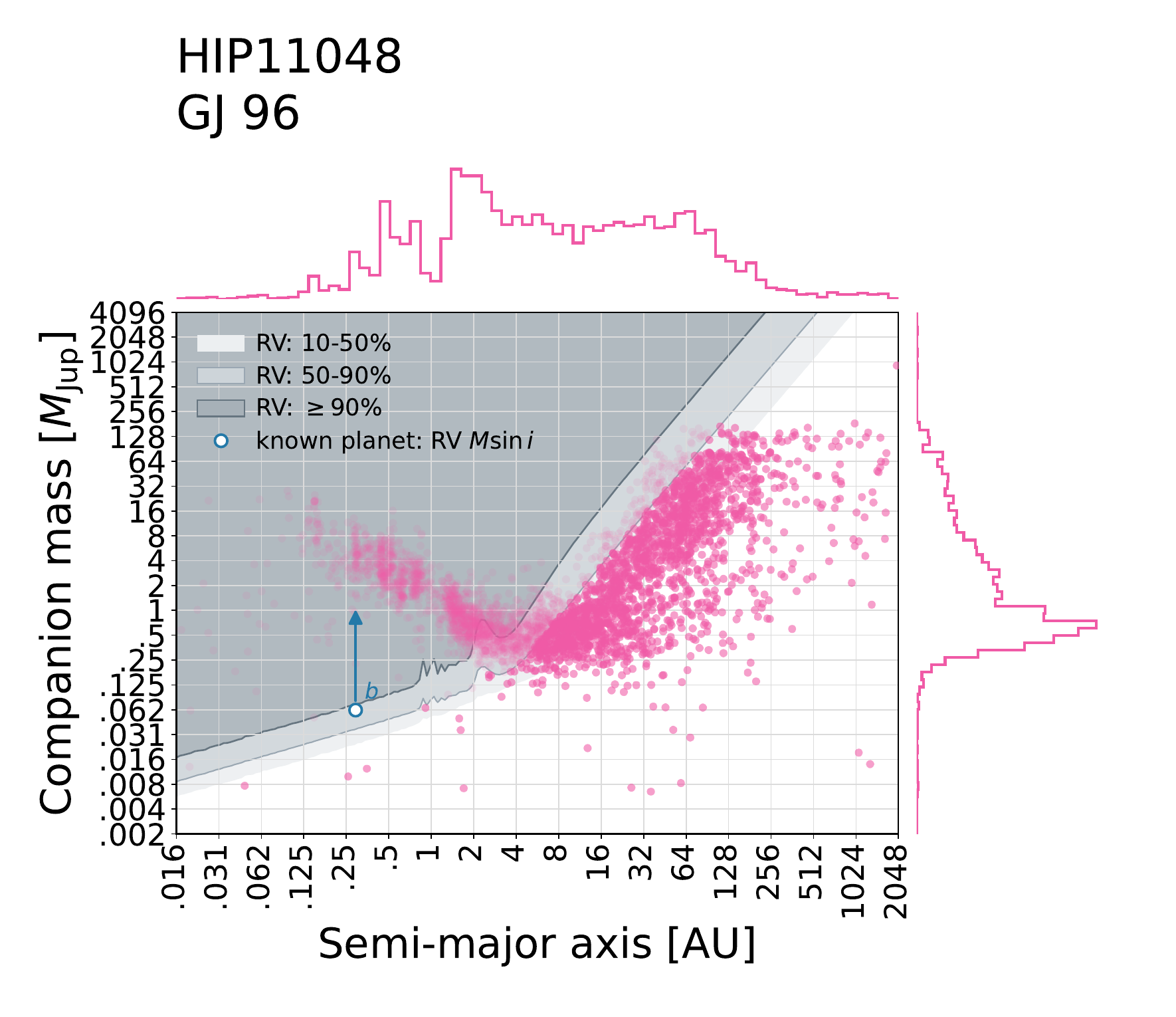}
    \caption{Posterior distribution of companion mass and semi-major axis for HIP 11048 (GJ 96). The gray shaded regions show our RV sensitivity proxy based on the published observations.}
    \label{fig:HIP11048}
\end{figure}

Figure~\ref{fig:HIP11048} shows a broad astrometric posterior for an additional companion, with mass $M = 1.28^{+10.17}_{-0.81}\,M_{\rm J}$ and semi-major axis
$a = 6.27^{+55.06}_{-5.43}$~AU, where the uncertainties denote the
16th--84th percentile interval. The 72 tabulated SOPHIE+ velocities span 2018.63~days (5.53~yr) and have a post-fit residual dispersion of $\sigma_{\rm eff}=3.37$~m\,s$^{-1}$. Under our sensitivity calculation, approximately 54\% of the astrometric posterior falls in regions with at least 50\% RV sensitivity, while 35.5\% falls in regions
with at least 90\% sensitivity.

\subsubsection{$\lambda^2$ For}
\label{obj:HIP12186}

The HIP~12186 ($\lambda^2$~Fornacis; HD~16417) system comprises a G1V star of mass $\sim1.2\,M_\odot$ at a distance of 25.5~pc hosting one confirmed exoplanet, HD~16417~b. The planet has a
measured minimum mass of $M\sin i = 22.1 \pm 2.0\,M_\oplus$ and
follows a mildly eccentric orbit with $e = 0.20 \pm 0.09$ and
semi-major axis $a_b = 0.14 \pm 0.01$~AU
\citep{2009ApJ...697.1263O}. It was discovered using radial
velocities obtained with UCLES at the AAT and independently confirmed with Keck/HIRES observations. \citet{2009ApJ...697.1263O} noted that the apparently non-circular orbit could result from an external perturbation by a potentially undetected outer planet \citep{2008ApJ...686L..29M}. However, they cautioned that the eccentricity is not strongly constrained ($\sigma_e \simeq 0.1$) and that low SNR radial velocity measurements are biased against recovering nearly circular orbits \citep{2008ApJ...685..553S}. 
\par
\begin{figure}[htbp]
    \centering
    \includegraphics[width=1.15\linewidth]{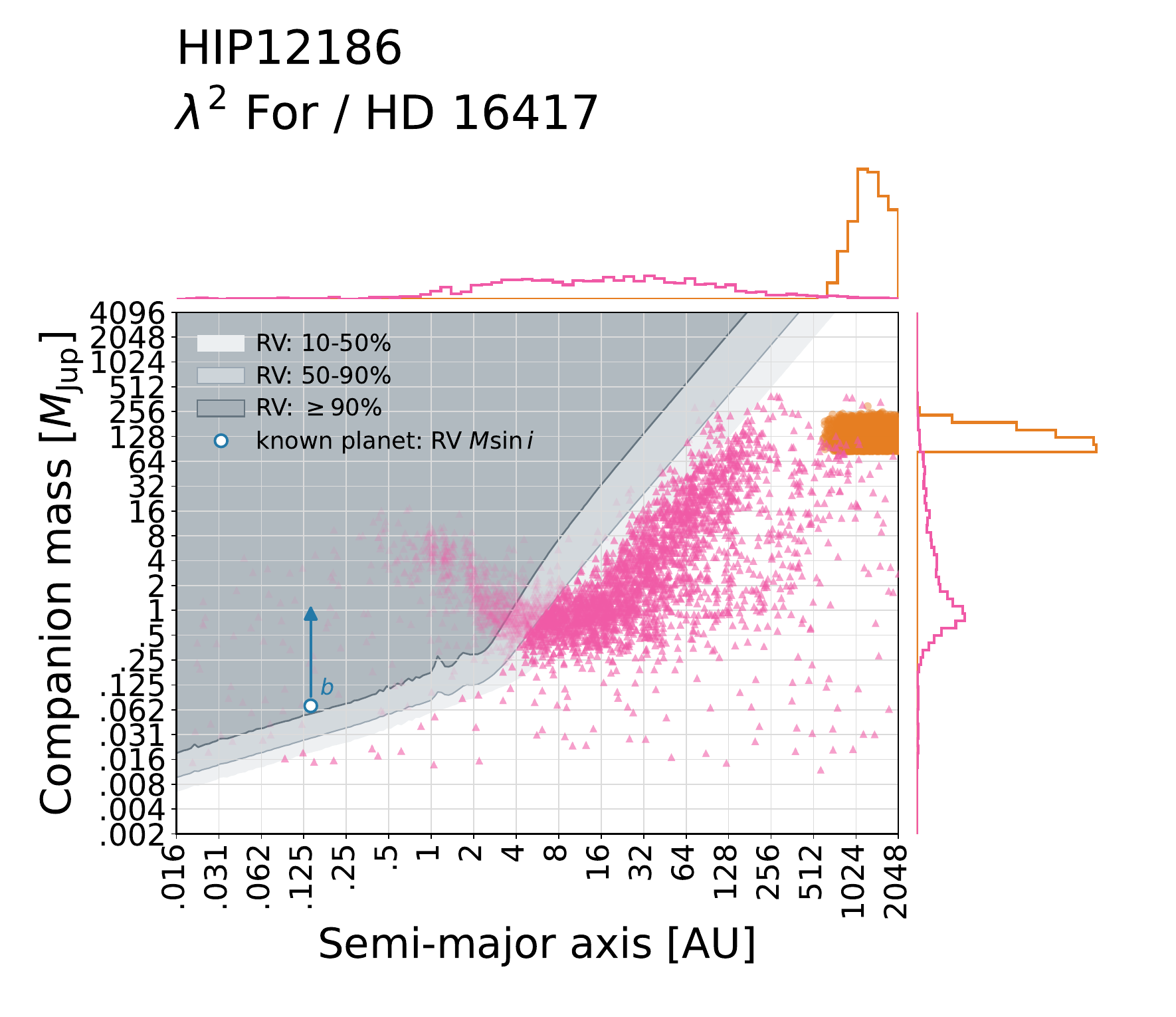}
    \caption{Companion mass and semi-major axis posteriors
for HIP~12186 ($\lambda^2$~Fornacis / HD~16417). The orange component, with median mass $M\simeq120\,M_{\rm J}$ and semi-major axis $a\simeq1340$~AU, is consistent with the known stellar
companion HD~16417~B imaged at a projected separation
of 1159~AU \citep{2014MNRAS.439.1063M}.
The blue circle marks the published semi-major
axis and minimum mass of the inner planet
HD~16417~b. Progressively darker gray regions
correspond to RV sensitivities of 10--50\%,
50--90\%, and $\geq90\%$.}
    \label{fig:HIP12186}
\end{figure}

Figure~\ref{fig:HIP12186} contains two astrometric components. The
outer stellar component has mass
$M = 119.5^{+45.1}_{-25.8}\,M_{\rm J}$ and semi-major axis $a = 1338^{+600}_{-365}$~AU. This component is identified at a projected separation of
1159~AU by \citet{2014MNRAS.439.1063M}. Its mass estimate
of approximately $0.11\,M_\odot$ ($115\,M_{\rm J}$) agrees closely with the G23H posterior. The additional giant planet component has mass $M = 1.59^{+12.37}_{-0.95}\,M_{\rm J}$ and semi-major axis $a = 17.0^{+75.5}_{-14.1}$~AU, where the uncertainties denote the
16th--84th percentile intervals.
\par
The NTT/SofI observations presented in \citet{2014MNRAS.439.1063M} reached an $H$-band limiting magnitude of 18.6, corresponding to approximately $45\,M_{\rm J}$, and were complete to stellar companions over projected separations of 140--3800~AU (see their Table 4). These observations confirm the outer stellar component and exclude additional stellar companions over this wide separation range. Because the study reports
only a specific limiting magnitude and an average
sensitivity curve for their entire survey, and not a specific curve for HD 16417, we do not plot
an imaging sensitivity contour for this system. Regardless, the inner limit of approximately 140~AU lies well outside the median semi-major axis of the giant planet posterior, so the SofI observations would not provide a meaningful constraint on that component.
\par
For the RV sensitivity calculation,
we use the 60 measurements entering the published combined fit: 50 AAT/UCLES measurements and 10 Keck/HIRES measurements spanning
3.10~yr, with a published post-fit RMS of
$\sigma_{\rm eff}=2.6$~m\,s$^{-1}$. Approximately 30\% of the giant planet posterior falls in regions with at least 50\% RV sensitivity, while 19.6\% falls in regions with at least 90\% sensitivity.

\subsubsection{HD 20794}
\label{obj:HIP15510}

The HIP~15510 (HD~20794) system comprises a G6V star with
a published mass of $0.79\pm0.01\,M_\odot$ at a distance
of 6.04~pc. The combined HARPS and ESPRESSO analysis of
\citet{2025AA...693A.297N} identifies three planets with
minimum masses of $2.15\pm0.17$, $2.98\pm0.29$, and
$5.82\pm0.57\,M_\oplus$, at semi-major axes of
$0.12570^{+0.00052}_{-0.00053}$,
$0.3625^{+0.0015}_{-0.0016}$, and
$1.3541\pm0.0068$~AU, respectively. The outermost
planet follows an eccentric orbit with
$e=0.45^{+0.10}_{-0.11}$ that crosses the habitable zone.
The authors do not find an additional long period Keplerian signal, and their fitted acceleration is consistent with zero.
\par
\begin{figure}[htbp]
    \centering
    \includegraphics[width=1.15\linewidth]{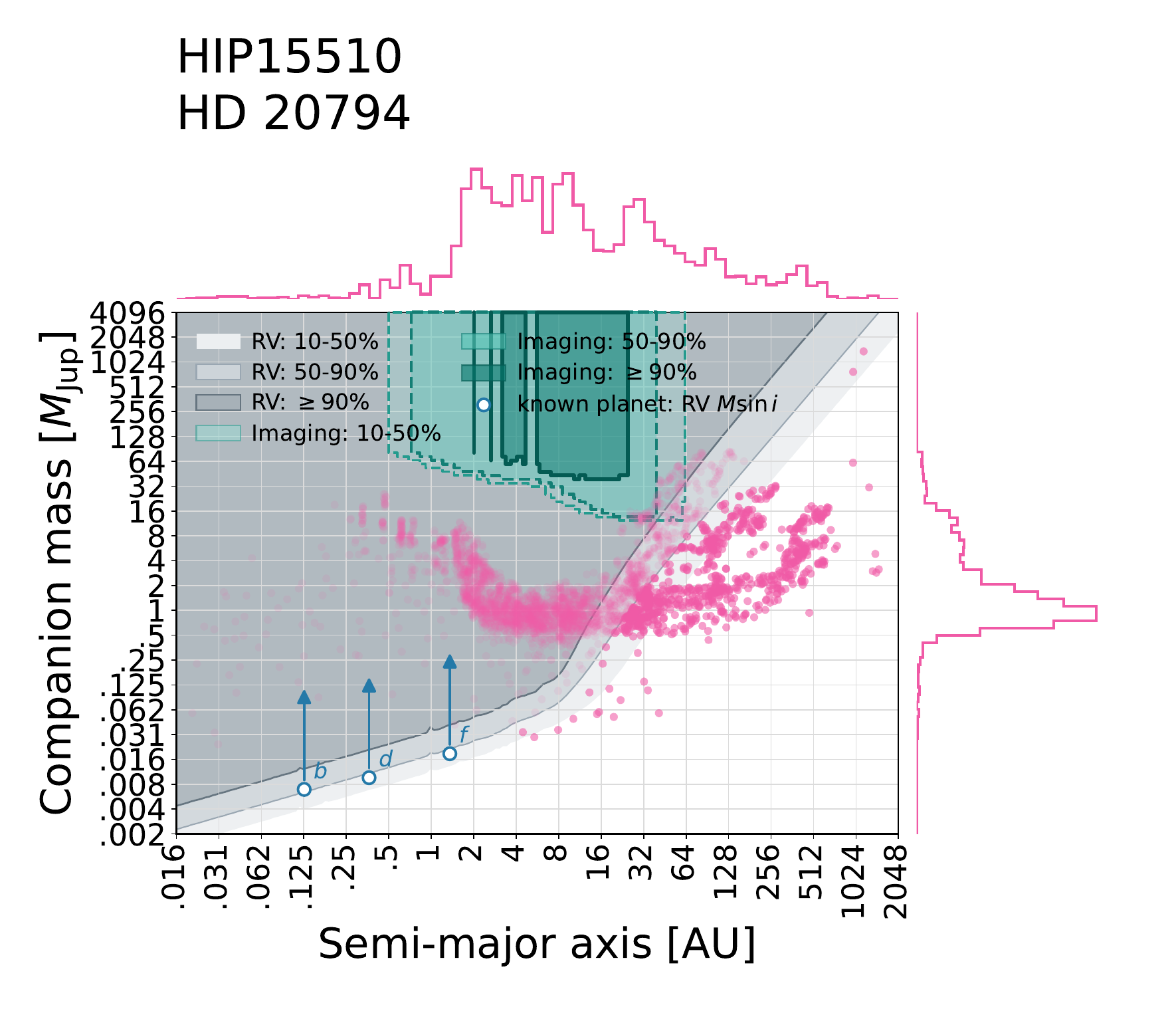}
    \caption{Posterior distribution of mass and semi-major
    axis for HIP~15510 (HD~20794). Gray shaded regions
    correspond to our RV sensitivity proxy. Teal regions correspond to our imaging limits proxy.}
    \label{fig:HIP15510}
\end{figure}

Figure~\ref{fig:HIP15510} shows a broad astrometric posterior
for an additional companion with true mass
$M=1.37^{+5.27}_{-0.64}\,M_{\rm J}$ and semi-major axis
$a=8.31^{+47.22}_{-6.22}$~AU. For this system, the evidence for a companion comes from the calibrated Gaia DR3 UEVA likelihood, which compares the star's reported RUWE with the expectation for single stars of similar brightness and color \citep{2025AA...702A..76K,Thompson2026}.
The proper motion measurements from Hipparcos,
Gaia DR2, and Gaia DR3 constrain the allowed orbital
solutions but do not independently support the
companion detection. For the RV sensitivity
calculation, we used 806 nightly epochs spanning 20.52~yr:
512 pre-upgrade HARPS, 231 post-upgrade HARPS, and
63 ESPRESSO observations. The observing dates were obtained from the machine-readable data in
\citet{2025AA...693A.297N}. The retained ESPRESSO nights were identified by matching the released observations to their Figures~1 and~3. 
\par
The three instrumental regimes were evaluated using the published post-fit residual RMS values of $\sigma_{\rm eff}=0.93$, 0.99, and
0.72~m\,s$^{-1}$, respectively. 
Approximately $81.3\%$ of the astrometric posterior lies in regions with $f_{\rm sens}\geq0.5$, while $66.1\%$ has $f_{\rm sens}\geq0.9$. 
\par
HD~20794 is also constrained by VLT/SPHERE imaging
\citep{2023AA...680A..64D}. We adopted the imaging
MESS3 completeness contours from their Figure~H.2,
retaining the authors' model mass conversion. These observations primarily probe the brown dwarf and stellar mass regime. The posterior samples covered by imaging already fall within the RV-sensitive regions, so the combined coverage fractions remain $81.3\%$ and $66.1\%$. 

\subsubsection{HD 28471}
\label{obj:HIP20625}
The HD~28471 system comprises a G5V star of mass
$0.98\pm0.04\,M_\odot$ at a distance of 43.7~pc hosting
a compact system of three super-Earths and sub-Neptunes.
The planets b, c, and d have semi-major axes of
$0.042\pm0.001$, $0.065\pm0.001$, and
$0.100\pm0.002$~AU and minimum masses of
$3.72^{+0.40}_{-0.43}$, $5.72^{+0.57}_{-0.72}$, and
$4.91^{+0.82}_{-0.77}\,M_\oplus$, respectively.
Their orbital periods are close to a 1:2:4 ratio
\citep{2025MNRAS.543...28S}. The system was characterized using 122 HARPS radial
velocities spanning 19.3~yr. A strong long period signal was also detected, with $P=1496\pm12$~d,
$M\sin i=0.372\pm0.02\,M_{\rm J}$ and
$a=2.54\pm0.04$~AU. If real, this signal would
correspond to a cold Saturn mass planet. However, its correlations with several activity indicators led
\citet{2025MNRAS.543...28S} to caution that it may instead
trace a short stellar magnetic cycle.
\par
\begin{figure}[htbp]
    \centering
    \includegraphics[width=1.15\linewidth]{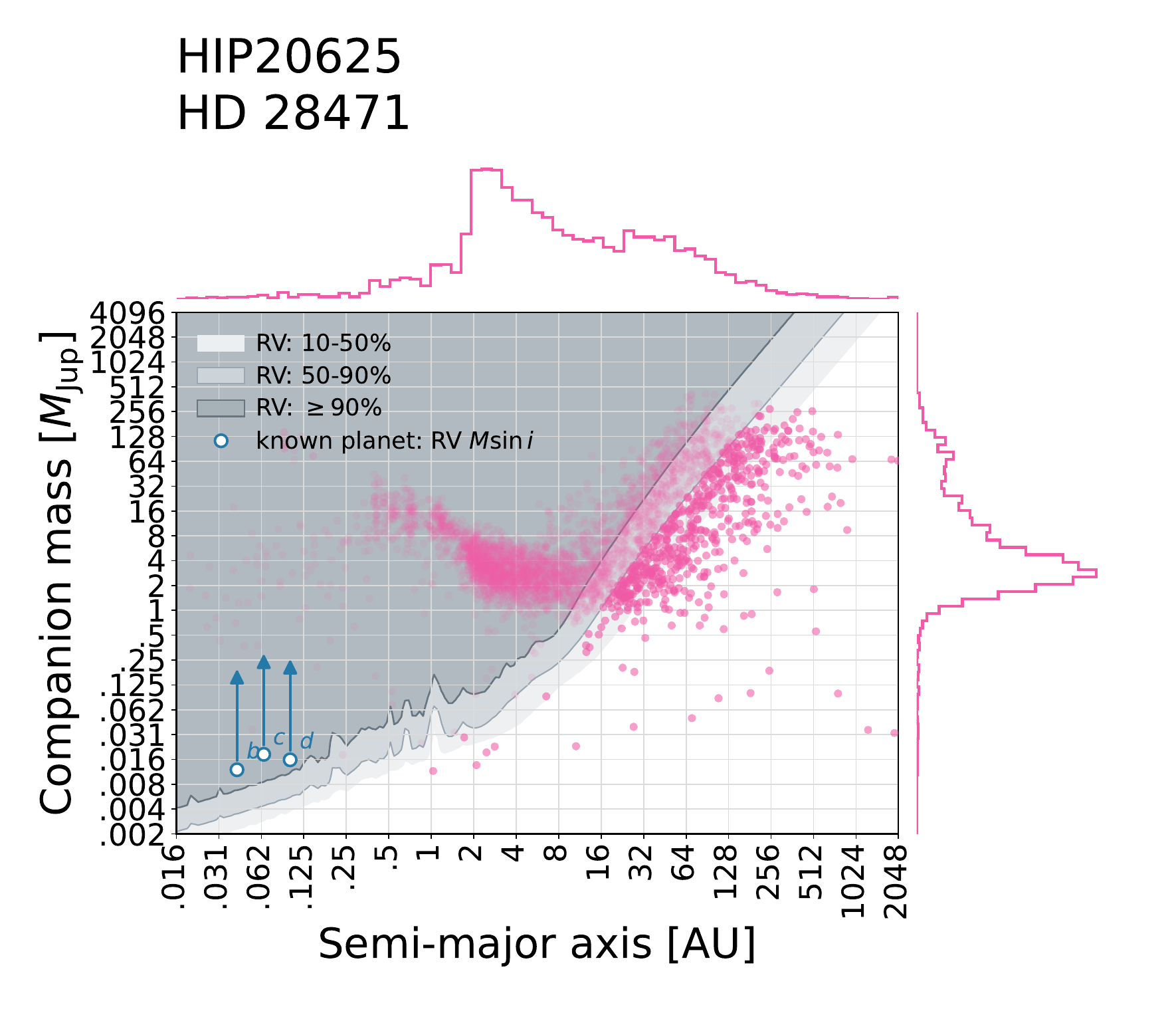}
    \caption{Posterior distribution of companion mass and semi-major axis for HIP 20625 (HD 28471). The gray shaded regions show our RV sensitivity proxy based on the published observations.}
    \label{fig:HIP20625}
\end{figure}

Figure~\ref{fig:HIP20625} shows a broad astrometric
posterior for a companion with true mass
$M=4.2^{+19.1}_{-2.2}\,M_{\rm J}$ and semi-major axis
$a=6.3^{+41.7}_{-4.3}$~AU. To compare
the published minimum mass with the astrometric
solution, we projected each G23H mass using its
fitted inclination. The RV solution lies within
the resulting 95\% posterior region, with
$C\simeq0.88$, suggesting a possible association
with the astrometric signal.
\par
The RV sensitivity shading was calculated using the
observing epochs of all 122 measurements spanning
19.3~yr \citep{2025MNRAS.543...28S}, with 23 pre-upgrade, 81 post-upgrade, and 18 post-COVID
HARPS measurements. For each regime, the reported formal RV precision was combined in quadrature with the fitted noise term, giving effective uncertainties of 2.31, 0.63, and 1.49~m\,s$^{-1}$ respectively.
Approximately 86.3\% of the G23H posterior lies
in regions with at least 50\% RV sensitivity,
and 69.6\% lies in regions with at least 90\%
sensitivity.

\subsubsection{Kapteyn's Star} 
\label{obj:HIP24186}

\begin{figure}[htbp]
    \centering
    \includegraphics[width=1.15\linewidth]{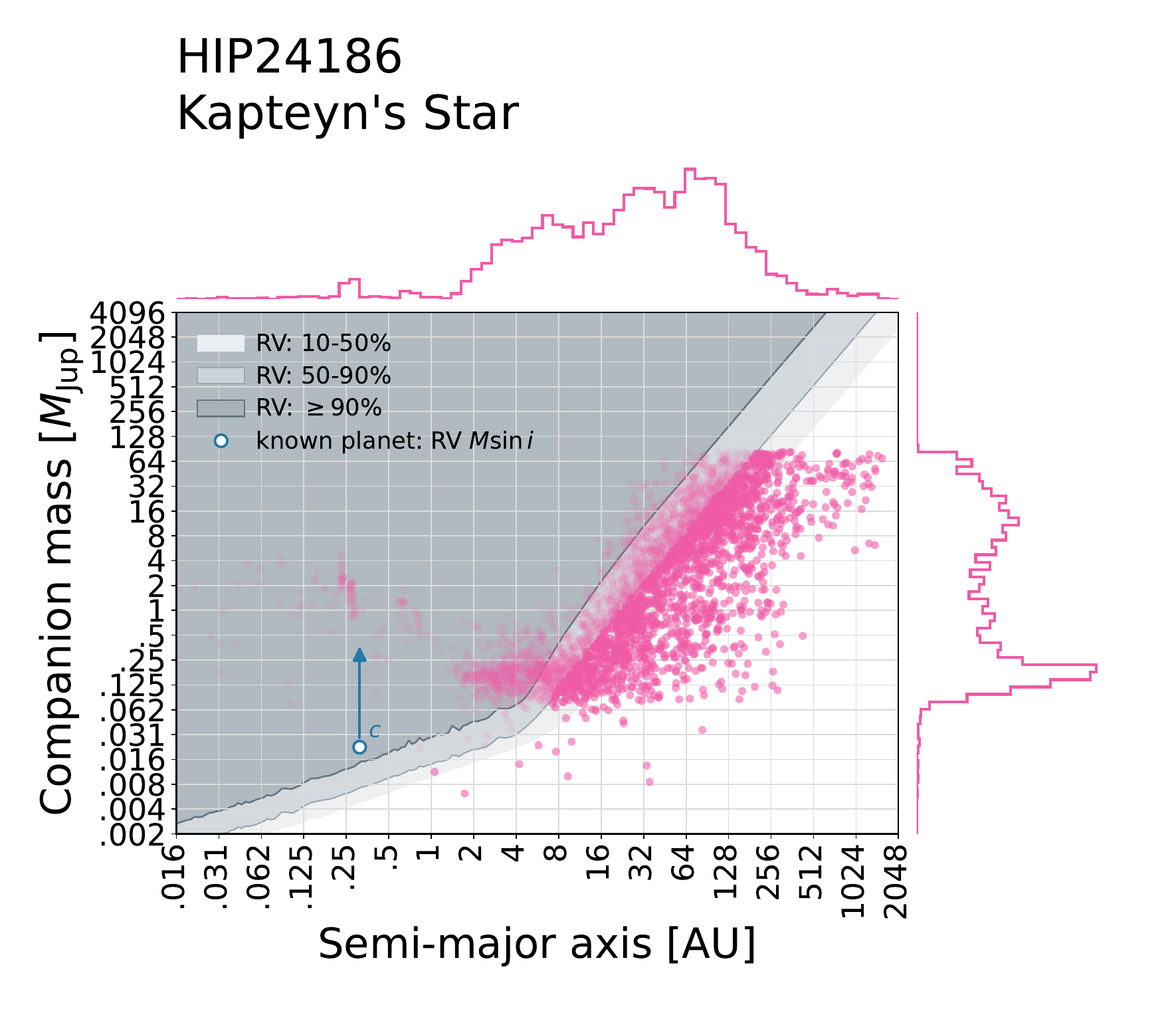}
    \caption{Posterior distribution of companion mass and
semi-major axis for HIP~24186 (Kapteyn's Star / GJ~191).
Pink points and marginal histograms show the G23H
astrometric posterior. The blue open circle marks the
reported semi-major axis and minimum mass of the disputed
candidate Kapteyn~c, shown for comparison as it is flagged as a controversial planet. The upward arrow indicates its unknown
true mass if the RV signal is planetary. Kapteyn~b is
omitted since it is flagged as a false positive. Progressively darker gray regions indicate
10--50\%, 50--90\%, and $\geq90\%$ of trial orbital
configurations satisfying our RV sensitivity criterion.
Posterior points within regions of sensitivity
$\geq50\%$ are shown more transparently. The shading
represents an approximate sensitivity proxy, not a formal
exclusion limit; the astrometric posterior has not been
updated using an RV likelihood.}
    \label{fig:HIP24186}
\end{figure}

Kapteyn's Star (GJ~191, HD~33793) is a nearby M1 halo star with a mass of approximately $0.28\,M_\odot$ at a distance of 3.91~pc. Two super-Earth candidates, Kapteyn~b and c, were initially reported from HARPS, HIRES, and PFS radial velocities \citep{2014MNRAS.443L..89A}. Subsequent analyses associated the signal attributed to Kapteyn~b with stellar activity \citep{2015ApJ...805L..22R}. Kapteyn~c is flagged as controversial \citep{2016ApJ...830...74A}. A later joint Gaussian process analysis of the radial velocities and H$\alpha$ measurements found no convincing evidence for either planet and attributed the apparent periodicities to stellar rotation, with $P_{\rm rot}\simeq125$~d \citep{2021AJ....161..230B}. For that reason, we omit Kapteyn~b from Figure~\ref{fig:HIP24186} and mark the reported $M\sin i$ of Kapteyn~c only. 
\par
For the RV sensitivity calculation, we used the internally consistent data set published by \citet{2014MNRAS.443L..89A}, comprising 95 HARPS, 32 HIRES, and 8 PFS measurements spanning 15.24~yr. The adopted effective RV scatters were 0.88, 2.06, and 0.68~m\,s$^{-1}$ for HARPS, HIRES, and PFS, respectively, obtained by combining the median formal uncertainty and instrumental jitter in quadrature.
\par
Figure~\ref{fig:HIP24186} shows a broad astrometric posterior for a companion with true mass
$M=1.5^{+17.8}_{-1.4}\,M_{\rm J}$ and semi-major axis
$a=31.4^{+79.8}_{-26.1}$~AU, where the uncertainties represent the 16th and 84th percentiles. Kapteyn's Star shows a component of its Hipparcos--Gaia proper motion anomaly perpendicular to its direction of motion \citep{2026arXiv260114459B}. This component cannot be explained by an error in the systemic RV used to correct for perspective acceleration. Approximately 58.2\% of the posterior samples fall in regions  with 50\% of RV sensitivity, while 20.7\% fall in regions with modeled sensitivity of at least 90\%. 
\par

\subsubsection{LHS 1903}
\label{obj:HIP34730}
The HIP~34730 (LHS~1903, TOI-1730) system comprises an
M dwarf of mass $0.538^{+0.039}_{-0.030}\,M_\odot$ at a
distance of $\sim$36 pc hosting four transiting planets.
LHS~1903~b, c, d, and e have measured masses of
$3.28\pm0.42$, $4.55^{+0.73}_{-0.69}$,
$5.96^{+1.15}_{-1.13}$, and
$5.79^{+1.60}_{-1.61}\,M_\oplus$, respectively, with
semi-major axes of
$0.02656^{+0.00055}_{-0.00058}$,
$0.05387^{+0.00112}_{-0.00117}$,
$0.08604^{+0.00178}_{-0.00186}$, and
$0.15135^{+0.00314}_{-0.00338}$~AU
\citep{2026Sci...392l2348W}.
\par
\begin{figure}[htbp]
    \centering
    \includegraphics[width=\linewidth]{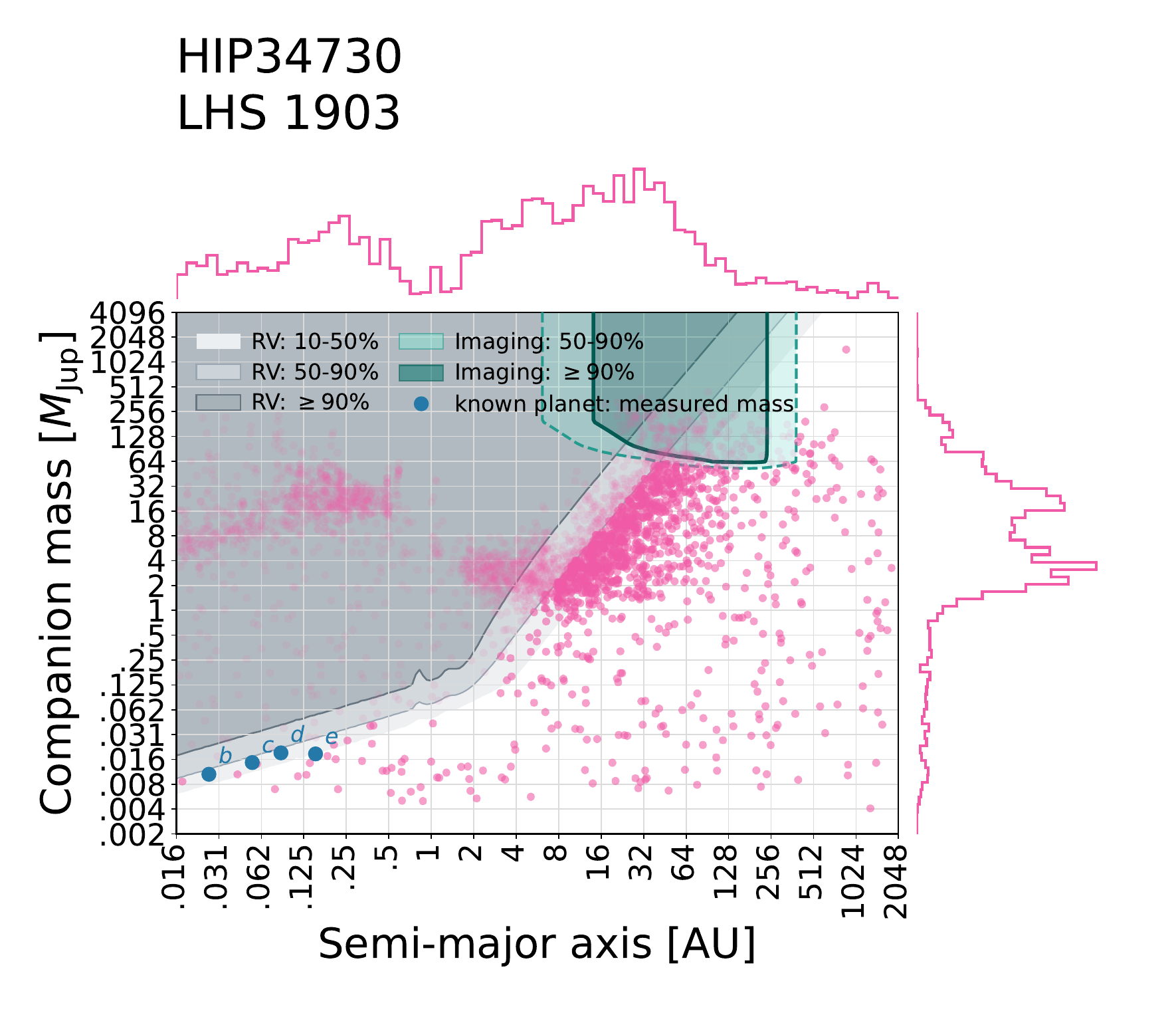}
    \caption{Posterior distribution of mass and semi-major axis for HIP~34730 (LHS~1903). Gray and aqua shaded regions correspond to our RV and imaging sensitivity proxies, respectively. Posterior samples within the $\geq50\%$ sensitivity region of either technique are shown with reduced opacity. Blue points mark the measured masses and semi-major axes of the four known planets.}
    \label{fig:HIP34730}
\end{figure}

Figure~\ref{fig:HIP34730} shows a broad astrometric posterior
for an additional companion with true mass
$M=7.55^{+31.32}_{-5.70}\,M_{\rm J}$ and semi-major axis
$a=6.89^{+39.24}_{-6.72}$~AU. For the RV sensitivity
calculation, we used all 91 published HARPS-N measurements, with a baseline of 869.86~days, or 2.38~yr. We combined the fitted GP amplitude of
3.67~m\,s$^{-1}$ and additive jitter of
1.08~m\,s$^{-1}$ from Table~S7 of
\citet{2026Sci...392l2348W} in quadrature, obtaining
$\sigma_{\rm eff}=3.83$~m\,s$^{-1}$.
Approximately $60.3\%$ of the astrometric posterior lies
in regions with $f_{\rm sens}\geq0.5$, while $41.5\%$
has $f_{\rm sens}\geq0.9$. The published analysis finds
no significant linear RV trend.
\par
LHS~1903 was also observed with Palomar/PHARO on
2020 November 5 \citep{2026Sci...392l2348W}.
We used the machine-readable $5\sigma$ Br$\gamma$
contrast curve shown in their Figure~S8, adopting
$K_s=8.21$ and the published isochronal age estimate of
$4.9\pm4.0$~Gyr. Ages were sampled within the supported
model range, with an upper bound of 10~Gyr, and model
$K_s$ magnitudes were used as a proxy for Br$\gamma$.
We applied the curve only over its nonzero tabulated
separations, $0.156''$--$9.837''$ without extrapolating toward the unresolved stellar image.
\par
The imaging primarily constrains the brown dwarf and
stellar mass portion of the posterior. Approximately
$8.9\%$ and $6.4\%$ of the samples lie in regions with imaging sensitivities of at least $50\%$ and $90\%$, respectively. Considering both techniques, $64.9\%$ of the posterior lies within the $50\%$ sensitivity region of at least one technique, and $47.0\%$ lies within the corresponding $90\%$ region. 

\subsubsection{HD 63433}
\label{obj:HIP38228}

The HD 63433 (TOI-1726) system comprises a young Sun-like star with mass $M_\star=0.99\pm0.03\,M_\odot$ and age $414\pm23$~Myr at a distance of $22.32\pm0.06$~pc. The star is a member of the Ursa Major moving group
and hosts three small transiting planets, HD~63433~d,
b, and c, with semi-major axes of
$0.0503^{+0.0025}_{-0.0027}$,
$0.0714^{+0.0036}_{-0.0038}$, and
$0.1448^{+0.0073}_{-0.0077}$~AU, respectively
\citep{2024AJ....167...54C}.
For planets b and c, \citet{2023AA...671A.163M}
report a $3\sigma$ upper limit of
$M_b<21.8\,M_\oplus$ and a measured mass of
$M_c=15.54^{+3.86}_{-3.80}\,M_\oplus$.
Planet d does not yet have a measured mass.
\par
\begin{figure}[htbp]
    \centering
    \includegraphics[width=1.15\linewidth]{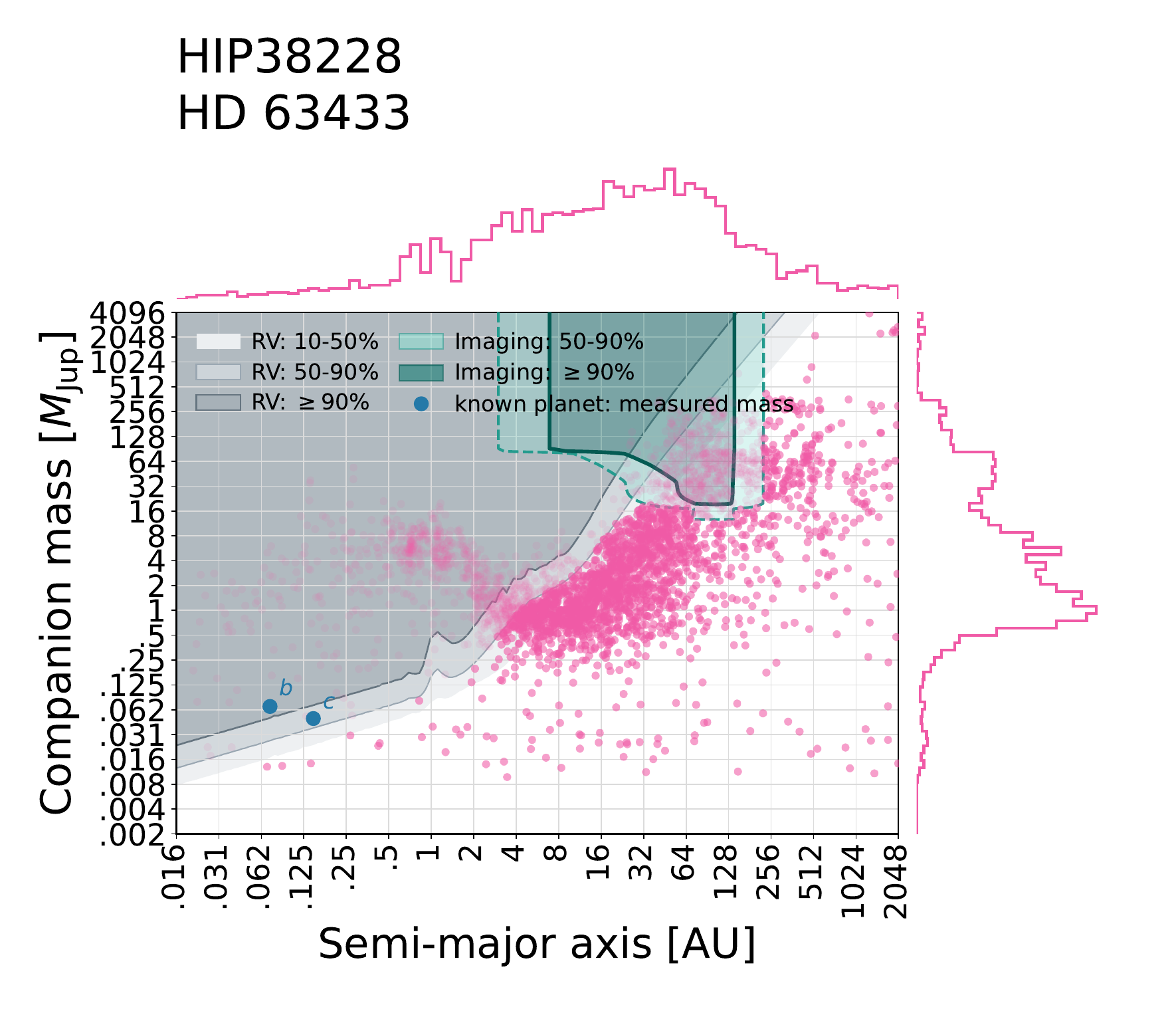}
    \caption{Posterior distribution of companion mass and semi-major axis for HIP 38228 (HD 63433 / TOI-1726). The gray shaded regions show our RV sensitivity proxy based on the published observations.}
    \label{fig:HIP38228}
\end{figure}

Figure~\ref{fig:HIP38228} shows the astrometric posterior
for a candidate outer companion with mass
$M=3.5^{+39.0}_{-2.7}\,M_{\rm J}$ and semi-major axis
$a=21.0^{+93.3}_{-18.7}$~AU. For the RV sensitivity calculation,
we used 26 epochs from
\citet{2020AJ....160..179M}, comprising three ELODIE,
12 SOPHIE, and 11 Lick/Hamilton observations. The two SOPHIE
measurements obtained on the same night were treated as a single
cadence epoch. We assigned an effective
scatter of 37~m\,s$^{-1}$ to the ELODIE and SOPHIE measurements
and 24~m\,s$^{-1}$ to the Hamilton measurements.

We also included the 150 VIS RV data from CARMENES \citet{2023AA...671A.163M}. We adopted $\sigma_{\rm eff}=3.95$~m\,s$^{-1}$, which was obtained by combining their reported mean VIS uncertainty of 3.9~m\,s$^{-1}$ with the fitted VIS jitter of 0.64~m\,s$^{-1}$ in quadrature. Finally, we included 103 HARPS-N epochs obtained over 781 days by \citet{2023AA...672A.126D}. The DRS and TERRA velocities are alternative reductions of the same spectra and for that reason were not treated as separate observations. We used the TERRA reduction with its published raw RMS of 24.1~m\,s$^{-1}$ as $\sigma_{\rm eff}$. The resulting calculation
uses 279 cadence epochs spanning 25.13~yr. 
\par
HD~63433 is also constrained by published high-resolution imaging. We digitized the $5\sigma$ VLT $K$-band contrast curve from Figure~5 of \citet{2024AJ....167...54C} and converted
the contrast limits into companion masses using the Sonora Red Diamondback v2 and BHAC15 models
\citep{2025ApJ...994..198D,2015AA...577A..42B}, adopting the system age of $414\pm23$~Myr and propagating its uncertainty throughout the calculation. Model $K_s$ magnitudes were used as a proxy for the published $K$-band limits. 
\par
Approximately $19.1\%$ and $12.5\%$ of the G23H posterior
samples lie in regions with imaging sensitivities of at
least $50\%$ and $90\%$, respectively. Including the RV
sensitivity map, these fractions increase to $44.6\%$
and $29.8\%$ for sensitivity at the corresponding level
from at least one technique. The imaging primarily
constrains the higher mass portion of the posterior at
wider orbits.
\par

\subsubsection{HD 67200}
\label{obj:HIP39242}

HD~67200 (DMPP-6) is an F-type star of mass
$1.079\pm0.045\,M_\odot$ at a distance of
$54.438\pm0.041$~pc. \citet{2026MNRAS.547ag370S}
initially reported two Keplerian radial velocity signals: DMPP-6~b, with $M\sin i=5.80^{+0.94}_{-0.80}\,M_\oplus$
and $a=0.0779^{+0.0016}_{-0.0012}$~AU, and DMPP-6~c,
with $M\sin i=13.47^{+1.60}_{-1.97}\,M_\oplus$ and
$a=0.2206^{+0.0033}_{-0.0039}$~AU. They also found
moderate evidence for a third signal near 3~d. A later analysis of the same HARPS observations that included Gaussian process models for stellar activity strongly favored an activity model over the purely Keplerian interpretation and considered the previously reported DMPP-6~b and c signals to be superseded by the models \citep{2026MNRAS.tmp.1422B}. That analysis retained only a tentative candidate with
$P=2.67^{+0.31}_{-0.20}$~d,
$M\sin i=2.07^{+0.52}_{-0.47}\,M_\oplus$, and
$a=0.0409^{+0.0030}_{-0.0021}$~AU. 
\par
\begin{figure}[htbp]
    \centering
    \includegraphics[width=1.15\linewidth]{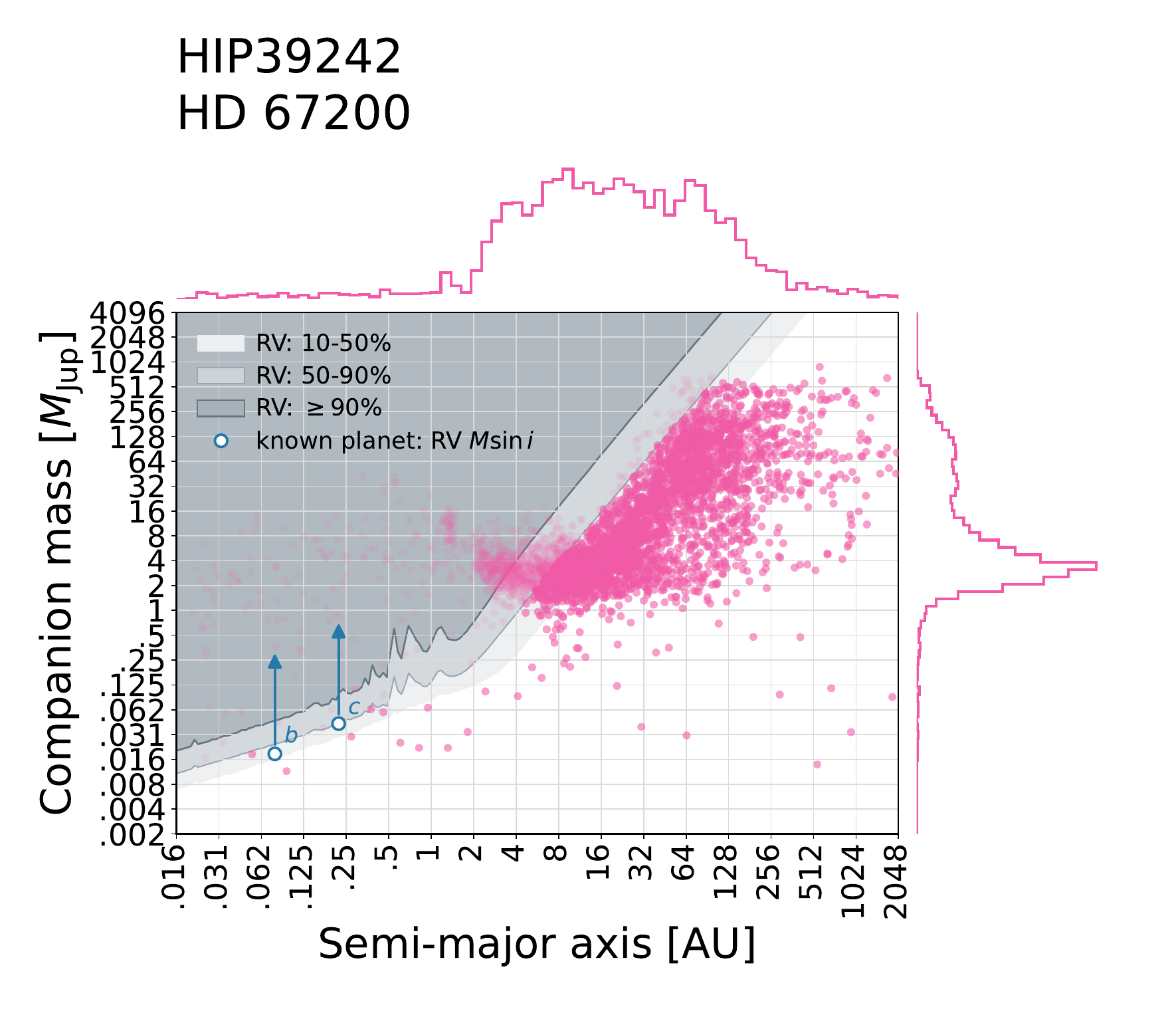}
    \caption{Posterior distribution of companion mass and semi-major axis for HIP 39242 (HD 67200 / DMPP-6). The gray shaded regions show our RV sensitivity proxy based on the published observations.}
    \label{fig:HIP39242}
\end{figure}

Figure~\ref{fig:HIP39242} shows a broad astrometric
posterior for an outer companion with true mass
$M=5.23^{+55.78}_{-2.89}\,M_{\rm J}$ and semi-major axis $a=19.44^{+76.34}_{-15.28}$~AU. This parameter space is clearly distinct from the tentative short period RV candidate. The published data set contains 114 HARPS observations spanning 6.25~yr
\citep{2026MNRAS.547ag370S}. 
\par
The updated activity analysis reports jitter terms of
$0.78^{+0.11}_{-0.09}$ and
$1.24^{+0.36}_{-0.30}$~m\,s$^{-1}$ for the post-2015 and post-2020 HARPS regimes,
respectively, together with a correlated GP amplitude of
$2.91^{+0.99}_{-0.62}$~m\,s$^{-1}$
\citep{2026MNRAS.tmp.1422B}. A post-fit residual
RMS is not reported, so we combined the median GP amplitude and each jitter term in quadrature, obtaining effective scatter proxies of $\sigma_{\rm eff}=3.01$ and $3.16$~m\,s$^{-1}$. 

\subsubsection{HD 90156}
\label{obj:HIP50921}

The HIP~50921 (HD~90156) system comprises a G5V star at a
distance of approximately 22~pc hosting one known planet,
HD~90156~b, discovered with HARPS
\citep{2011AA...526A.111M}. The subsequent HIRES analysis of
\citet{2021ApJS..255....8R} gives a stellar mass of
$0.86\pm0.04\,M_\odot$ and a planetary minimum mass of
$M\sin i=11.8^{+2.0}_{-1.9}\,M_\oplus$, with
$a_b=0.2509\pm0.0037$~AU and
$e_b=0.16^{+0.18}_{-0.11}$.
\par
\begin{figure}[htbp]
    \centering
    \includegraphics[width=1.15\linewidth]{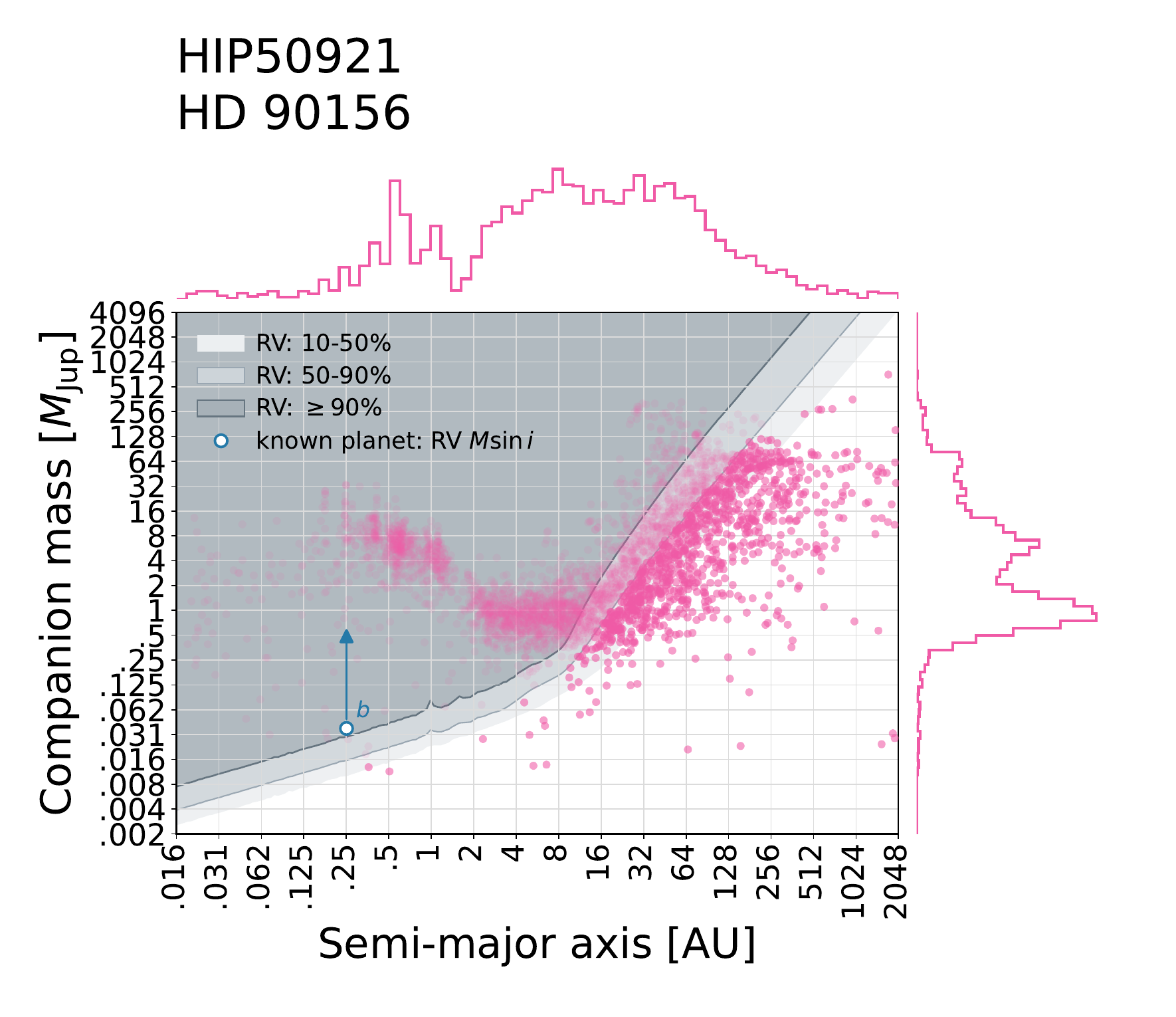}
    \caption{Posterior distribution of mass and semi-major
    axis for HIP~50921 (HD~90156). Gray shaded regions
    correspond to our RV sensitivity proxy.}
    \label{fig:HIP50921}
\end{figure}

Figure~\ref{fig:HIP50921} shows a broad astrometric posterior
for an additional companion with true mass
$M=2.6^{+15.4}_{-1.9}\,M_{\rm J}$ and semi-major axis
$a=11.6^{+61.6}_{-10.6}$~AU. For the RV sensitivity
calculation, we used the 66 HARPS measurements published by \citet{2011AA...526A.111M} and 167 HIRES measurements from
the data release of \citet{2021ApJS..255....8R}. The latter comprise 16 pre-upgrade and 151 post-upgrade measurements. Together, these measurement epochs span 23.17~yr.
The sensitivity grid uses the median G23H stellar mass of
$0.95\,M_\odot$.
\par
For HARPS, we adopted the post-fit residual RMS of
$1.23$~m\,s$^{-1}$ reported in Table~3 of
\citet{2011AA...526A.111M}. For HIRES, we combined the median
formal uncertainties of 1.26 and 1.25~m\,s$^{-1}$ with the
corresponding fitted jitter terms of 2.12 and
1.85~m\,s$^{-1}$ from the machine-readable fit table of
\citet{2021ApJS..255....8R}. Adding these terms in
quadrature gives effective scatter proxies of
$\sigma_{\rm eff}=2.46$ and 2.23~m\,s$^{-1}$ for the
pre- and post-upgrade measurements.
Approximately $76.5\%$ of the astrometric posterior lies in regions with $f_{\rm sens}\geq0.5$, while $53.8\%$ has $f_{\rm sens}\geq0.9$. 
\par
HD~90156 was also observed with VLT/NaCo by
\citet{2015MNRAS.450.3127M}, who did not find a companion candidate. Their Table~7 reports sensitivity to stellar companions at projected separations of 5--177~AU and a minimum detectable mass of approximately
$37\,M_{\rm J}$ assuming an age of 5~Gyr and a detection
threshold of $S/N=3$. These observations constrain the high mass portion of the posterior. The paper provides a detection curve averaged from the survey rather than a specific  curve for this target, so we do not plot this imaging sensitivity contour.

\subsubsection{HD 96735}
\label{obj:HIP54491}

The HIP~54491 (HD~96735, TOI-1799) system comprises a
G-type star of mass $M_\star=0.957\pm0.046\,M_\odot$ at a
distance of 62.1~pc hosting one transiting exoplanet,
TOI-1799~b. The planet has radius $R_b=1.64\,R_\oplus$ and
measured mass
$M_b=4.0^{+1.7}_{-1.8}\,M_\oplus$. It was modeled on a
circular orbit with semi-major axis
$a_b=0.071\pm0.001$~AU and was characterized using
Keck/HIRES and APF radial velocities
\citep{2024ApJS..272...32P,2025AJ....169...89C}.
\par
\begin{figure}[htbp]
    \centering
    \includegraphics[width=1.15\linewidth]{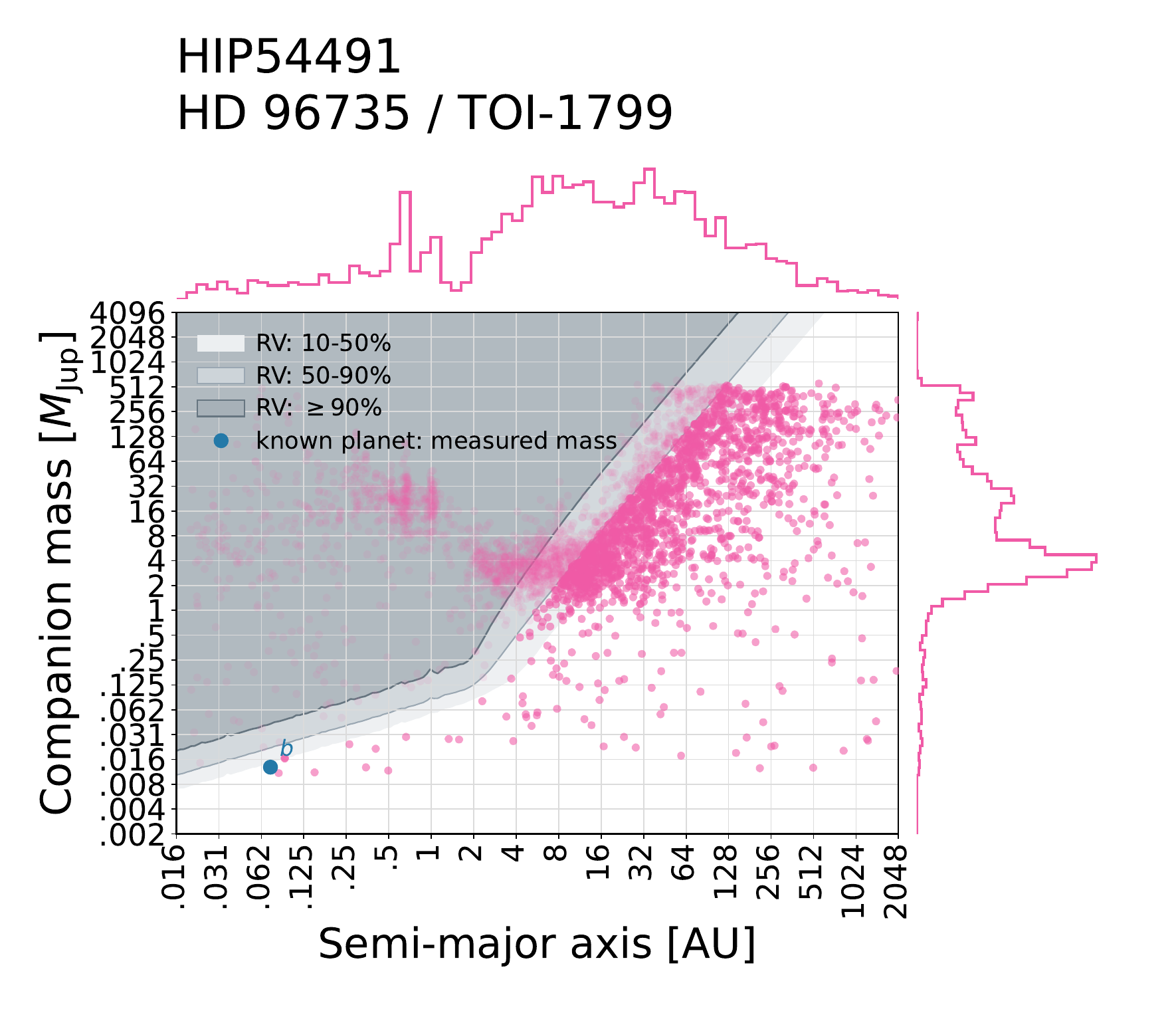}
    \caption{Posterior distribution of companion mass and semi-major axis for HIP 54491 (HD 96735 / TOI-1799). The gray shaded regions show our RV sensitivity proxy based on the published observations.}
    \label{fig:HIP54491}
\end{figure}

Figure~\ref{fig:HIP54491} shows a broad astrometric
posterior for an additional companion with true mass
$M=9.0^{+90.8}_{-6.4}\,M_{\rm J}$ and semi-major axis
$a=13.1^{+79.2}_{-12.1}$~AU. For the RV sensitivity
calculation, we used the 78 usable velocities contained in
the machine-readable data release accompanying
\citet{2024ApJS..272...32P}. These comprise 66 HIRES and 12 APF measurements spanning 781.99~days, or 2.14~yr. We adopted the fitted instrumental jitter terms reported by \citet{2024ApJS..272...32P},
$\sigma_{\rm eff}=3.16$~m\,s$^{-1}$ for HIRES and
$8.0$~m\,s$^{-1}$ for APF, as noise proxies. Under this proxy, 53.5\% of the astrometric posterior
falls within the region having $f_{\rm sens}\geq0.5$, while
31.0\% has $f_{\rm sens}\geq0.9$.

\subsubsection{GJ 514}
\label{obj:HIP65859}
The GJ~514 system comprises an M0.5--M1 dwarf of mass
$0.510\pm0.051\,M_\odot$ at a distance of 7.6~pc hosting
one confirmed planet, GJ~514~b. The planet has
$M\sin i=5.2\pm0.9\,M_\oplus$ and follows an eccentric orbit
with $e=0.45^{+0.15}_{-0.14}$ and semi-major axis
$a_b=0.422^{+0.014}_{-0.015}$~AU
\citep{2022AA...666A.187D}. Its orbit places it within the
conservative habitable zone for approximately 34\% of its
orbital period.
\par
\begin{figure}[htbp]
    \centering
    \includegraphics[width=1.15\linewidth]{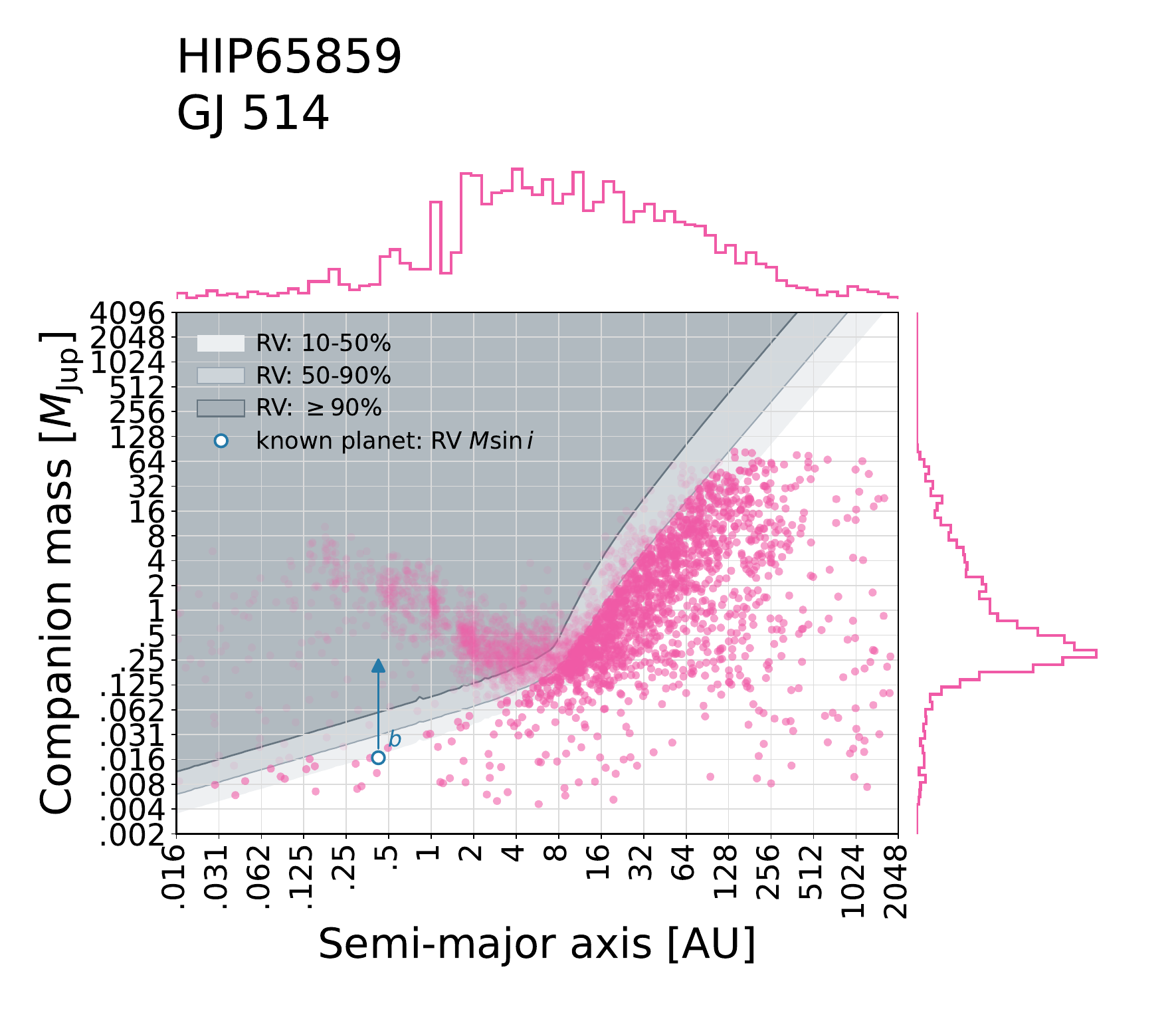}
    \caption{Posterior distribution of companion mass and semi-major axis for HIP 65859 (GJ 514). The gray shaded regions show our RV sensitivity proxy based on the published observations.}
    \label{fig:HIP65859}
\end{figure}

Figure~\ref{fig:HIP65859} shows a broad astrometric
posterior for an additional companion with true mass
$M=0.55^{+3.60}_{-0.34}\,M_{\rm J}$ and semi-major axis
$a=8.89^{+58.18}_{-7.26}$~AU. For the RV sensitivity
calculation, we used all 540 measurements analyzed by
\citet{2022AA...666A.187D}: 104 HIRES, 142 pre-upgrade
HARPS, 20 post-upgrade HARPS, and 274 CARMENES-VIS
velocities. The combined cadence spans 8761~days, or
23.99~yr. We adopted the RMS values reported in their
Table~2 as the noise proxies for each instrument:
3.8~m\,s$^{-1}$ for HIRES, 2.8 and 2.5~m\,s$^{-1}$ for
pre- and post-upgrade HARPS, respectively, and
2.6~m\,s$^{-1}$ for CARMENES-VIS. 
\par
Under our proxy, 53.8\% of the astrometric
posterior has $f_{\rm sens}\geq0.5$, while 38.8\% has
$f_{\rm sens}\geq0.9$. The published RV analysis did not find a significant long period Keplerian
signal nor a significant acceleration. The proper motion anomaly analysis of \citet{2022AA...666A.187D} additionally places a 1$\sigma$ limit of approximately
$0.2\,M_{\rm J}$ between 3 and 10~AU. 




\subsubsection{HD 149143}
\label{obj:HIP81022}

The HIP 81022 (HD 149143) system comprises a G0 star of mass
$1.20\pm0.20\,M_\odot$ hosting one known exoplanet,
HD 149143 b. The planet is a hot Jupiter with
$M\sin i=1.33\pm0.15\,M_{\rm J}$ on a nearly circular orbit
with $e=0.0167\pm0.0040$ and semi-major axis
$a_b=0.0530\pm0.0029$~AU
\citep{2006ApJ...637.1094F,2006AA...446..717D,
2018AJ....156..213M}. The planet was discovered independently
in the N2K and ELODIE radial velocity programs.
\par
\begin{figure}[htbp]
    \centering
    \includegraphics[width=1.15\linewidth]{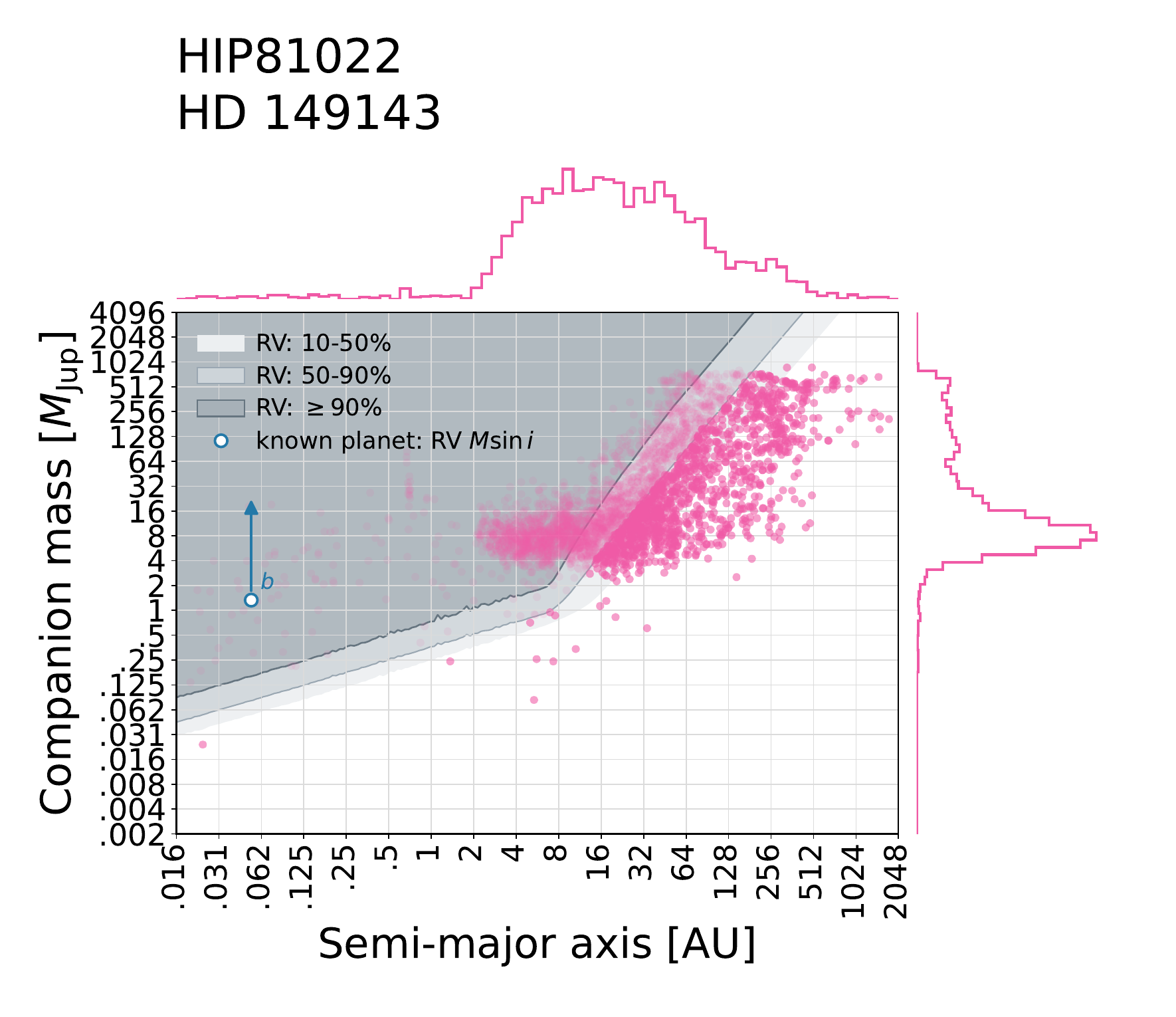}
    \caption{Posterior distribution of companion mass and semi-major axis for HIP 81022 (HD 149143). The gray shaded regions show our RV sensitivity proxy based on the published observations.}
    \label{fig:HIP81022}
\end{figure}

Figure~\ref{fig:HIP81022} shows a broad astrometric posterior
for an additional companion with true mass
$M=12.5^{+109.1}_{-6.5}\,M_{\rm J}$ and semi-major axis
$a=19.2^{+60.6}_{-13.7}$~AU. For the RV sensitivity
calculation, we used the 58 measurements included in the
combined analysis of \citet{2018AJ....156..213M}. Their
Table~2 contains 50 Keck/HIRES epochs, while
the remaining eight measurements correspond to the ELODIE
velocities from \citet{2006AA...446..717D}. Together,
the observations span 4593.26~days, or 12.58~yr. We
applied the post-fit residual RMS of
11.80~m\,s$^{-1}$ reported by
\citet{2018AJ....156..213M} to the HIRES block and the
13.3~m\,s$^{-1}$ post-fit RMS reported by
\citet{2006AA...446..717D} to the ELODIE block.
\par
The resulting sensitivity calculation assigns at least 50\% RV sensitivity to 70.9\% of the G23H posterior samples
and at least 90\% sensitivity to 39.5\%. 

\subsubsection{GJ 674}
\label{obj:HIP85523}
The GJ~674 system comprises an M2.5 dwarf of mass
$0.35\,M_\odot$ hosting one known planet, GJ~674~b
\citep{2007AA...474..293B}. The latest analysis finds $M\sin i=10.95\pm0.14\,M_\oplus$, semi-major axis $a_b=0.03867087\pm0.00000015$~AU, and eccentricity $e=0.242^{+0.012}_{-0.013}$ \citep{2024AA...690A.234L}.
The injection recovery analysis of \citet{2024AA...690A.234L}
finds a mean detection boundary of $K=0.29$~m\,s$^{-1}$ over
periods of 1--100~days. Their
$N$-body integrations show that undetected planets
with periods of 6--30~days, and masses consistent with these RV limits, cannot excite the measured eccentricity of GJ~674~b. Those integrations did not test distant Jovian companions at the
semi-major axes considered here.
\par
\begin{figure}[htbp]
    \centering
    \includegraphics[width=1.15\linewidth]{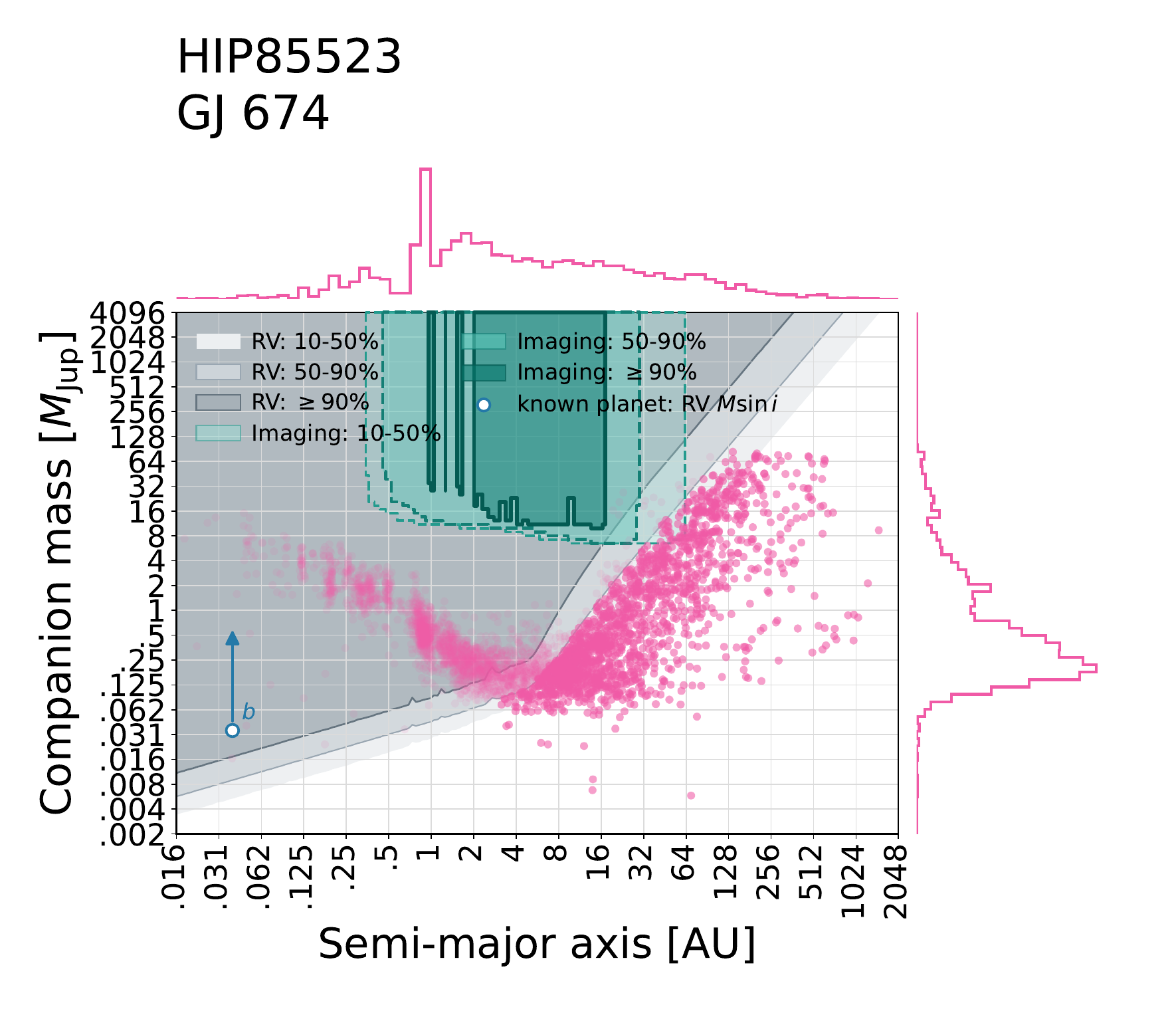}
    \caption{Posterior distribution of companion mass and semi-major axis for HIP 85523 (GJ 674). The gray shaded regions show our RV sensitivity proxy based on the published observations.}
    \label{fig:HIP85523}
\end{figure}

Figure~\ref{fig:HIP85523} shows a broad astrometric posterior for
an additional companion with true mass
$M=0.39^{+2.10}_{-0.23}\,M_{\rm J}$ and semi-major axis
$a=3.26^{+31.51}_{-2.42}$~AU. For our uniform RV sensitivity
calculation, we used the 256 HARPS spectra listed by
\citet{2024AA...690A.234L}. The corresponding observation
timestamps comprise 180 pre-upgrade and 76 post-upgrade
measurements spanning 15.42~yr. Because the authors do not report a
single residual RMS, we combined their fitted GP amplitudes of
2.75 and 2.80~m\,s$^{-1}$ with the corresponding additive
jitter terms of 0.48 and 0.60~m\,s$^{-1}$ in quadrature. This
gives effective scatter proxies of 2.79 and 2.86~m\,s$^{-1}$
for the pre- and post-upgrade measurements, respectively.
Approximately 61.1\% of the astrometric posterior falls in a
region with $f_{\rm sens}\geq0.5$, while 48.0\% has
$f_{\rm sens}\geq0.9$. GJ~674 was also observed with VLT/SPHERE by
\citet{2023AA...680A..64D}. We used their target specific MESS3
direct imaging detection map directly in companion mass and
semi-major axis. Only approximately 0.1\% of the astrometric
posterior falls in a region with at least 50\% imaging sensitivity,
and none of the posterior samples reaches 90\% imaging sensitivity.

\subsubsection{HD 177565}
\label{obj:HIP93858}
The HIP~93858 (HD~177565) system comprises a G6V star of mass
$0.99^{+0.03}_{-0.04}\,M_\odot$ at a distance of 16.9~pc hosting
one known exoplanet, HD~177565~b. The planet has a measured
$M\sin i=15.1^{+6.4}_{-6.1}\,M_\oplus$ and semi-major axis
$a_b=0.246\pm0.019$~AU. It was detected in HARPS radial
velocities with a semi-amplitude of
$K=2.71^{+1.12}_{-0.99}$~m\,s$^{-1}$ 
\citep{2017MNRAS.470.4794F}.
\par
\begin{figure}[htbp]
    \centering
    \includegraphics[width=1.15\linewidth]{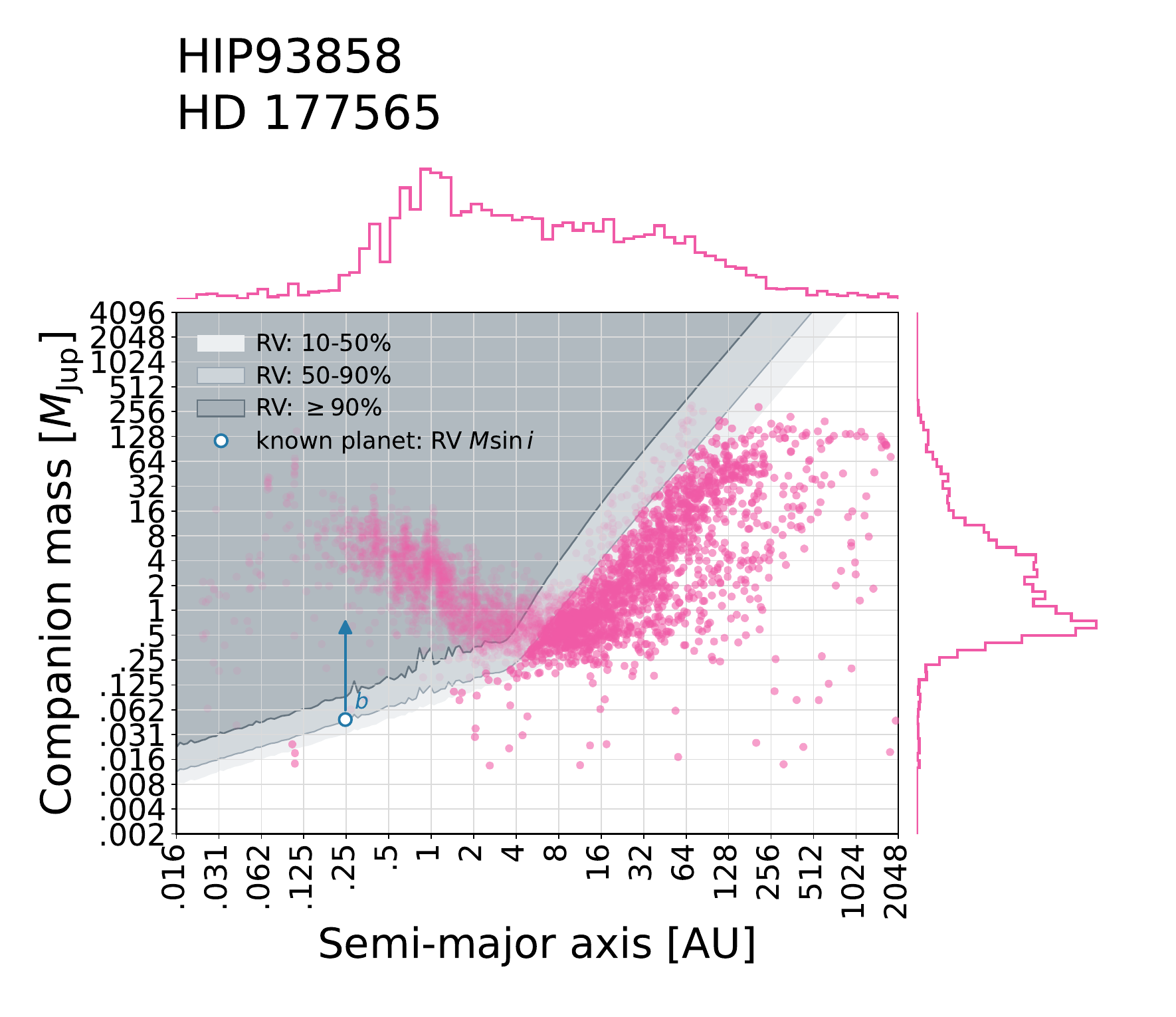}
    \caption{Posterior distribution of companion mass and semi-major axis for HIP 93858 (HD 177565). The gray shaded regions show our RV sensitivity proxy based on the published observations.}
    \label{fig:HIP93858}
\end{figure}

Figure~\ref{fig:HIP93858} shows a broad astrometric posterior
for an additional companion with true mass
$M=1.82^{+7.71}_{-1.25}\,M_{\rm J}$ and semi-major axis $a=4.12^{+43.41}_{-3.41}$~AU. For the RV sensitivity
calculation, we used the 68 HARPS measurements presented in
\citet{2017MNRAS.470.4794F}, which span 1684.39~days, or
4.61~yr. Because the authors do not report a single post-fit
residual RMS, we adopted the 3.34~m\,s$^{-1}$ RMS calculated
directly from their published velocities. Approximately 59\% of the astrometric posterior has $f_{\rm sens}\geq0.5$, while 48\% has
$f_{\rm sens}\geq0.9$. 

\subsubsection{HD 179079}
\label{obj:HIP94256}
The HIP 94256 (HD 179079) system comprises a G5 subgiant
of mass $1.25\pm0.09\,M_\odot$ and radius
$1.792\pm0.016\,R_\odot$ at a distance of
$69.848\pm0.395$~pc \citep{2020AJ....159..197H}. The
system hosts one known exoplanet, HD 179079 b. The
discovery analysis measured
$M\sin i=27.5\pm2.5\,M_\oplus$, an eccentricity
$e=0.115\pm0.087$, semi-major axis
$a_b=0.1216\pm0.0010$~AU, and RV semiamplitude
$K=6.64\pm0.60$~m\,s$^{-1}$ \citep{2009ApJ...702..989V}.
An extended Keck/HIRES analysis favored a circular orbit and  obtained $M\sin i=0.076\pm0.012\,M_{\rm J}$  \citep{2020AJ....159..197H}.
\par
\begin{figure}[htbp]
    \centering
    \includegraphics[width=1.15\linewidth]{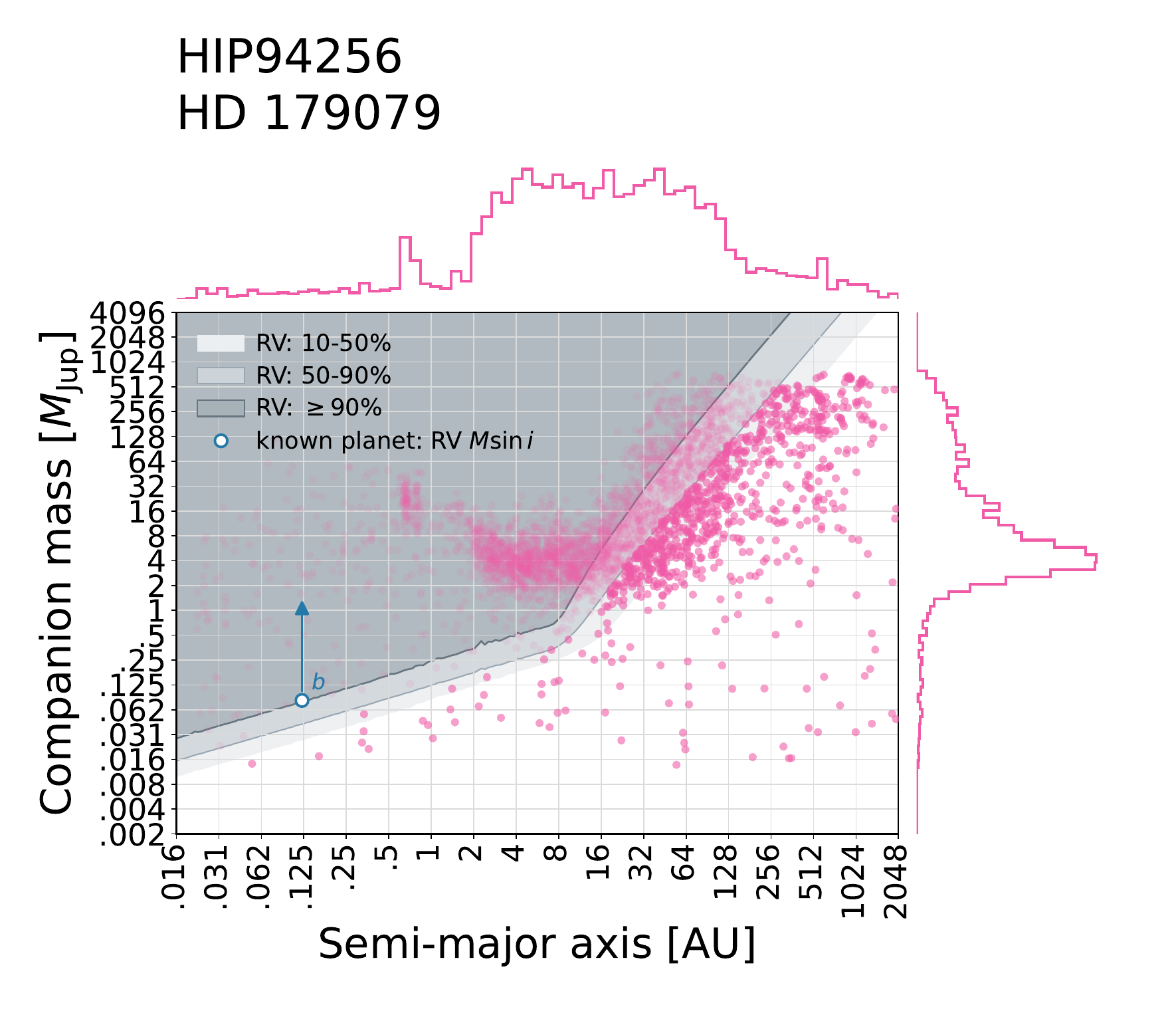}
    \caption{Posterior distribution of companion mass and semi-major axis for HIP 94256 (HD 179079). The gray shaded regions show our RV sensitivity proxy based on the published observations.}
    \label{fig:HIP94256}
\end{figure}

Figure~\ref{fig:HIP94256} shows an astrometric
posterior for an additional companion with true mass
$M=7.2^{+71.7}_{-4.4}\,M_{\rm J}$ and semi-major axis
$a=16.1^{+73.7}_{-13.2}$~AU. For the RV sensitivity
calculation, we used all 93 Keck/HIRES measurements
from \citet{2020AJ....159..197H}, spanning
15.13~yr. We adopted the published post-fit residual RMS
of $4.19$~m\,s$^{-1}$ for the favored one-planet model
without a linear trend. Approximately 79.5\% of the
astrometric posterior falls in a region with
$f_{\rm sens}\geq0.5$, while 55.7\% has
$f_{\rm sens}\geq0.9$. 

\subsubsection{HIP 109384}
\label{obj:HIP109384}

The HIP~109384 (BD$+70\,1218$) system comprises a
G5V star of mass $0.78\pm0.06\,M_\odot$ at
a distance of approximately 56~pc. It hosts one known
giant planet, HIP~109384~b, with
$M\sin i=1.56\pm0.08\,M_{\rm J}$,
$a_b=1.134\pm0.029$~AU, and
$e_b=0.549\pm0.003$, discovered with SOPHIE
\citep{2016AA...588A.145H}.
\par
\citet{2016AA...588A.145H} noted that the velocity offset
between the pre- and post-upgrade observations could
indicate an additional drift, but their data did not
confirm this interpretation. The system therefore presents a tentative indication
of an outer companion rather than a significant RV trend.
\par
\begin{figure}[htbp]
    \centering
    \includegraphics[width=1.15\linewidth]{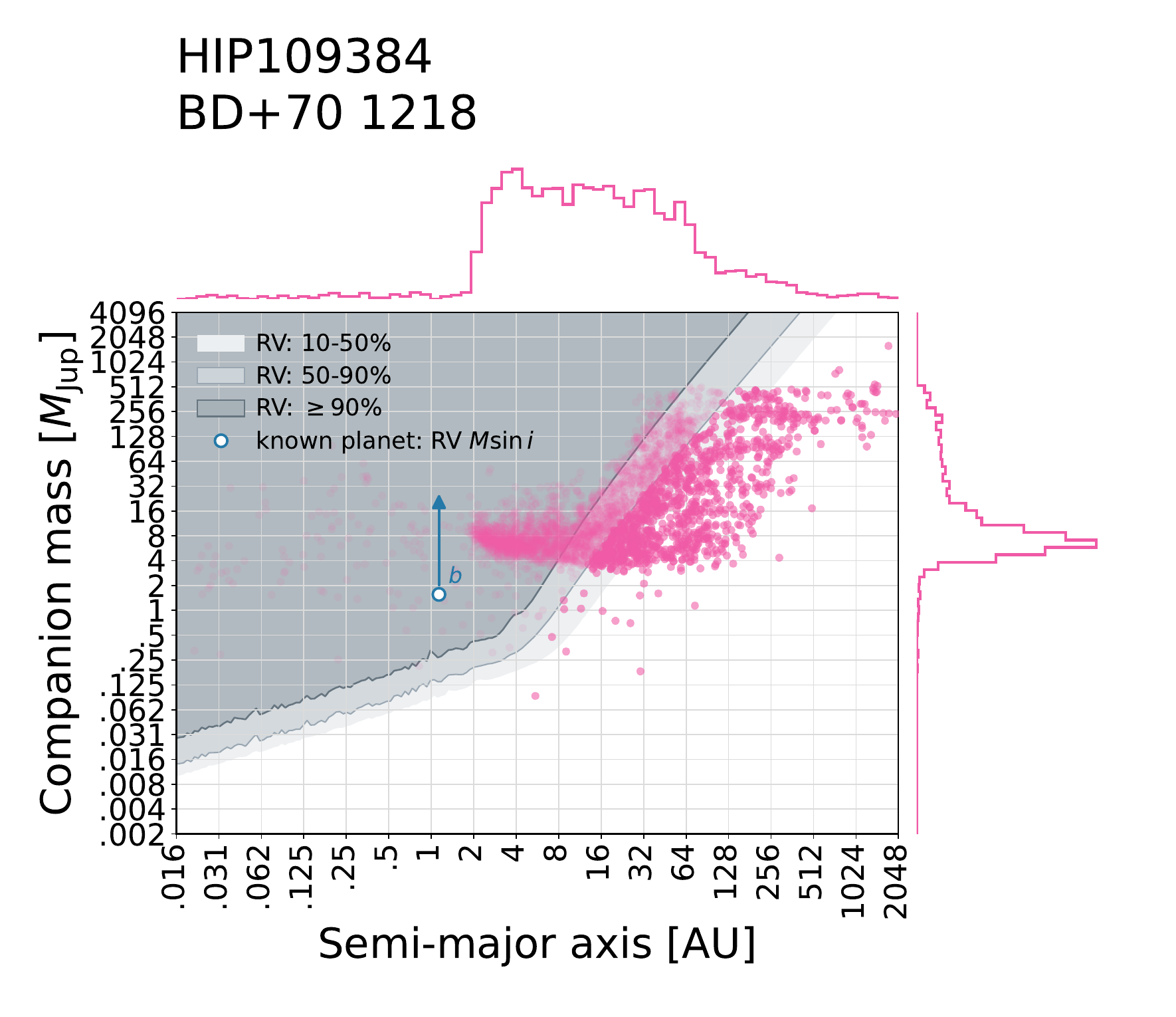}
    \caption{Posterior distribution of mass and semi-major
    axis for HIP~109384. Gray shaded regions correspond
    to our RV sensitivity proxy.}
    \label{fig:HIP109384}
\end{figure}

Figure~\ref{fig:HIP109384} shows a broad astrometric posterior for an additional companion with true mass
$M=9.2^{+54.6}_{-3.9}\,M_{\rm J}$ and semi-major axis
$a=13.5^{+44.6}_{-9.9}$~AU. For the RV sensitivity
calculation, we used all 42 published measurement epochs, comprising seven SOPHIE and 35 SOPHIE+ observations spanning
7.60~yr. The two instrumental regimes were evaluated using the post-fit residual dispersions of
3.7 and 5.8~m\,s$^{-1}$ reported in Table~3 of
\citet{2016AA...588A.145H}. The sensitivity grid uses
the median G23H stellar mass of $0.88\,M_\odot$.
\par
Approximately $73.6\%$ of the astrometric posterior lies in regions with $f_{\rm sens}\geq0.5$, while $44.5\%$ has $f_{\rm sens}\geq0.9$. The astrometric signal could potentially be associated with the tentative RV drift. A joint astrometric and RV fit along with an extended observing baseline would help determine whether the two signals
have a common origin.

\subsubsection{HD 216770}
\label{obj:HIP113238}

The HIP~113238 (HD~216770) system comprises a K0V star
of mass $0.74\pm0.07\,M_\odot$ at a distance of 36.7~pc hosting one exoplanet, HD~216770~b. The planet has $M\sin i=0.57\pm0.05\,M_{\rm J}$ and follows an eccentric orbit with $e=0.37\pm0.06$ and semi-major axis $a_b=0.46\pm0.01$~AU
\citep{2017AJ....153..136S}. It was detected at a semi-amplitude of $K=30.9\pm1.9$~m\,s$^{-1}$
with the CORALIE spectrograph
\citep{2004AA...415..391M}.
\par
\begin{figure}[htbp]
    \centering
    \includegraphics[width=1.15\linewidth]{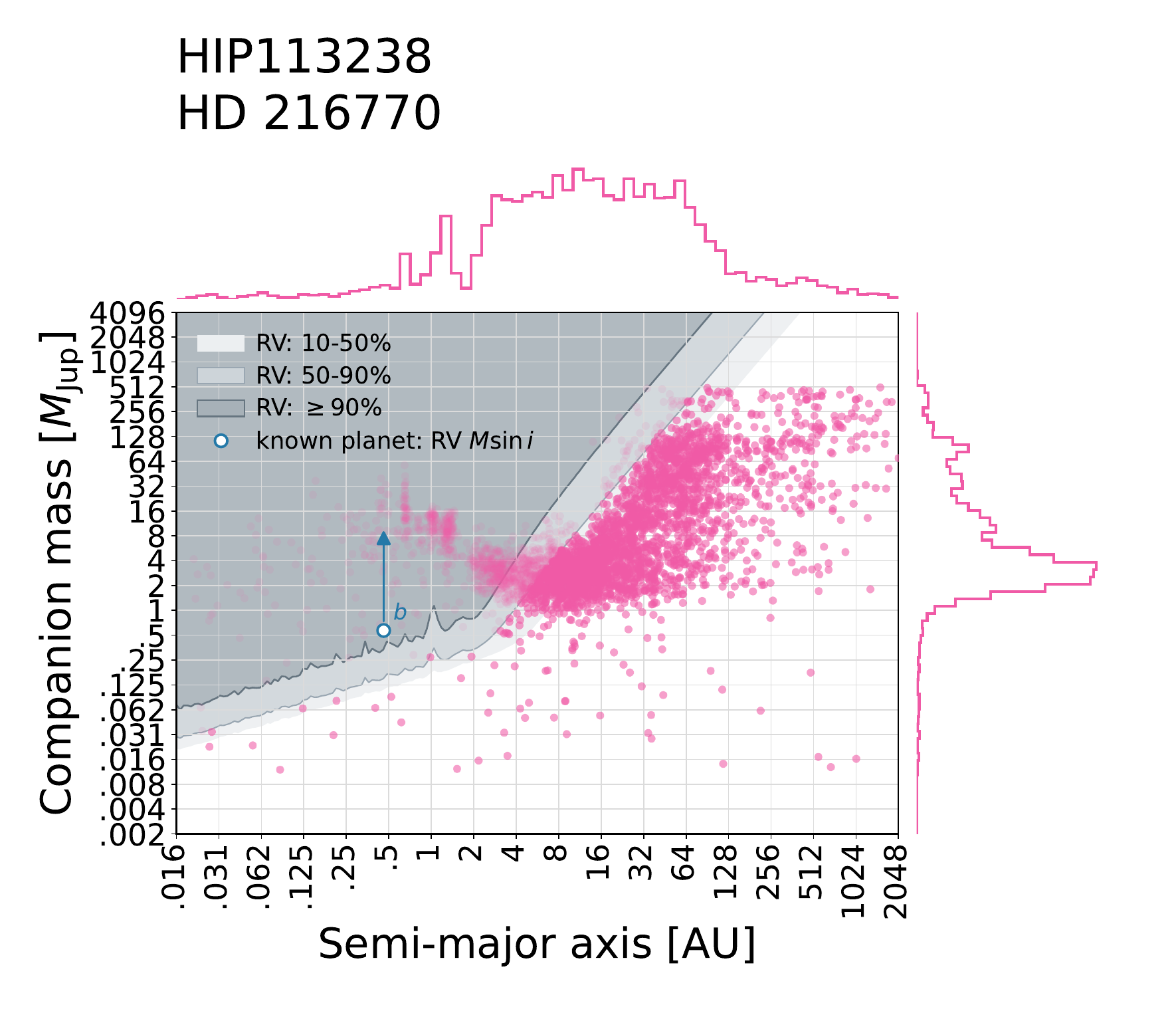}
    \caption{Posterior distribution of companion mass and semi-major axis for HIP 113238 (HD 216770). The gray shaded regions show our RV sensitivity proxy based on the published observations.}
    \label{fig:HIP113238}
\end{figure}

Figure~\ref{fig:HIP113238} shows a broad astrometric posterior
for an additional companion with true mass
$M=4.6^{+33.3}_{-2.6}\,M_{\rm J}$ and semi-major axis
$a=13.4^{+52.0}_{-10.6}$~AU. For the RV sensitivity
calculation, we used the 16 CORALIE measurements from
\citet{2004AA...415..391M}. Because the electronic table
contains one exactly repeated observing epoch, the calculation
uses 15 unique timestamps spanning 827~days, or 2.264~yr. We adopted the published post-fit residual RMS of $\sigma_{\rm eff}=7.8$~m\,s$^{-1}$. Approximately 32.9\% of the astrometric posterior lies in a region with $f_{\rm sens}\geq0.5$, while 19.4\% has $f_{\rm sens}\geq0.9$. 

\subsubsection{GJ 887}
\label{obj:HIP114046}

The HIP 114046 (GJ 887, Lacaille 9352) system comprises a M1V dwarf of mass
$0.495\pm0.049\,M_\odot$ at a distance of 3.29~pc. Two
super-Earths, GJ~887~b and c, were discovered from HARPS
radial velocities \citep{2020Sci...368.1477J}, together with
an unconfirmed signal near 50~d. A subsequent analysis
recovered both planets, confirmed the 50-d signal as a
habitable zone planet, and identified an additional
Earth-mass planet with a sub-m\,s$^{-1}$ RV semi-amplitude
\citep{2026AA...707A..93H}. The resulting four-planet system
comprises GJ~887~e
($M\sin i=1.46^{+0.19}_{-0.18}\,M_\oplus$,
$a_e=0.0417\pm0.0014$~AU), b
($3.9\pm0.5\,M_\oplus$,
$a_b=0.0683^{+0.0022}_{-0.0024}$~AU), c
($6.5^{+1.0}_{-0.9}\,M_\oplus$,
$a_c=0.121^{+0.004}_{-0.005}$~AU), and d
($6.1\pm1.4\,M_\oplus$,
$a_d=0.212^{+0.007}_{-0.008}$~AU). An additional signal at
2.2~d, corresponding to $M\sin i\simeq0.47\,M_\oplus$ if
planetary, remains unconfirmed.
\par
\begin{figure}[htbp]
    \centering
    \includegraphics[width=1.15\linewidth]{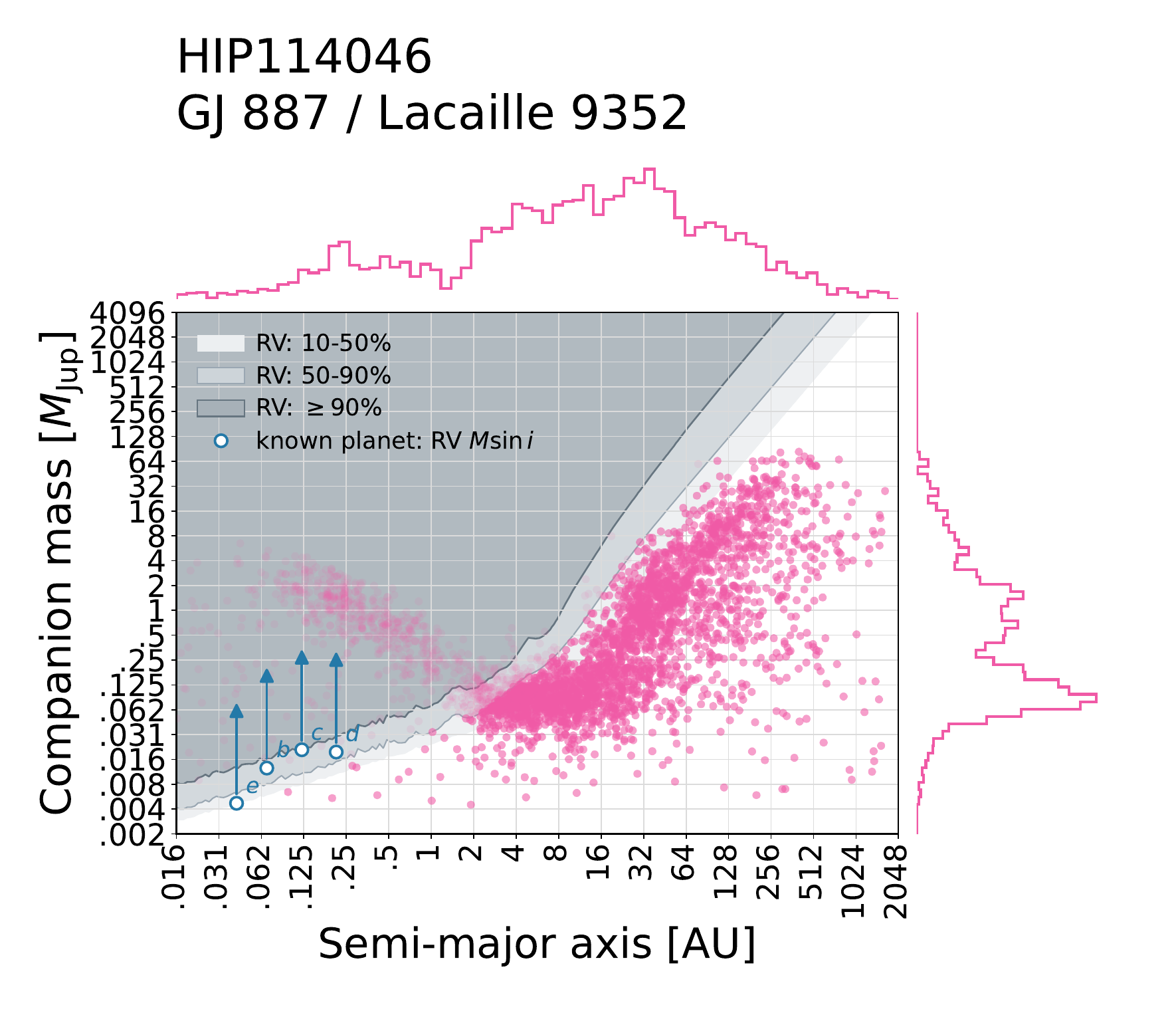}
    \caption{Posterior distribution of companion mass and semi-major axis for HIP 114046 (GJ 887 / Lacaille 9352). The gray shaded regions show our RV sensitivity proxy based on the published observations.}
    \label{fig:HIP114046}
\end{figure}

Figure~\ref{fig:HIP114046} shows a broad astrometric
posterior for an additional companion with true mass
$M=0.34^{+2.52}_{-0.27}\,M_{\rm J}$ and semi-major axis
$a=14.1^{+81.6}_{-13.2}$~AU. There is no single
reported post-fit residual RMS, so we combined the fitted Gaussian process amplitude and jitter of each block in quadrature, obtaining respectively,
$\sigma_{\rm eff}=3.66$, 4.00, 2.60, and
1.41~m\,s$^{-1}$ for HARPSpre, HARPSpost, HARPSpw, and ESPRESSO.
\par
Approximately 24.8\% of the astrometric posterior lies in a region with $f_{\rm sens}\geq0.5$, while 18.8\% has $f_{\rm sens}\geq0.9$. 
The mean detection limit of $25\pm3$~cm\,s$^{-1}$ reported by
\citet{2026AA...707A..93H} applies to periodic signals with
periods between 0.5 and 100~d and was for that reason not applied
directly to the multi-year companions considered here.

\subsubsection{HD 224693}
\label{obj:HIP118319}

The HIP~118319 (HD~224693) system comprises a G2V star of
mass $1.31\pm0.29\,M_\odot$ hosting one known exoplanet,
HD~224693~b. The planet has a measured
$M\sin i=0.70\pm0.12\,M_{\rm J}$ and is on a nearly circular
orbit with eccentricity $e=0.104\pm0.017$ and semi-major axis
$a_b=0.191\pm0.014$~AU. It was discovered with Keck/HIRES as
part of the N2K survey \citep{2006ApJ...647..600J}. An updated
analysis using a longer HIRES baseline refined its RV
semi-amplitude to $K=39.96\pm0.68$~m\,s$^{-1}$ and found no
additional significant signal in the residual velocities
\citep{2018AJ....156..213M}.
\par
\begin{figure}[htbp]
    \centering
    \includegraphics[width=1.15\linewidth]{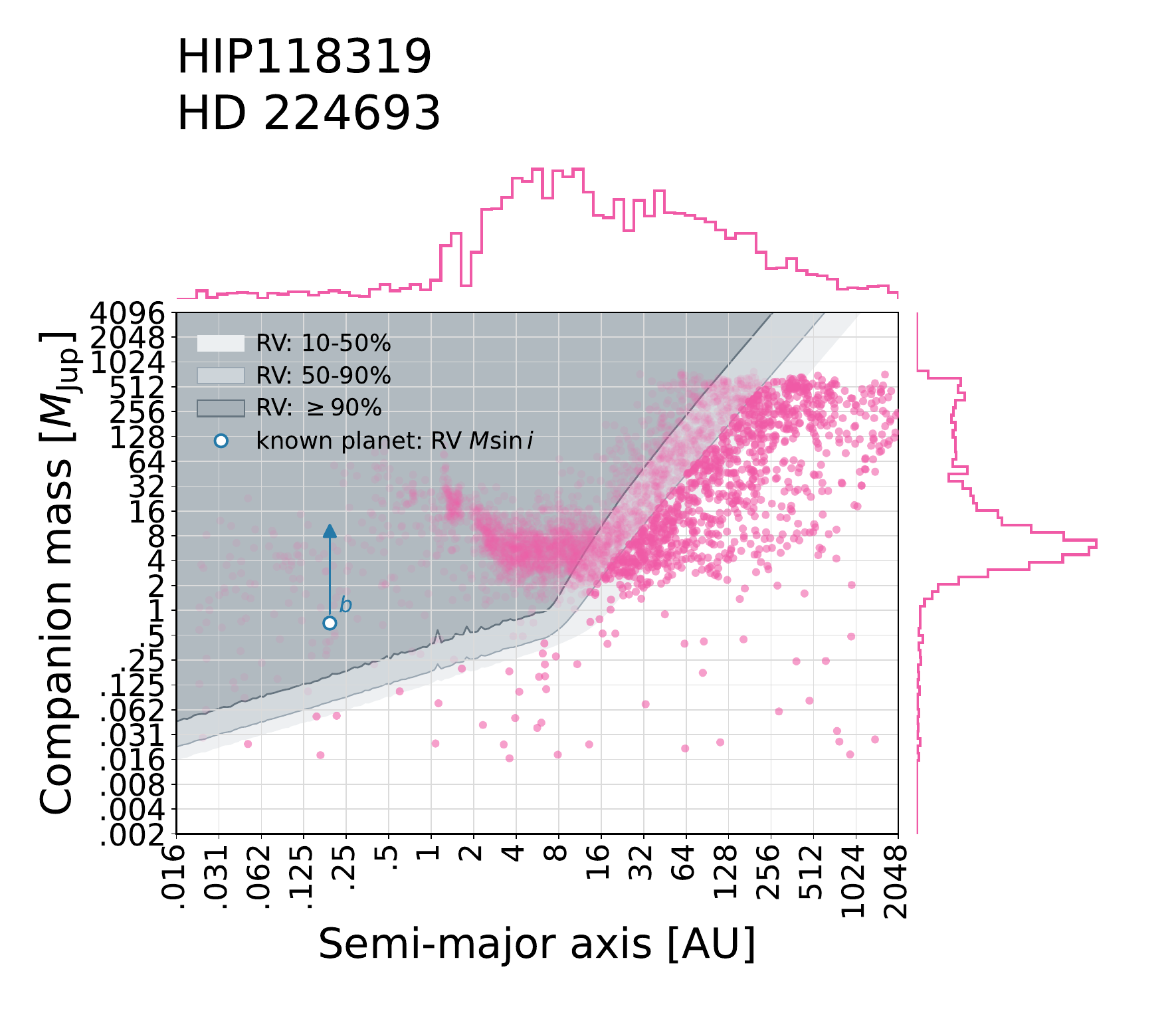}
    \caption{Posterior distribution of companion mass and semi-major axis for HIP 118319 (HD 224693). The gray shaded regions show our RV sensitivity proxy based on the published observations.}
    \label{fig:HIP118319}
\end{figure}

Figure~\ref{fig:HIP118319} shows a broad astrometric
posterior for an additional companion with true mass
$M=9.4^{+120.1}_{-5.5}\,M_{\rm J}$ and semi-major axis
$a=14.4^{+105.1}_{-11.4}$~AU. For the RV sensitivity
calculation, we used all 40 Keck/HIRES measurements tabulated
by \citet{2018AJ....156..213M}. These observations span
12.07~yr, and we adopted the published post-fit residual RMS
of 5.73~m\,s$^{-1}$ as $\sigma_{\rm eff}$. Approximately
74.8\% of the astrometric posterior lies in a region with
$f_{\rm sens}\geq0.5$, while 51.0\% has
$f_{\rm sens}\geq0.9$. 

\section{Discussion}
\label{sec:discussion}
\subsection{Interpreting the Astrometric Candidates}

Our comparison to published observations emphasizes the need to interpret the G23H posteriors in conjunction with existing RV and imaging constraints. Our astrometric solutions cover a large range of companion masses and orbital scales, including regions where our sensitivity proxy indicates detectable RV variations or trends. Eight of the 29 systems have known or tentative long-term RV variations reported in the literature (Section~\ref{sec:rv_trends}). \par
For systems without a reported residual RV signal, the reported G23H signals motivate their further monitoring and analysis with RVs and with GDR4. They may be indicative of an additional companion with low orbit coverage or arise from stellar activity or instrumental effects. 
\par
The astrometric posteriors provided here do not establish that the RV trends and our G23H signals share a common origin. Our goal here is simply to compare the G23H posteriors to published constraints using a common proxy for sensitivity. A joint fit with all available datasets for a given system (imaging, RV, astrometry) could establish whether the signals share a common origin. We leave these particular fits for future work. Continued RV monitoring would extend the observed orbital arcs and help to disentangle companion signals from other sources of long-term variation.

\subsection{Orbital Scales and Comparison with RV Surveys}
\label{sec:occurrence_comparison}

For the 21 systems without reported RV trends,
the sensitivity screen shifts the retained
astrometric samples toward wider orbits
(Figure~\ref{fig:sample_sma_comparison}).
The number of systems with median semi-major axes
above 10~AU increases from 13/21 before screening to
20/21 afterward, with the screened medians spanning
approximately 9--138~AU. This shift reflects the preferential sensitivity
of the published RV observations to closer companions.
\par
These orbital scales are of interest in the context of
RV occurrence measurements. Using the CLS, \citet{2021ApJS..255...14F} measured
$14.1^{+2.0}_{-1.8}$ planets per 100 stars at 2--8~AU
and $8.9^{+3.0}_{-2.4}$ at 8--32~AU, for minimum masses
of $30$--$6000\,M_\oplus$. Their results favor a decline
at wider separations, but the occurrence beyond 10~AU
remains uncertain because of low completeness.  However, a more recent joint RV and astrometric analysis
finds a roughly constant occurrence rate density over
2--20~AU for massive giant planets and low-mass brown
dwarfs \citep{2026PNAS..12324764C}. The
background bands in Figure~\ref{fig:sample_sma_comparison}
mark these comparison intervals and separations beyond
32~AU. Our candidates could help investigate this less
constrained population if their planetary masses and
wide orbits are confirmed with GDR4. However, the current sample alone cannot establish whether they are rare outliers or indicate a different occurrence distribution than what is reported in the previous literature. 

\subsection{Prospects for Direct Imaging}
\label{sec:imaging_prospects}

HD~63433 (Section~\ref{obj:HIP38228}) and
HD~63454 (Section~\ref{obj:HIP37284}) are of particular interest for further investigation with direct imaging. The relevant quantities for planning imaging observations are the predicted contrast and projected angular separation at a given observing date.
\par
HD~63433 is a relatively young ($\sim$400~Myr), nearby system. Approximately 45\% of its astrometric posterior lies in regions with at least 50\% sensitivity from RVs
or VLT imaging. Outside these regions,
the 16th--84th percentile intervals span companion
masses of $0.64$--$12.0\,M_{\rm J}$ and semi-major axes
of $8$--$214$~AU. Therefore, its young age and remaining wide orbit solutions motivate deeper imaging datasets.
\par
HD~63454 is approximately 1~Gyr old and has a reported
RV trend. Its unfiltered astrometric posterior
spans companion masses of $3.4$--$61.8\,M_{\rm J}$ and semi-major axes of $6.1$--$62.1$~AU over the 16th--84th percentile intervals. Imaging could test whether the signal arises from the wider or more massive posterior solutions. Continued RV monitoring would help connect any detected companion to the reported trend.

\section{Conclusion}

In this work, we used the G23H catalogue to search for astrometric evidence of additional outer companions in systems with known inner planets. Identifying outer giant planets is important for understanding their influence on
the orbital architecture and dynamical evolution of these systems. The main findings of this work are:
\begin{enumerate}
    \item From an initial sample of 170 systems, we
    identify 29 systems as candidates for additional
    outer companions (Section~\ref{sec:results}).
    In each system, at least one fitted component has
    posterior support for masses between $0.3$ and
    $13\,M_{\rm J}$ and a median semi-major axis beyond
    that of the outermost known planet. Each qualifying
    component also satisfies $C_{\rm known}>0.95$:
    the published parameters of every assessed known
    planet lie outside the corresponding 95\% posterior
    region. These criteria identify candidates for
    further investigation rather than confirm the existence of giant
    planets. Brown dwarf or stellar companion solutions
    also remain possible in some systems.
    \item We made sensitivity maps from published
    RV observations and, where quantitative constraints were
    available, imaging observations from the literature
    (Sections~\ref{sec:rv_limits} and
    \ref{sec:imaging_limits}). These maps identify
    portions of the astrometric parameter space where
    companions would be expected to produce detectable
    signals. They provide a uniform basis for
    assessing the candidates and planning follow-up
    observations, but are approximate sensitivity
    estimates rather than formal exclusion limits.
    \item Eight systems have known or tentative long-term
    RV variations reported in the literature
    (Section~\ref{sec:rv_trends}). Their astrometric
    posteriors provide candidate counterparts to these
    variations and initial constraints on possible
    companion masses and orbital scales. Our sensitivity
    calculation does not establish that the astrometric
    and RV signals have a common origin. Joint
    astrometric and RV fits are required to test whether
    the same companion can reproduce both signals.
\end{enumerate}
Future astrometric measurements from \textit{Gaia} DR4,
combined with continued RV monitoring and targeted
imaging, will help test these candidates and refine
their companion masses and orbital parameters. For confirmed companions, these constraints will enable dynamical studies of their influence on the
formation and evolution of the known inner planets.

\begin{acknowledgments}
C.D.O acknowledges support from the B. Thomas Soifer Fellowship at the California Institute of Technology.
\par
A.S acknowledges support from the National Science Foundation Graduate Research Fellowship under Grant No.~2139433.
\par
J.W.X is grateful for support from the Heising-Simons Foundation 51 Pegasi b Fellowship (grant \#2025-5887). 
J.S. is grateful for support from the Heising–Simons Foundation through the 51 Pegasi b Fellowship (grant \#2025-5885).
K.F. acknowledges support through the NASA Hubble Fellowship grant HST-HF2-51574.001-A, awarded by the Space Telescope Science Institute, which is operated by the Association of Universities for Research in Astronomy, Inc., for NASA under contract NAS5-26555.
This research has made use of the NASA Exoplanet Archive, which is operated by the California Institute of Technology, under contract with the National Aeronautics and Space Administration under the Exoplanet Exploration Program.
\end{acknowledgments}

\clearpage
\appendix
\section{Posterior Contours and Known Planet Associations}
\label{app:known_planet_association}

Here we illustrate how we compare the G23H posterior with the
known planets in each system. These contours describe the
astrometric posterior density and its overlap with the known planets' parameters. For a planet with a published minimum mass, we
perform the comparison in $(\log_{10}a,\log_{10}M\sin i)$
space. Each G23H mass is projected using the inclination from
the same posterior draw. When a true mass
is available, we use $(\log_{10}a,\log_{10}M)$ space.
The calculation uses the published mean estimates for the known planets and does not include their measurement uncertainties.
\par
We estimate posterior density by measuring how closely
the samples are clustered. Before measuring distances,
we rescale and rotate the logarithmic mass and
semi-major axis coordinates to account for their
different spreads and correlation. For each sample,
we then measure the distance to its
$n_{\rm nn}$th nearest neighbor: shorter distances
indicate higher local density. We adopt
$n_{\rm nn}=\max[30,\mathrm{round}(\sqrt{N})]$,
where $N$ is the number of valid posterior draws
containing the selected companion. For example,
4096 draws give a neighbor count of 64.
\par
We define
\begin{equation}
    C_k=\frac{1}{N}\sum_{j=1}^{N}\mathbf{1}(r_j\leq r_k),
    \label{eq:known_planet_density_rank}
\end{equation}
where $\mathbf{1}$ is one when the condition is satisfied and
zero otherwise. This estimates the probability content of the
smallest highest posterior density region containing the known
planet's location. A small $C_k$ places the known planet in a
high density part of the posterior. Conversely, a value near one places it in the outskirts.
\par
To display the contours, we evaluate the same distance estimator on a $200\times200$ logarithmic grid. The 68\%, 95\%, and 99.7\% boundaries correspond to the respective
percentiles of the sample distances $r_j$. These are approximate joint credible regions in the logarithmic comparison space. The classification is evaluated directly at the known planet's location.
\par
For each fitted component, we adopt
$C_{\rm known}=\min_k C_k$, taking the minimum over
the assessed known planets. A component satisfies
our association criterion when $C_{\rm known}>0.95$,
meaning that every assessed known planet lies outside
its estimated 95\% posterior region.
\par
\begin{figure*}[htbp]
    \centering
    \includegraphics[width=\textwidth]{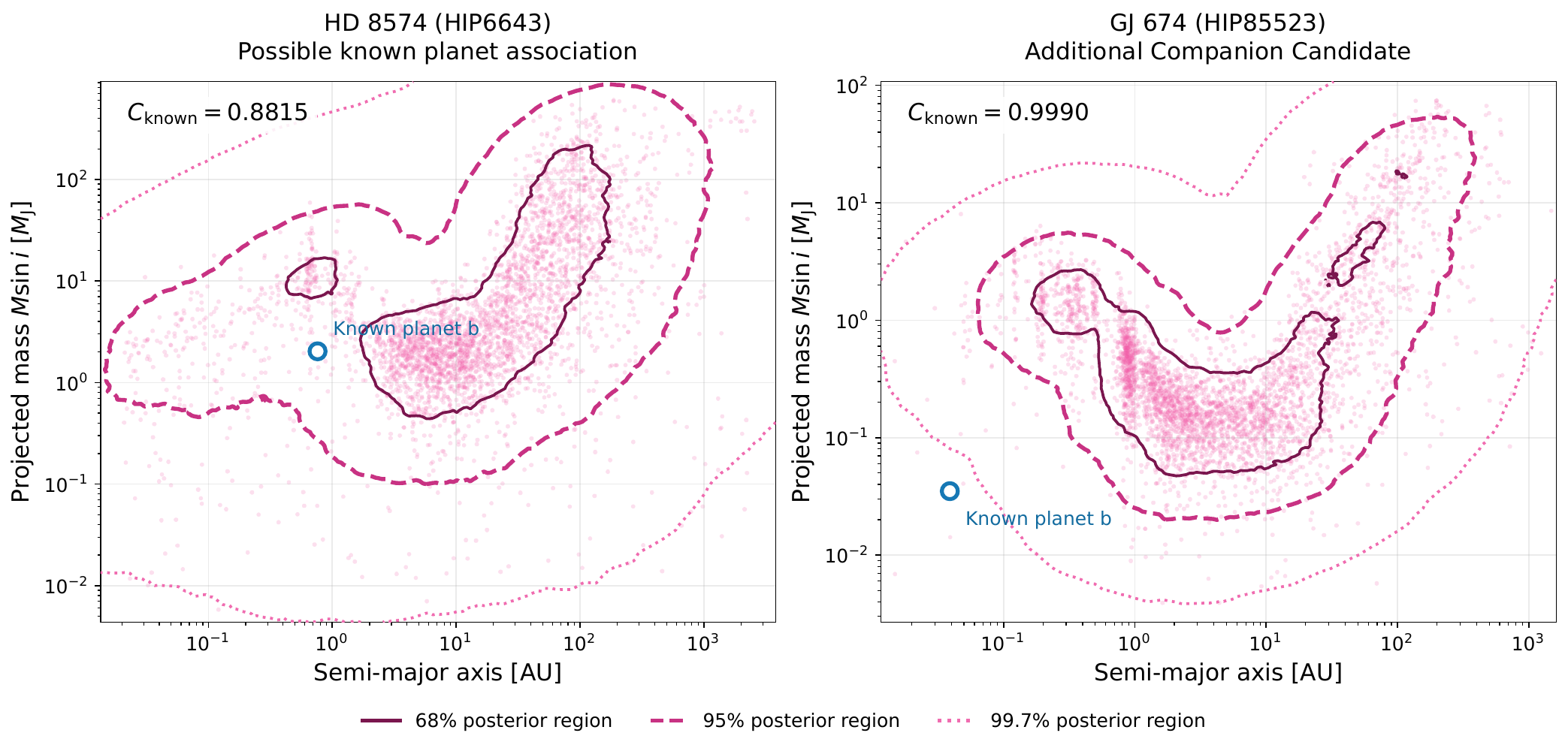}
    \caption{Examples of the posterior comparison with known
    planets for HD~8574 (left) and GJ~674 (right). Pink points
    show the G23H draws projected into $M\sin i$ using their
    individual inclinations. Blue open circles mark the
    published parameters of the known planets. Solid, dashed,
    and dotted curves show the approximate 68\%, 95\%, and
    99.7\% highest posterior density regions. HD~8574~b lies inside the 95\% region,
    allowing a possible association of its presence with the detected signal, whereas
    GJ~674~b lies outside the 99.7\% region.}
    \label{fig:known_planet_association_examples}
\end{figure*}

Sample membership is determined using the  neighbor count defined above. To assess sensitivity
to the density estimate, we repeat the calculation
using 30 and 80 neighbors. Three systems
(BD$+55\,362$, BD$+45\,564$, and LHS~1903) fall below
$C_{\rm known}=0.95$ under at least one alternative
choice. However, they pass the nominal criterion and therefore are retained in our sample.
Figure~\ref{fig:known_planet_association_examples} illustrates the two different outcomes. For HD~8574, $C_{\rm known}=0.8815$, and the known planet lies inside the 95\% region. For GJ~674,
$C_{\rm known}=0.9990$, and the known planet lies outside
even the 99.7\% region. GJ~674 therefore passes the
known planet separation criterion for an additional companion.
This does not establish that the astrometric signal is a new
planet. Likewise, an overlap does not
identify the known planet as the source. A joint astrometric
and RV fit would be required to test either association.

\onecolumngrid
\section{RV Sensitivity Contour: An Example}
\label{app:rv_sensitivity_example}
Here we illustrate the computation of the presented RV sensitivity contours. We
compute the stellar radial velocity for a companion mass at the observing epochs present on the Table in Appendix \ref{app:rv_inputs}. The baseline $T_{\rm RV}$ specifies the interval
covered by these epochs and $N_{\rm RV}$ gives the number of measurements. We use the literature observation dates when available. The stellar mass used for each target's sensitivity grid is the median of its G23H stellar mass posterior. Given a semi-major axis and companion mass, we consider 10,000 trial orbital
configurations with uniformly drawn orbital parameters (isotropic in inclination). We then compute the stellar radial velocity at the adopted
observing epochs for each configuration. We then measure the difference between the largest and smallest sampled velocities within each
instrumental regime and compare this range with the quadrature sum of the effective uncertainties assigned to the two epochs. A configuration meets our sensitivity criterion if its velocity range exceeds this threshold
in at least one instrumental regime.
\par
\begin{figure*}[htbp]
    \centering
    \includegraphics[width=\textwidth]{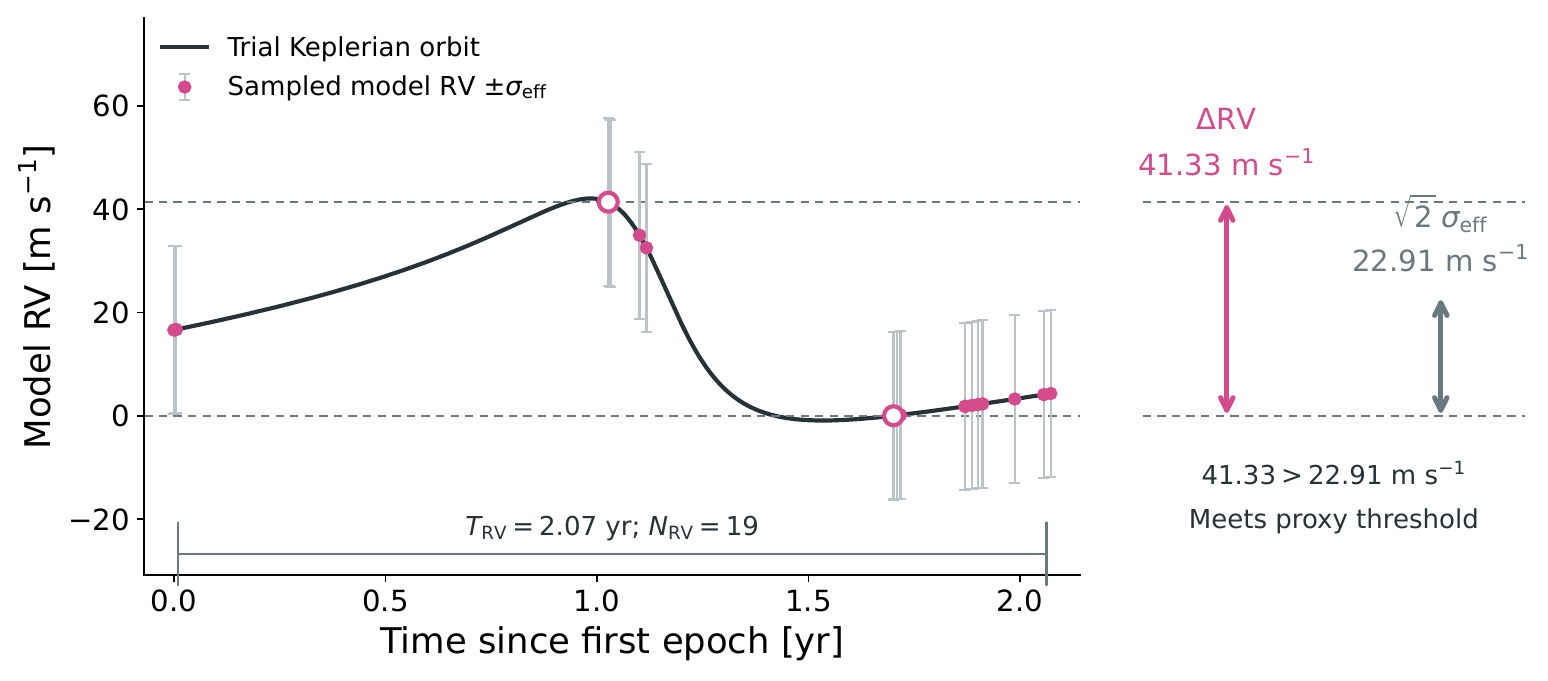}
    \caption{Illustration of the RV sensitivity proxy for a single
    instrumental regime for HIP 5763 using the published CHIRON epochs reported in \citealt{2021AJ....162..176P}. Left panel: The solid line shows a trial Keplerian
    orbit and the pink points show the predicted velocities at 19
    reported observing epochs spanning $\sim$2.07 yr. Error bars
    represent the adopted effective scatter of
    $\sigma_{\rm eff}=16.2$~m\,s$^{-1}$.
    The dashed lines mark the maximum and minimum sampled
    velocities whose difference is
    $\Delta{\rm RV}=41.33$~m\,s$^{-1}$. Right panel:
    The arrows compare this variation with the adopted threshold
    of $\sqrt{2}\sigma_{\rm eff}=22.91$~m\,s$^{-1}$ on the same
    velocity scale. This realization meets the proxy criterion
    because $\Delta{\rm RV}>\sqrt{2}\sigma_{\rm eff}$.}
    \label{fig:rv_proxy_delta_panel}
\end{figure*}

The fraction of configurations satisfying this criterion
defines the RV sensitivity at that mass and semi-major
axis. We construct the sensitivity map on a logarithmically
spaced grid with 200 values along each axis. We visually show this process in Figure \ref{fig:rv_proxy_delta_panel} for one of our targets (HIP 5763).

\section{Imaging Sensitivity Contour: An Example}
\label{app:imaging_sensitivity_example}
Here we illustrate the computation of the presented imaging
sensitivity contours. We use the published contrast curves
listed in Table~\ref{tab:imaging_inputs} along with the
Sonora Red Diamondback and BHAC15 evolutionary models
\citep{2025ApJ...994..198D,2015AA...577A..42B} to estimate
the brightness of trial companions. We consider 10,000 trial orbital
configurations using a given semi-major axis and companion mass. We draw orbital parameters uniformly (isotropic in inclination). The mean anomaly specifies the orbital phase
at the imaging observation. We then compute the projected
physical separation for each orbital configuration and divide this
separation in AU by the system distance in pc to obtain
the angular separation $\rho$ in arcseconds.
\par
For each orbital realization, we also draw a system age from the
adopted age distribution accounting for its uncertainty. The companion mass and sampled age determine its model absolute magnitude.
We convert the apparent stellar magnitude to an absolute
magnitude using the system distance and subtract it from
the companion magnitude to obtain the predicted
$\Delta\mathrm{mag}$. We use model $K_s$ magnitudes as a
proxy where the observed filter is not provided directly. We interpolate the published contrast curve
at the trial angular separation and compare its limiting
$\Delta\mathrm{mag}$ with the predicted value. A realization
meets our imaging sensitivity criterion when the companion
is brighter than this limit and lies within the adopted
angular coverage. We do not assess sensitivity below the supported model mass range.
\par
\begin{figure*}[htbp]
    \centering
    \includegraphics[width=\textwidth]{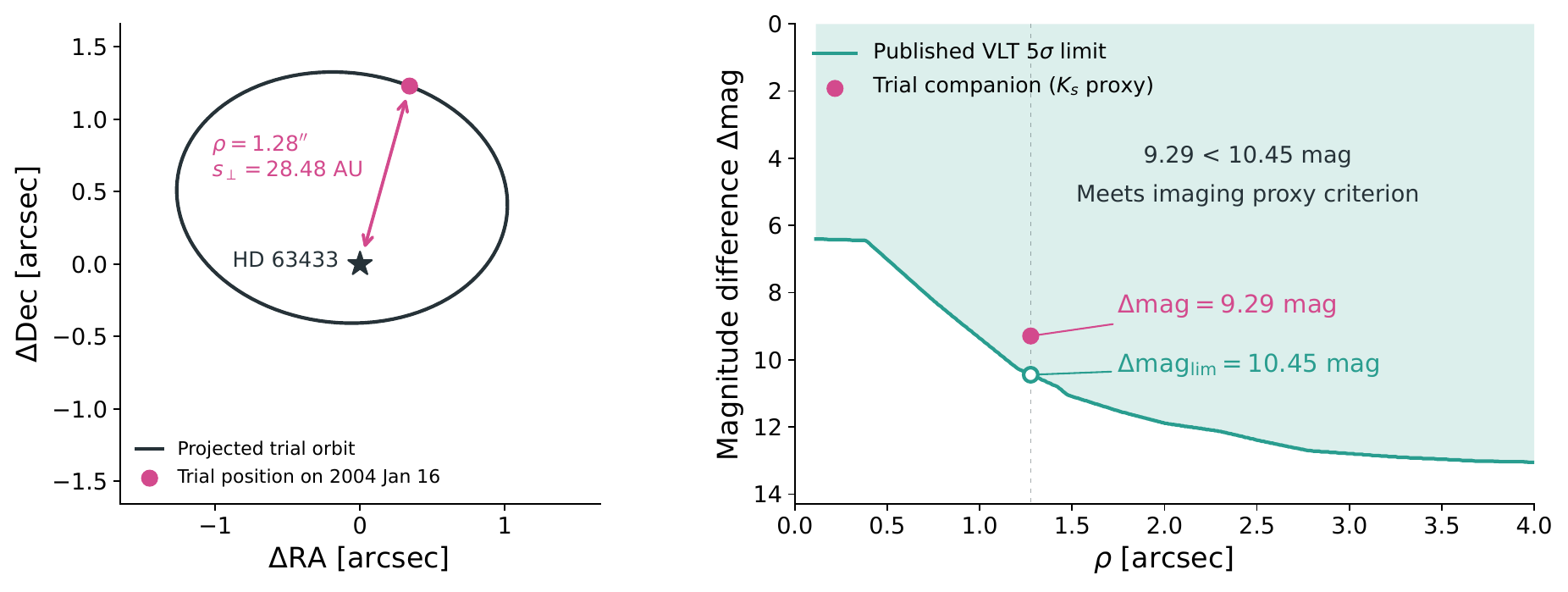}
    \caption{Illustration of the imaging sensitivity proxy
    for HD~63433 using the VLT $5\sigma$ contrast curve in
    Figure~5 of \citealt{2024AJ....167...54C}.
    Left panel: The solid line shows the projected
    orbit of a trial companion with $M=30\,M_{\rm J}$ and
    $a=30$~AU. The pink point marks its simulated position
    on 2004 January 16, the adopted observation date
    \citep{2020AJ....160..179M}. Its projected physical
    separation is 28.48~AU, corresponding to
    $\rho=1.28''$ at a distance of 22.32~pc.
    Right panel: At the sampled age of 412~Myr, the models
    predict $\Delta\mathrm{mag}=9.29$, using $K_s$ as a
    proxy for the observed $K$ band. The published limit
    at this separation is
    $\Delta\mathrm{mag}_{\rm lim}=10.45$. This realization
    meets the imaging proxy criterion because
    $\Delta\mathrm{mag}<\Delta\mathrm{mag}_{\rm lim}$.
}
    \label{fig:imaging_proxy_hd63433}
\end{figure*}

The fraction of realizations satisfying this criterion
defines the imaging sensitivity at that mass and
semi-major axis. We evaluate this fraction on the same
logarithmically spaced grid used for the RV calculation
with 200 values along each axis. This procedure applies
to contrast curves only. The published mass limits do not require
a new photometric conversion. Published completeness
maps retain the original authors' orbital and evolutionary
model assumptions. We illustrate the contrast curve
calculation in Figure~\ref{fig:imaging_proxy_hd63433}
using the VLT observations of HD~63433.

\section{Published Radial Velocity Inputs}
\label{app:rv_inputs}
Table~\ref{tab:rv_inputs_two_systems} summarizes the published radial velocity data adopted for the sensitivity calculations, including the measurements used, observing baseline, effective scatter, and literature source for each system.

\startlongtable
\begin{deluxetable*}{llcccc}
\tablecaption{Published radial velocity inputs adopted for the sensitivity calculations\label{tab:rv_inputs_two_systems}}
\tablewidth{0pt}
\tabletypesize{\footnotesize}
\tablehead{
\colhead{System} &
\colhead{RV data} &
\colhead{$N_{\rm RV}/N_{\rm used}$\tablenotemark{a}} &
\colhead{$T_{\rm RV}$ (yr)} &
\colhead{$\sigma_{\rm eff}$ (m\,s$^{-1}$)\tablenotemark{b}} &
\colhead{Ref.}
}
\startdata
HIP~5763
& CHIRON
& 19/19
& 2.07
& 16.2
& 1 \\
HIP~7441 (BD$+55\,362$)
& SOPHIE+
& 22/26\tablenotemark{c}
& 8.35
& 3.75
& 2 \\
HIP~10245 (BD$+45\,564$)
& SOPHIE+
& 14/16\tablenotemark{d}
& 8.36
& 3.30
& 2 \\
HIP~11048 (GJ~96)
& SOPHIE+
& 72/72\tablenotemark{e}
& 5.53
& 3.37
& 3 \\
HIP~12186 ($\lambda^2$~For)
& AAT/UCLES; Keck/HIRES
& 60/60\tablenotemark{f}
& 3.10
& 2.6
& 4 \\
HIP~20625 (HD~28471) &
HARPS (3 regimes) &
122/122\tablenotemark{g} &
19.33 &
2.31/0.63/1.49\tablenotemark{h} &
5 \\
HIP~24186 (Kapteyn's Star) &
HARPS/HIRES/PFS &
135/135 &
15.24 &
0.88/2.06/0.68\tablenotemark{i} &
6 \\
HIP~30905 (HD~45652) &
ELODIE/CORALIE/SOPHIE/HIRES &
49/50\tablenotemark{j} &
9.87 &
18.16 &
7 \\
HIP~37284 (HD~63454) &
HARPS &
34/34 &
6.064 &
6.84\tablenotemark{k} &
8 \\
HIP~38228 (HD~63433) &
ELODIE/SOPHIE/Ham./CARMENES/HARPS-N &
279/279\tablenotemark{l} &
25.13 &
37/24/3.95/24.1\tablenotemark{m} &
9--11 \\
HIP~39242 (HD~67200) &
HARPS (2 regimes) &
114/115\tablenotemark{n} &
6.25 &
3.01/3.16\tablenotemark{o} &
12,13 \\
HIP~54491 (HD~96735) &
HIRES/APF &
78/78\tablenotemark{p} &
2.14 &
3.16/8.0\tablenotemark{q} &
14 \\
HIP~65859 (GJ~514) &
HIRES/HARPS/CARM. &
540/540\tablenotemark{r} &
23.99 &
3.8/2.8/2.5/2.6 &
15,16 \\
HIP~75181 ($\nu^2$~Lupi) &
HARPS/HIRES/PFS/UCLES &
463/463\tablenotemark{s} &
21.9 &
1.19--3.89\tablenotemark{t} &
17 \\
HIP~81022 (HD~149143) &
HIRES/ELODIE &
58/58\tablenotemark{u} &
12.58 &
11.80/13.3\tablenotemark{v} &
7,18 \\
HIP~85523 (GJ~674) &
HARPS (pre/post) &
256/256\tablenotemark{w} &
15.42 &
2.79/2.86\tablenotemark{x} &
19 \\
HIP~86214 (GJ~682) & UVES/HARPS &
61/61\tablenotemark{y} & 6.16 &
4.00/2.28\tablenotemark{z} & 20,21 \\
HIP~91258 (BD$+61\,1762$) &
SOPHIE+ &
27/27\tablenotemark{aa} &
0.399 &
5.97 &
22 \\
HIP~93858 (HD~177565) &
HARPS &
68/68\tablenotemark{ab} &
4.61 &
3.34 &
23 \\
HIP~94256 (HD~179079) &
Keck/HIRES &
93/93\tablenotemark{ac} &
15.13 &
4.19 &
24 \\
HIP~104780 (BD$+14\,4559$) &
HET/HRS &
43/43\tablenotemark{ad} &
3.46 &
11.43 &
25 \\
HIP~108513 (HD~208897) &
TUG/CES; HIDES-F1/F2 &
86/86\tablenotemark{ae} &
11.85 &
12.776\tablenotemark{af} &
26 \\
HIP~113238 (HD~216770) &
CORALIE &
16/15\tablenotemark{ag} &
2.264 &
7.8 &
27 \\
HIP~114046 (GJ~887) &
HARPS (3 regimes)/ESPRESSO &
289/293\tablenotemark{ah} &
18.06 &
3.66/4.00/2.60/1.41\tablenotemark{ai} &
28 \\
HIP~118319 (HD~224693) &
Keck/HIRES &
40/40\tablenotemark{aj} &
12.07 &
5.73 &
7 \\
HIP~15510 (HD~20794) &
HARPS/ESPRESSO &
806/806\tablenotemark{ak} &
20.52 &
0.93/0.99/0.72 &
30 \\
HIP~34730 (LHS~1903) &
HARPS-N &
91/91\tablenotemark{al} &
2.38 &
3.83 &
31 \\
HIP~50921 (HD~90156) &
HARPS/HIRES (pre/post) &
\nodata/233\tablenotemark{am} &
23.17 &
1.23/2.46/2.23\tablenotemark{an} &
29,32 \\
HIP~109384 &
SOPHIE/SOPHIE+ &
42/42\tablenotemark{ao} &
7.60 &
3.7/5.8 &
33 \\
\enddata

\end{deluxetable*}

\begingroup
\footnotesize
\setlength{\parindent}{0pt}
\setlength{\parskip}{2pt plus 1pt}
\def\tablenotetext#1#2{%
  \par\noindent
  \hangindent=2.5em
  \hangafter=1
  \makebox[2.5em][l]{\textsuperscript{\normalfont\itshape #1}}#2\par
}
\def\tablerefs#1{%
  \par\medskip\noindent\textbf{References.}---#1\par
}
\noindent\textit{Notes to Table~\ref{tab:rv_inputs_two_systems}.}\par

\tablenotetext{a}{$N_{\rm RV}$ is the number of measurements reported for the published analysis. $N_{\rm used}$ is the number of archived literature measurements used in our cadence calculation.}

\tablenotetext{b}{Published post-fit residual RMS, residual dispersion, or fitted jitter adopted as the effective RV scatter, as specified in the individual notes from what was available in the literature.}

\tablenotetext{c}{Reference 2 reports 22 fitted measurements, but its electronic table contains 26 timestamps over the same 3050.81-day baseline. Because the four excluded measurements are not identified, all 26 timestamps were used as a cadence proxy.}

\tablenotetext{d}{Reference 2 reports 14 fitted measurements, but electronic table contains 16 timestamps over the same 3052.81-day baseline. Because the two excluded measurements are not identified, all 16 timestamps were used as a cadence proxy.}
\tablenotetext{e}{Reference 3 reports that 79 spectra were obtained
and eight observations were rejected, whereas its published Table~A.1 contains 72 radial velocities. We use all 72 explicitly tabulated epochs in the  calculation.}
\tablenotetext{f}{Reference 4 tabulates 88 velocities spanning 10.61~yr. However, its combined orbital solution and reported $2.6$~m\,s$^{-1}$ residual RMS use 60 measurements: 50 AAT/UCLES
observations obtained after 2005 July 19 and 10 Keck/HIRES observations. We use these 60 self-consistent epochs in the
sensitivity calculation.}
\tablenotetext{g}{
All 122 published S-BART measurements were used. Following
\citet{2025MNRAS.543...28S}, the observations were divided
into 23 pre-upgrade, 81 post-upgrade, and 18 post-COVID
HARPS measurements. 
}

\tablenotetext{h}{
The three values correspond to the pre-upgrade,
post-upgrade, and post-COVID HARPS regimes. For each regime,
$\sigma_{\rm eff}$ was calculated by adding in quadrature
the representative S-BART formal uncertainty from Table~1
of \citet{2025MNRAS.543...28S} (0.54, 0.41, and
0.58~m\,s$^{-1}$) and the fitted additive white noise term
from their Table~6 (2.25, 0.48, and 1.37~m\,s$^{-1}$).
The resulting values are 2.31, 0.63, and
1.49~m\,s$^{-1}$, respectively. 
}
\tablenotetext{i}{Values of $\sigma_{\rm eff}$ are listed in
HARPS/HIRES/PFS order. The published data comprise 95 HARPS,
32 HIRES, and 8 PFS measurements. For each instrument, we
combined the median formal RV uncertainty (0.59, 1.695, and
0.605~m\,s$^{-1}$) with the fitted additional white noise
term from Table~2 of \citet{2014MNRAS.443L..89A} (0.65,
1.17, and 0.32~m\,s$^{-1}$) in quadrature.}
\tablenotetext{j}{Reference 7 reports 49 measurements in the combined fit. Its electronic table contains 13 Keck/HIRES epochs, while the earlier ELODIE, CORALIE, and SOPHIE tables reconstruct 37 observing nights. Because the one omitted historical epoch cannot be identified, all 50 epochs were used as a cadence proxy. The resulting timestamps span 9.87~yr. Reference 7 reports a post-fit residual RMS of
18.16~m\,s$^{-1}$.}
\tablenotetext{k}{Reference 8 combines the 26 discovery measurements with eight additional HARPS observations. The fit includes a linear trend of
$-3.95\pm0.95$~m\,s$^{-1}$\,yr$^{-1}$. }
\tablenotetext{l}{The adopted data comprise 26 ELODIE, SOPHIE, and Hamilton measurements from Reference 9, 150 CARMENES VIS measurements from Reference 10, and 103 HARPS-N measurements from Reference 11.  Reference 10 reports that 157 CARMENES spectra were collected. Seven were rejected, leaving 150 VIS radial
velocities.}
\tablenotetext{m}{Values of $\sigma_{\rm eff}$ are listed in
ELODIE/SOPHIE/Hamilton/
CARMENES VIS/HARPS-N order.
For CARMENES, the published mean formal uncertainty
of 3.9~m\,s$^{-1}$ was combined in quadrature with the fitted 0.637~m\,s$^{-1}$ jitter yielding
3.95~m\,s$^{-1}$. For HARPS-N, we adopted the published 24.1~m\,s$^{-1}$ raw RMS of the
velocities.}
\tablenotetext{n}{Reference 12 reports 114 HARPS radial velocity measurements. We recovered 115 usable
\texttt{STAR,WAVE} exposure times from the ESO archive over the same observing interval. Because the excluded exposure is not identified, all 115 timestamps were used as a cadence
proxy, with 95 post-2015 and 20 post-2020
measurements.}
\tablenotetext{o}{Values of $\sigma_{\rm eff}$ are listed for
the post-2015 and post-2020 HARPS regimes, respectively.
From the GP-only Model 3a in Reference 13, we combined the
published RV GP amplitude of 2.91~m\,s$^{-1}$ with the
corresponding additive white noise terms of 0.78 and
1.24~m\,s$^{-1}$ in quadrature, obtaining 3.01 and
3.16~m\,s$^{-1}$.}
\tablenotetext{p}{Reference 14 does not explicitly state a value of $N_{\rm RV}$ for this system. Its machine-readable
data release contains 79 records for CPS~96735. One HIRES
record contains no radial velocity or uncertainty, leaving
78 usable measurements: 66 HIRES and 12 APF velocities.}
\tablenotetext{q}{Values are listed as HIRES/APF and are the
fitted instrumental jitter terms reported in Table~6 of
Reference 14.}
\tablenotetext{r}{Reference 15 reports 104 HIRES,
142 pre-upgrade HARPS, 20 post-upgrade HARPS, and
274 CARMENES-VIS velocities. The listed noise values are
the corresponding RMS values from its Table~2. The HIRES epochs
were recovered from Reference 16.}
\tablenotetext{s}{Reference 17 reports 463 nightly binned
velocities: 7 HARPS-post, 234 HARPS-pre, 29 HIRES-post,
3 PFS-post, 21 PFS-pre, and 169 UCLES measurements. All
463 tabulated epochs were used in the calculation.}
\tablenotetext{t}{
We combined the jitter from Table~2 of the HD~136352 supplementary report
in Reference~17 with the median formal RV uncertainty
for each instrumental regime from Table~6 in quadrature.
The resulting effective scatters are 1.19, 1.28, 2.55,
1.43, 1.27, and 3.89~m\,s$^{-1}$ for post-upgrade HARPS,
pre-upgrade HARPS, post-upgrade HIRES, post-upgrade PFS,
pre-upgrade PFS, and UCLES, respectively.}
\tablenotetext{u}{Reference 7 reports 58 measurements in the
combined fit. Its Table~2 contains 50 HIRES
epochs, with the remaining 8 ELODIE measurements
tabulated in Reference 18. All 58 epochs were used.}
\tablenotetext{v}{The system-level post-fit residual RMS of 11.80~m\,s$^{-1}$ from Reference 7 was applied to HIRES. The ELODIE block was assigned the published post-fit RMS of 13.3~m\,s$^{-1}$ from Reference 18.}
\tablenotetext{w}{Reference 19 reports 256 HARPS spectra. We recovered the corresponding observation timestamps from ESO archive metadata using the observing programs listed in its Table~2. The adopted cadence comprises 180 pre-upgrade and 76 post-upgrade measurements spanning 15.42~yr.}
\tablenotetext{x}{Values are listed for the pre- and
post-upgrade HARPS regimes. Because Reference 19 does not report a single post-fit residual RMS, we combined its fitted
GP amplitudes of 2.75 and 2.80~m\,s$^{-1}$ with the
corresponding jitter terms of 0.48 and
0.60~m\,s$^{-1}$ in quadrature, obtaining
$\sigma_{\rm eff}=2.79$ and 2.86~m\,s$^{-1}$, respectively.}
\tablenotetext{y}{The calculation uses 49 UVES exposure epochs and
12 HARPS epochs.}
\tablenotetext{z}{For UVES, we adopted the published
4.0~m\,s$^{-1}$ RV scatter from Reference 21. The HARPS value is calculated from the velocities in Table~B17 of Reference~20. Reference~20 reports a significant linear trend but does not report a slope.}
\tablenotetext{aa}{Reference 22 reports 27 SOPHIE+
measurements spanning 145.8~days. All 27 epochs in
their Table~7 were used. We adopted the published
5.97~m\,s$^{-1}$ post-fit residual RMS.}
\tablenotetext{ab}{Reference 23 tabulates 68
HARPS velocities spanning 1684.39~days. Because it does not
report a single post-fit residual RMS, we adopted the
3.34~m\,s$^{-1}$ RMS calculated directly from the published
velocities.}
\tablenotetext{ac}{Reference 24 tabulates 93 Keck/HIRES
velocities obtained between 2004 July 11 and 2019 August 28. All 93 exact epochs from its machine-readable Table~A3 were used. We adopted the 4.19~m\,s$^{-1}$ residual RMS of the AICc-favored one planet model without
a linear trend from its Table~13. Reference 24 also
provides a formal RVSearch injection recovery map.}
\tablenotetext{ad}{Reference 25 tabulates all 43 HET/HRS
radial velocities and their observing epochs in its
Table~2. The observations span MJD 53546.31624 to
54811.05137, about 3.46~yr. All 43 epochs
were used in the cadence calculation, with the
published post-fit residual RMS of 11.43~m\,s$^{-1}$
adopted as $\sigma_{\rm eff}$. }
\tablenotetext{ae}{Reference 26 reports 86 nightly binned
radial velocities used in the updated fit: 39 TUG/CES,
30 HIDES-F1, and 17 HIDES-F2 epochs. The supplementary
data contain 129 individual exposures, which we binned by
observing night/instrument to reproduce the
published sample size. The resulting cadence spans
11.85~yr.}
\tablenotetext{af}{Reference 26 reports a global post-fit
residual RMS of 12.776~m\,s$^{-1}$ in its supplementary
orbital parameter table. We adopted this published value as
$\sigma_{\rm eff}$ for all three instrumental regimes.}
\tablenotetext{ag}{Reference 27 reports 16 CORALIE
measurements spanning 827~days and a post-fit residual RMS of
7.8~m\,s$^{-1}$. The electronic RV table contains one exactly
duplicated observing epoch, so the cadence calculation
uses 15 unique timestamps.}
\tablenotetext{ah}{Reference 28 reports 277 nightly binned
HARPS velocities and 12 ESPRESSO velocities, for a total of
289 measurements. Its electronic tables contain 356 unbinned
HARPS measurements, corresponding to 282 observing nights,
and 11 ESPRESSO measurements. Because the measurements
excluded from the published analysis are not identified, we
used the 293 recoverable observing epochs as a cadence proxy.
The cadence spans 18.06~yr.}
\tablenotetext{ai}{Values of $\sigma_{\rm eff}$ are listed in
HARPSpre/HARPSpost/
HARPSpw/ESPRESSO order. Because Reference
28 does not report a single residual RMS, we combined
the fitted Gaussian process amplitude and jitter for
each data set in quadrature, obtaining 3.66, 4.00, 2.60, and
1.41~m\,s$^{-1}$, respectively.}
\tablenotetext{aj}{Reference 7 provides 40 Keck/HIRES radial
velocities in its machine-readable Table~2. These include all
24 measurements originally published by
\citet{2006ApJ...647..600J} and 16 subsequent measurements,
extending the cadence to 12.07~yr. All 40 epochs were used.
The published post-fit residual RMS of
5.73~m\,s$^{-1}$ was adopted as $\sigma_{\rm eff}$.}

\tablenotetext{ak}{The adopted cadence comprises 512
pre-upgrade HARPS (H03), 231 post-upgrade HARPS (H15),
and 63 ESPRESSO (E19) nightly measurements from
Reference~30. The retained ESPRESSO nights were identified
by matching the released observations to Figures~1 and~3a
of that reference. The combined baseline is 20.52~yr.
Values of $\sigma_{\rm eff}$ are the published post-fit
residual RMS values in H03/H15/E19 order.}
\tablenotetext{al}{Reference 31 reports 91 retained HARPS-N
measurements from 108 acquired spectra. We used all 91
published epochs, spanning 869.86 days. This agrees with
the approximately 870 day baseline in the supplementary
observing section, although the main text states 769 days.
We combined the fitted GP amplitude of 3.67~m\,s$^{-1}$
and additive jitter of 1.08~m\,s$^{-1}$ from Table~S7
in quadrature, obtaining $\sigma_{\rm eff}=3.83$~m\,s$^{-1}$.}
\tablenotetext{am}{We used 66 HARPS measurements from
\citet{2011AA...526A.111M} and 167 HIRES measurements
from the data release of \citet{2021ApJS..255....8R},
comprising 16 pre-upgrade and 151 post-upgrade
measurements. The combined cadence spans 23.17~yr.
The 233 measurements are our combined literature input. }

\tablenotetext{an}{Values are listed in
HARPS/HIRES-pre/HIRES-post order. For HARPS, we adopted
the post-fit residual RMS of 1.23~m\,s$^{-1}$ from
Table~3 of \citet{2011AA...526A.111M}. For HIRES, we
combined the median formal uncertainties of 1.26 and
1.25~m\,s$^{-1}$ with the fitted jitter values of
2.12 and 1.85~m\,s$^{-1}$ in quadrature, obtaining
2.46 and 2.23~m\,s$^{-1}$, respectively. The jitter
values are the \texttt{jit\_k} and \texttt{jit\_j}
entries for HD~90156 in the released
\texttt{system\_props.csv} table of
\citet{2021ApJS..255....8R}.}
\tablenotetext{ao}{We used all 42 measurements in the
electronic Table~1 of \citet{2016AA...588A.145H},
comprising seven SOPHIE and 35 SOPHIE+ observations.
The published epochs span 2775.66~days (7.60~yr).
$\sigma_{\rm eff}$ are the post-fit residual
dispersions reported in their Table~3.}
\tablerefs{
(1) \citealt{2021AJ....162..176P};
(2) \citealt{2021AA...651A..11D};
(3) \citealt{2018AA...618A.103H};
(4) \citealt{2009ApJ...697.1263O};
(5) \citealt{2025MNRAS.543...28S};
(6) \citealt{2014MNRAS.443L..89A};
(7) \citealt{2018AJ....156..213M};
(8) \citealt{2011ApJ...737...58K};
(9) \citealt{2020AJ....160..179M};
(10) \citealt{2023AA...671A.163M};
(11) \citealt{2023AA...672A.126D};
(12) \citealt{2026MNRAS.547ag370S};
(13) \citealt{2026MNRAS.tmp.1422B};
(14) \citealt{2024ApJS..272...32P};
(15) \citealt{2022AA...666A.187D};
(16) \citealt{2019MNRAS.484L...8T};
(17) \citealt{2025AJ....170..343H};
(18) \citealt{2006AA...446..717D};
(19) \citealt{2024AA...690A.234L};
(20) \citealt{2014MNRAS.441.1545T};
(21) \citealt{2009AA...505..859Z};
(22) \citealt{2014AA...563A..22M};
(23) \citealt{2017MNRAS.470.4794F};
(24) \citealt{2020AJ....159..197H};
(25) \citealt{2009ApJ...707..768N};
(26) \citealt{2023PASJ...75.1030T};
(27) \citealt{2004AA...415..391M};
(28) \citealt{2026AA...707A..93H};
(29) \citealt{2011AA...526A.111M};
(30) \citealt{2025AA...693A.297N};
(31) \citealt{2026Sci...392l2348W};
(32) \citealt{2021ApJS..255....8R};
(33) \citealt{2016AA...588A.145H}.
}
\endgroup

\clearpage
\section{Published Imaging Inputs}
\label{app:imaging_inputs}
Here we present the input contrast curves or detection limits
for targets where quantitative imaging data were available
in the literature. The inputs are listed in
Table~\ref{tab:imaging_inputs}.

\begin{deluxetable*}{lllllc}
\tablecaption{Published imaging inputs adopted for the sensitivity calculations
\label{tab:imaging_inputs}}
\tablewidth{0pt}
\tabletypesize{\footnotesize}
\tablehead{
\colhead{System} &
\colhead{Imaging data} &
\colhead{Epoch(s)} &
\colhead{Published product} &
\colhead{Mass treatment} &
\colhead{Ref.}
}
\startdata
HIP~75181 ($\nu^2$~Lupi) &
VLT/SPHERE &
2017, 2020 &
MESS3 map &
Published\tablenotemark{a} &
1 \\
HIP~85523 (GJ~674) &
VLT/SPHERE &
2017, 2021 &
MESS3 map &
Published\tablenotemark{a} &
1 \\
HIP~86214 (GJ~682) &
VLT/SPHERE &
2017, 2019 &
MESS3 map &
Published\tablenotemark{a} &
1 \\
HIP~38228 (HD~63433) &
VLT/NaCo $K$ &
2004-01-16 &
$5\sigma$ contrast curve &
Models\tablenotemark{b} &
2 \\
HIP~15510 (HD~20794) &
VLT/SPHERE &
2017 &
MESS3 map &
Published\tablenotemark{a} &
1 \\
HIP~34730 (LHS~1903) &
Palomar/PHARO Br$\gamma$ &
2020-11-05 &
$5\sigma$ contrast curve &
Models\tablenotemark{c} &
3 \\
\enddata

\tablenotetext{a}{
For these three systems, we used the target
10\%, 50\%, and 90\% completeness contours from
Figure~H.2 of Reference~1. These published maps already
incorporate the imaging data and evolutionary model
conversion. For HD~20794, the published calculation adopts an age of 6.0~Gyr; Table~D.1 lists observations on 2017 September 3 and 4.
}

\tablenotetext{b}{
We digitized only the VLT curve from Figure~5 of Reference~2.
We used the Sonora Red Diamondback v2 and BHAC15 models
\citep{2025ApJ...994..198D,2015AA...577A..42B},
adopting an age of $414\pm23$~Myr and model $K_s$ as a
proxy for the published $K$ band. The observing date
was adopted from \citet{2020AJ....160..179M}.
}

\tablenotetext{c}{
We used the machine-readable PHARO contrast curve shown
in Figure~S8 of Reference~3. We adopted a system distance of 35.68~pc and apparent stellar magnitude $K_s=8.21$~mag from their Table~S3.
Contrasts were converted using Red Diamondback v2 and
BHAC15, with model $K_s$ as a proxy for Br$\gamma$.
Gaussian draws from the published isochronal age estimate
of $4.9\pm4.0$~Gyr were restricted to the adopted
0.0005--10~Gyr model range.
}

\tablerefs{
(1) \citealt{2023AA...680A..64D};
(2) \citealt{2024AJ....167...54C};
(3) \citealt{2026Sci...392l2348W}.
}
\end{deluxetable*}

\software{Octofitter \citep{2023AJ....166..164T}, \href{https://github.com/jay3332/pilmoji}{Pilmoji},
\href{https://pypi.org/project/plotdigitizer/}{plotdigitizer}, \texttt{astroquery}
\citep{2019AJ....157...98G}.}

\bibliography{sample701}{}
\bibliographystyle{aasjournalv7}

\end{document}